\documentclass[pre,preprint,showpacs,groupedaddress,repreint]{revtex4-1}

\usepackage{graphicx}% Include figure files
\usepackage{dcolumn}% Align table columns on decimal point
\usepackage{bm}% bold math
\usepackage{amsmath}
\usepackage{hyperref}% add hypertext capabilities
\usepackage{color}
\usepackage{amssymb}
\makeatother
\usepackage{tikz}
\begin{document}

%\preprint{}

\title{Transport and diffusion of paramagnetic ellipsoidal particles in a rotating magnetic field}


%\homepage[]{}

%\thanks{}

\author{Jing-jing Liao$^{1,2}$}
%\author{Xiao-qun Huang$^{1}$}
\author{Wei-jing Zhu$^{1}$}
\author{Bao-quan  Ai$^{1}$}\email[Email: ]{aibq@scnu.edu.cn}
 %\email[Email: ]{wrzhong@jnu.edu.cn}
%\homepage[]{}

%\thanks{}
\affiliation{$^{1}$ Guangdong Provincial Key Laboratory of Quantum Engineering and Quantum Materials, School of Physics and Telecommunication
Engineering, South China Normal University, Guangzhou 510006, China.}
\affiliation{$^{2}$ College of Applied Science, Jiangxi University of Science and Technology, Ganzhou 341000, China.}

\date{\today}
\begin{abstract}
\indent Transport and diffusion of paramagnetic ellipsoidal particles under the action
of a rotating magnetic field are numerically investigated in a
two-dimensional channel. It is found that paramagnetic ellipsoidal particles in a rotating magnetic field can be rectified in the upper-lower asymmetric channel. The transport and the effective diffusion coefficient are much more different and complicated for active
particles, while they have similar behaviors and change a little when
applying rotating magnetic fields of different frequencies for passive
particles. For active particles, the back-and-forth rotational motion
facilitates the effective diffusion coefficient and reduces the rectification,
whereas the rotational motion synchronous with the magnetic field
suppresses the effective diffusion coefficient and enhances the rectification.
There exist optimized values of the parameters (the anisotropic degree, the amplitude and frequency of magnetic field, the self-propelled velocity, and the rotational diffusion rate) at which the average velocity and diffusion take their
maximal values. Particles with different shapes, self-propelled speeds, or
rotational diffusion rates will move to the opposite directions and can be
separated by applying rotating magnetic fields of suitable strength and frequency. Our results can be used to separate particles, orient the particles
along any direction at will during motion, and control the particle diffusion.
\end{abstract}

%\pacs{05. 40. Fb, 02. 50. Ey, 05. 40. -a}
%\keywords{ chiral active particles, transport, V-shape barrier }



\maketitle
\section{Introduction}
\indent  Recently, transport and diffusion of Brownian particles in periodic structures have attracted considerable attention in biology, chemistry, and physics \cite{ref1,ref2,ref3,ref4,ref5,ref6,ref7,ref8,ref9,ref10,ref11}. Different from passive colloids, self-propelled Brownian particles can produce a force which pushes them forward due to an internal mechanism which may be based on a light stimulus (thermophoresis) or concentration gradients (diffusophoresis) \cite{ref12}. When colloidal particles are made using (or coated with) magnetic materials, they become responsive to external fields and show fascinating dynamical behavior which is of interest for both fundamental and technological aspects \cite{ref13,refAC,ref14,ref15,ref16,ref38,ref22,ref23,ref24,ref25,ref26,ref27,ref28}. For example, circular motion of an active magnetic particle in a rotating magnetic field \cite{refAC}; accumulating and clustering of magnetotactic bacteria under the effect of an external magnetic field \cite{ref14,ref15}; extracting work from magnetic-field-coupled Brownian particles \cite{ref16}; fabrication and actuation of composites materials with magnetic torque \cite{ref38}; manipulating and detecting biomaterials with magnetic torque \cite{ref38} and other interesting transport and diffusion phenomena \cite{ref22,ref23,ref24,ref25,ref26,ref27,ref28}.

\indent  However, there exit few natural particles which have the perfect spherical symmetry. In contrast to isotropic particles, anisotropic magnetic colloids are characterized by an induced or spontaneous magnetization which has a dependence on a particular direction, feature the advantages of being easily torqued by an external field, and attract considerable interest in theoretical \cite{ref18,ref29,ref30,ref31,ref39} and experimental studies \cite{ref17,ref21,ref40,ref32,ref33,ref34,ref19,ref20}. Martin \cite{ref18} studied a vortex magnetic field that can induce strong mixing in a magnetic particle suspension theoretically. Marino \textit{et al}. \cite{ref29} showed the rotational Brownian motion of colloidal particles in the overdamped limit generates an additional contribution to the "anomalous" entropy. G\"{u}ell and coworkers \cite{ref30} analyzed the diffusive properties of a paramagnetic passive particle torqued by a rotating magnetic field in the case of thermal noise considered. Fan and coworkers \cite{ref31} studied the diffusion of an ellipsoidal self-driven particle under the effect of magnetic field, self-propulsion, and the particle's shape by finding approximated analytical expressions. Matsunaga and coworkers \cite{ref39} showed how magnetic particles can be focused and segregated by size and shape by using a far-field hydrodynamic theory and simulations. Liu and their coworkers \cite{ref17} experimentally showed separating superparamagnetic particles with a size difference around 130 nm by using periodically switching magnetic fields and asymmetric sawtooth channel sidewalls. Gao \textit{et al}. \cite{ref21}. demonstrated that magnetically driven nanoswimmers provided a new approach for the rapid delivery of target-specific drug carriers to predetermined destinations. Hamilton \textit{et al}. \cite{ref40} demonstrated the experimental verification of a new class of autonomous ferromagnetic swimming devices, actuated and controlled solely by an oscillating magnetic field.

\indent  We are motivated by one of the previous experimental results describing the dynamics of paramagnetic ellipsoids in an external rotating magnetic field \cite{ref19}. The researchers of the experiment found the ellipsoid mean rotation frequency and orientational angle of the elongated particles can be controlled by changing the frequency or strength of the applied field. They reported on subsequently a new technique for orienting self-driven microellipsoids by using an external field and the magnetic anisotropy of the ellipsoids \cite{ref20}. In other studies related to the present work, how anisotropic magnetic particles under the action of an external field are rectified and diffuse in an asymmetric channel has not been considered yet, which results in peculiar behavior. In this paper, we consider paramagnetic ellipsoidal particles move in a two-dimensional asymmetric channel and are subject to an external magnetic field. We analyze transport and diffusion of paramagnetic ellipsoidal particles numerically. To focus on finding how the external magnetic field influences the transport and diffusion, we compare the transport and diffusion between the cases of active and passive particles. It is found that paramagnetic ellipsoidal particles in a rotating magnetic field can be rectified in the upper-lower asymmetric channel. The transport and the effective diffusion coefficient are much more different and complicated for active particles, while they have similar behaviors and change a little when applying rotating magnetic fields of different frequencies for passive particles. For active particles, there exist optimized values of the parameters (the anisotropic degree, the amplitude and frequency of magnetic field, the self-propelled velocity, and the rotational diffusion rate) at which the rectification and diffusion take their maximal values. Particles with different shapes, self-propelled speeds or rotational diffusion rates will move to the opposite directions and can be separated by applying rotating magnetic fields of suitable strength and frequency. Our results can be employed in several applications, such as particle separation, drug release and migration of contaminants in porous media.

\section{Model and Methods}
\indent  Inspired by finding transport and diffusion of paramagnetic ellipsoidal particles in the above experiment \cite{ref19,ref20}, we consider noninteracting paramagnetic ellipsoidal particles of semimajor axis $a$ and semiminor axis $b$, with preferred magnetization along semimajor axis. The particles move in a two-dimensional asymmetric channel and are subject to an external magnetic field of the form ${\bf{H}}(t)=H_{0} [\cos (\omega _{H} t),\sin (\omega _{H} t)]$ with angular frequency $\omega _{H} $ and amplitude $H_{0} $ (shown in Fig. 1). The particle at a given time $t$ can be described by the position vector ${\bf{R}}(t)$ of its center of mass, which can be decomposed as $(\delta \hat{x},\delta \hat{y})$ in the body frame and $(\delta x,\delta y)$ in the laboratory frame. $\theta (t)$ is the angle between the two frames. The swimming velocity (along its semimajor axis) of the self-propulsion particle is defined as the form ${\bf{v}}=v_{0} {\bf{e}}(t)$, where ${\bf{e}}(t)=[\cos \theta ({\it t}),\sin \theta ({\it t})]$ is the instantaneous unit vector in the direction of swimming with its origin at the center of the particle, and $v_{0} $ is the magnitude of the swimming velocity. To focus on finding the effect of external magnetic fields, we choose dilute particles, hydrodynamic interactions and particle interactions will be negligible. Rotational and translational motion in the body frame are always decoupled, so the dynamics of the paramagnetic particle can be governed by the Langevin equations in a low-Reynolds-number environment and in the body frame \cite{ref35,ref36,ref37},
\begin{figure*}[htbp]
\begin{center}
\includegraphics[width=16cm]{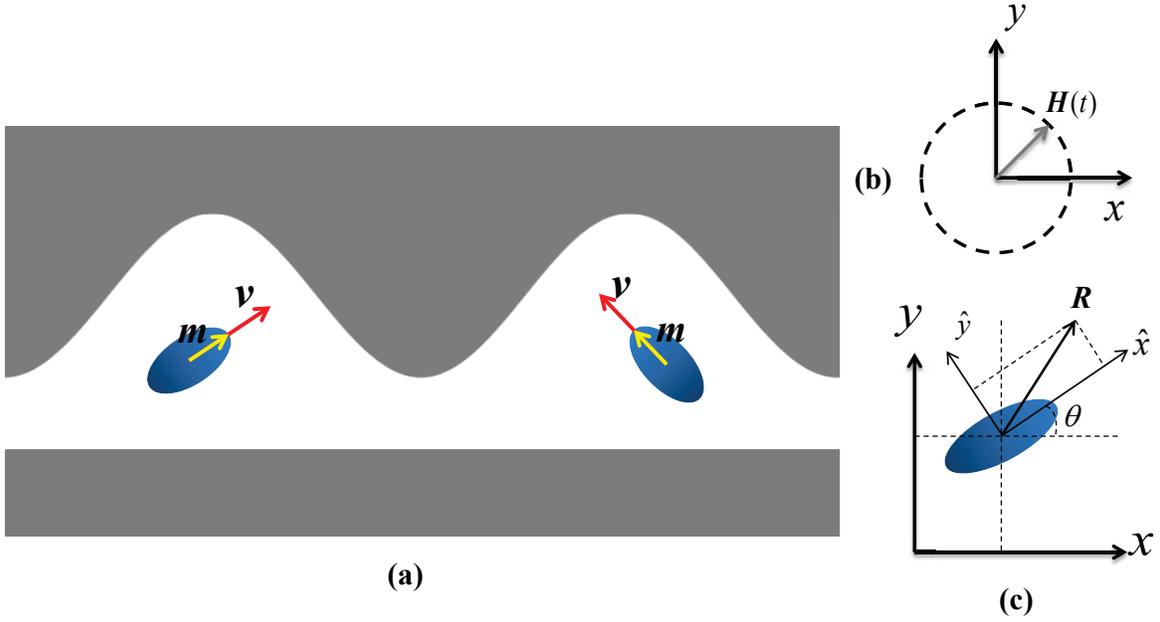}
\caption{Schematic of illustrating the system: (a) noninteracting paramagnetic Brownian swimmers with magnetic moment $\bf m$ and swimming velocity $\bf v$ moving in a two-dimensional asymmetric channel. (b) The swimmers are subject to a rotating magnetic field ${\bf{H}}(t)=H_{0} [\cos (\omega _{H} t),\sin (\omega _{H} t)]$. (c) Representation of an ellipsoid in the $x$-$y$ laboratory frame and the $\hat{x}$-$\hat{y}$ body frame. $\theta$ is the angle between two frames. $\bf{R}$ is the position of the particle.}
\end{center}
\end{figure*}
\begin{equation} \label{eq1}
\frac{d\hat{x}}{dt} =\Gamma _{x} [F_{x} \cos \theta (t)+F_{y} \sin \theta (t)+\hat{\xi }_{x} (t)]+v_{0} ,
\end{equation}
\begin{equation} \label{eq2}
\frac{d\hat{y}}{dt} =\Gamma _{y} [F_{y} \cos \theta (t)-F_{x} \sin \theta (t)+\hat{\xi }_{y} (t)],
\end{equation}
\begin{equation} \label{eq3}
\frac{d{\bf{e}}(t)}{dt} ={\bf \Omega }(t)\times {\bf e}(t),
\end{equation}
where $\Gamma _{x} $ and $\Gamma _{y} $ are the mobilities along its semimajor and semiminor axis, respectively. $F_{x} $ and $F_{y} $ are the forces along $x$ and $y$ direction of the laboratory frame. ${\bf \Omega }(t)$ is the swimmer's angular velocity. Equation (\ref{eq3}) can be further simplified by a torque-balance condition on the ellipse:
\begin{equation} \label{eq4}
{\boldsymbol{\tau}}_{m} +{\boldsymbol{\tau}}_{H} =\frac{\hat{{\boldsymbol{\xi}}}_{\theta } (t)}{\Gamma _{\theta } } ,
\end{equation}
where $\Gamma _{\theta } $ is the rotational mobility, ${\boldsymbol{\tau}}_{H} =-\frac{{\bf \Omega }(t)}{\Gamma _{\theta } } $ represents the hydrodynamic torque, ${\boldsymbol{\tau}}_{m} =\mu _{0} {\bf m}\times {\bf H}$ is the magnetic torque ($\mu _{0} $ is the magnetic susceptibility). The magnetic moment ${\bf{m}}$ is defined as ${\bf{m}}=V\underline{\chi }{ \bf{H}}$, where $V=\frac{4\pi }{3} ab^{2} $ and $\underline{\chi }$ is a second-order tensor representing the susceptibility of the ellipse. For the situation where there is a preferred magnetization direction, the susceptibility tensor may be expressed as $\underline{\chi }=\chi _{\bot } {\bf I}{\rm +}\Delta \chi {\bf{ee}}$, where $\Delta \chi =\chi _{\parallel } -\chi _{\bot } $ and $\chi _{\parallel } (\chi _{\bot } )$ is the susceptibility component parallel (normal) to ${\bf{e}}$. Equation (\ref{eq3}) finally leads to
\begin{equation} \label{eq5}
\frac{d\theta(t) }{dt} =\frac{\mu _{0} {\it V}\Delta \chi {\it H}_{0}^{2} \Gamma _{\theta } }{2} \sin [2(\omega _{H} t-\theta (t))]+\Gamma _{\theta } \hat{\xi }_{\theta } (t).
\end{equation}
Note that the noise $\hat{\xi }_{i} (t)$ has mean zero and satisfies
\begin{equation} \label{eq6}
\left\langle \hat{\xi }_{i} (t)\hat{\xi }_{j} (t')\right\rangle =\frac{2k_{B} T}{\Gamma _{i} } \delta _{i,j} \delta (t-t'),i,j=x,y,\theta ,
\end{equation}
where $T$ is the temperature and $k_{B} $ is the Boltzmann constant.

We now obtain these equations in the laboratory frame. For convenience, converting the equation of motion to the fixed laboratory frame by using the rotation matrix:
\begin{equation} \label{eq7}
R(\theta (t))\equiv \left(\begin{array}{l} {\cos \theta (t)\; \; \; -\sin \theta (t)} \\ {\sin \theta (t)\; \; \; \cos \theta (t)} \end{array}\right)\; ,
\end{equation}
The final set of equations describing the particle dynamics in the laboratory frame are:
\begin{equation} \label{eq8}
\frac{dx}{dt} =v_{0} \cos \theta (t)+F_{x} [\bar{\Gamma }+\Delta \Gamma \cos 2\theta (t)]+\Delta \Gamma F_{y} \sin 2\theta (t)+\xi _{x} (t),
\end{equation}
\begin{equation} \label{eq9}
\frac{dy}{dt} =v_{0} \sin \theta (t)+F_{y} [\bar{\Gamma }-\Delta \Gamma \cos 2\theta (t)]+\Delta \Gamma F_{x} \sin 2\theta (t)+\xi _{y} (t),
\end{equation}
\begin{equation} \label{eq10}
\frac{d\theta(t) }{dt} =\omega _{c} \sin {2[\omega _{H} t-\theta (t)]}+\xi _{\theta } (t),
\end{equation}
where the critical frequency $\omega _{c} =\frac{\mu _{0} {\it V}\Delta \chi {\it H}_{0}^{2} \Gamma _{\theta } }{2} $, the quantities $\bar{\Gamma }=\frac{1}{2} (\Gamma _{x} +\Gamma _{y} )$ and $\Delta \Gamma =\frac{1}{2} (\Gamma _{x} -\Gamma _{y} )$ are the average and difference mobilities of the body, respectively. The parameter $\Delta \Gamma $ determines the asymmetry of the body, the particle is a perfect sphere for $\Delta \Gamma =0$ and a very needlelike ellipsoid for $\Delta \Gamma \to \bar{\Gamma }$. $\xi _{\theta } (t)$ is a Gaussian random variable with zero mean and variance \cite{ref37}:
\begin{equation} \label{eq11}
\left\langle \xi _{\theta } (t)\xi _{\theta } (t')\right\rangle =2D_{\theta } \delta (t-t'),
\end{equation}
where $D_{\theta } =k_{B} T\Gamma _{\theta } $ is the rotational diffusion rate, which describes the nonequilibrium angular fluctuation. However, $\xi _{x} (t)$ and $\xi _{y} (t)$ are random Gaussian variables at fixed $\theta (t)$ with variance depending on the value of $\theta (t)$ \cite{ref37},
\begin{equation} \label{eq12}
\left\langle \xi _{i} (t)\xi _{j} (t')\right\rangle _{\theta (t)}^{} =2k_{B} T\Gamma _{ij} \delta (t-t'),
\end{equation}
with $i,j=x,y$ and
\begin{equation} \label{eq13}
\Gamma _{ij} [\theta (t)]=\bar{\Gamma }\delta _{ij} +\Delta \Gamma R[2\theta (t)]\cdot \left(\begin{array}{l} {1} \\ {0} \end{array}\right. \; \; \; \left. \begin{array}{l} {\; \; \; 0} \\ {-1} \end{array}\right),
\end{equation}
with $\Gamma _{ij} $ the mobility tensor.

Usually, one of the critical elements of ratchet setup in nonlinear systems is asymmetry (temporal and/or spatial), which can violate the left-right symmetry of the response \cite{ref41}. For our system, the asymmetry comes from the upper-lower asymmetry of the channel which is composed of the lower wall and the upper wall. Though the corrugated profile of the upper wall is left-right symmetric, the channel is upper-lower asymmetric. Coupling with the external field which induces the particles rotate, the upper-lower asymmetric channel can break the left-right symmetry and induce directed transport in the $x$ direction. The lower wall of the channel is fixed as $w_{l} (x)=0$, and the upper wall we choose corrugated structure [see Fig. 1(a)] can be described by
\begin{equation} \label{eq14}
w_{u} (x)=c[\sin (\frac{2\pi x}{L} )]+d,
\end{equation}
where $c$ and $d$ are the parameters that control the shape of the upper wall. $L$ is the periodicity of the channel. If we choose two simple walls, the channel is symmetric, and there is no directed transport.

In this paper, we use Brownian dynamic simulations performed by integration of the Langevin equations in the laboratory frame using the second-order stochastic Runge-Kutta algorithm. Because the particle along the $y$ direction is confined, we only calculate the $x$ direction average velocity based on Eqs. (\ref{eq8})-(\ref{eq10}),
\begin{equation} \label{eq15}
v_{\theta _{0} } =\mathop{\lim }\limits_{t\to \infty } \frac{\left\langle x(t)\right\rangle _{\theta _{0} }^{\xi _{x} ,\xi _{y} } }{t} ,
\end{equation}
where $\theta _{0} $ is initial angle of the trajectory. The full average velocity after another average over all $\theta _{0} $ is
\begin{equation} \label{eq16}
v=\frac{1}{2\pi } \int _{0}^{2\pi }d\theta _{0}  v_{\theta _{0} } ,
\end{equation}

The scaled average velocity $V_{s} $ is defined as $V_{s} =\frac{v}{v_{0} } $ for active particles, and the mobility $\mu$ is defined as $\mu =\frac{v}{f_{0} } $ for passive particles, where $f_{0}$ is the constant force applied at the particle along the $x$ direction of the laboratory frame. And the effective diffusion coefficient along $x$ direction  is
\begin{equation} \label{eq17}
D_{x} =\mathop{\lim }\limits_{t\to \infty } \frac{\left\langle x^{2} (t)\right\rangle -\left\langle x(t)\right\rangle ^{2} }{2t} ,
\end{equation}.
We use the scaled effective diffusion coefficient $D_{eff} =D_{x} /D_{free} $ for convenience, where $D_{free}=D_{0}+{v_0}^2/2D_{\theta}$.

\section{Numerical results and discussion}
In our numerical simulations, the total integration time was more than $10^{4} $ and the integration step time $dt$ was chosen to be smaller than $10^{-3} $. With these parameters, the simulation results are robust and do not depend on the time step and the integration time. Unless otherwise noted, we set $D_0=1.0$, $c=0.6$, $d=L=1.0$, $\bar{\Gamma }=0.02$, $D_{\theta } =0.1$, $v_{0} =3.5$ and $f_{0} =2.0$. Particle interactions are negligible. In the following discussion, we discuss the transport and diffusion for two cases: (A) rectification and diffusion of active particles ($v_{0} \ne 0$, $f_{0}=0$) and (B) mobility and diffusion of passive particles ($v_{0} =0$, $f_{0}\ne 0$).
\subsection{Rectification and diffusion of active particles ($v_{0} \ne 0$, $f_{0}=0$)}
The dependence of the scaled average velocity ${\it V}_{{\it s}} $ and the effective diffusion coefficient $D_{eff} $ on the anisotropic parameter $\Delta \Gamma $ of active particles are presented in Fig. 2. When the particles are without a magnetic field ($\omega _{c} =0$) or subject to a static magnetic field ($\omega _{H} =0,$ $\omega _{c} \ne 0$), shown in Fig. 2(a), ${\it V}_{{\it s}} $ decreases monotonically with the increase of the anisotropic parameter $\Delta \Gamma $. The larger $\omega _{c} $ is, the bigger ${\it V}_{{\it s}} $ is. When the particles are subject to a rotating magnetic field ($\omega _{H} \ne 0,$ $\omega _{c} \ne 0$), ${\it V}_{{\it s}} $ exhibits more complicated behavior. Figure 2(c) depicts the scaled average velocity ${\it V}_{{\it s}} $ as a function of $\Delta \Gamma $ for different values of $\omega _{c} $ and $\omega _{H} $ at $v_{0} =2.0$. Figure 2(e) shows ${\it V}_{{\it s}} $ as a function of $\Delta \Gamma $ for different values of $v_{0} $ at $\omega _{c} =1.5$ and $\omega _{H} =2.0$. We can find ${\it V}_{{\it s}} $ increases as $\Delta \Gamma $ increases for some values while ${\it V}_{{\it s}} $ is a peaked function of $\Delta \Gamma $ for other values. From Eq. (\ref{eq8}), the self-propelled velocity $v_{0} $, the critical frequency $\omega _{c} $, the magnetic frequency $\omega _{H} $ and the anisotropic parameter $\Delta \Gamma $ compete with each other. When $0.75\le \omega _{c} /\omega _{H} \le 1$ and $v_{0} =2.0$, $3.5$, ${\it V}_{{\it s}} $ is negative for suitable $\Delta \Gamma $. Therefore, for appropriate $\Delta \Gamma $, particles with different values of $v_{0} $ and subject to rotating magnetic fields of different frequencies and amplitudes move to different directions and can be separated. However, the effective diffusion coefficient $D_{eff} $ decreases with increasing $\Delta \Gamma $ for $\omega _{c} /\omega _{H} >1$ or $v_{0} <2.0$. And the effective diffusion coefficient is a peaked function of $\Delta \Gamma $ for the particles without a magnetic field , or subject to a static magnetic field, or a rotating magnetic field with $\omega _{c} /\omega _{H} \le 1$ and $v_{0} \ge 2.0$, shown in Figs. 2(b), 2(d), and 2(f).

\begin{figure*}[htb]
\centering
  \begin{tabular}{@{}cccc@{}}
    \includegraphics[width=.3\textwidth]{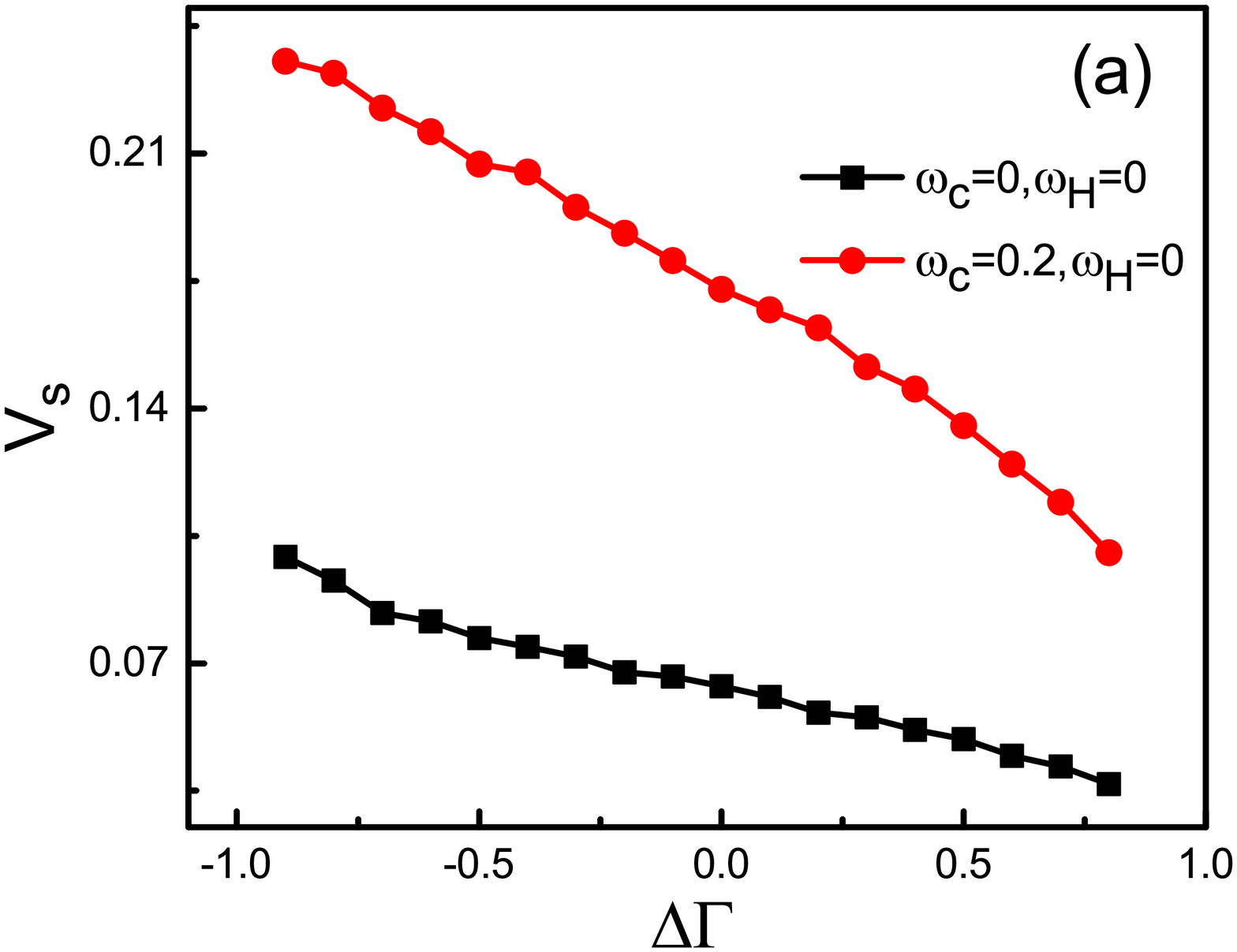} &
    \includegraphics[width=.3\textwidth]{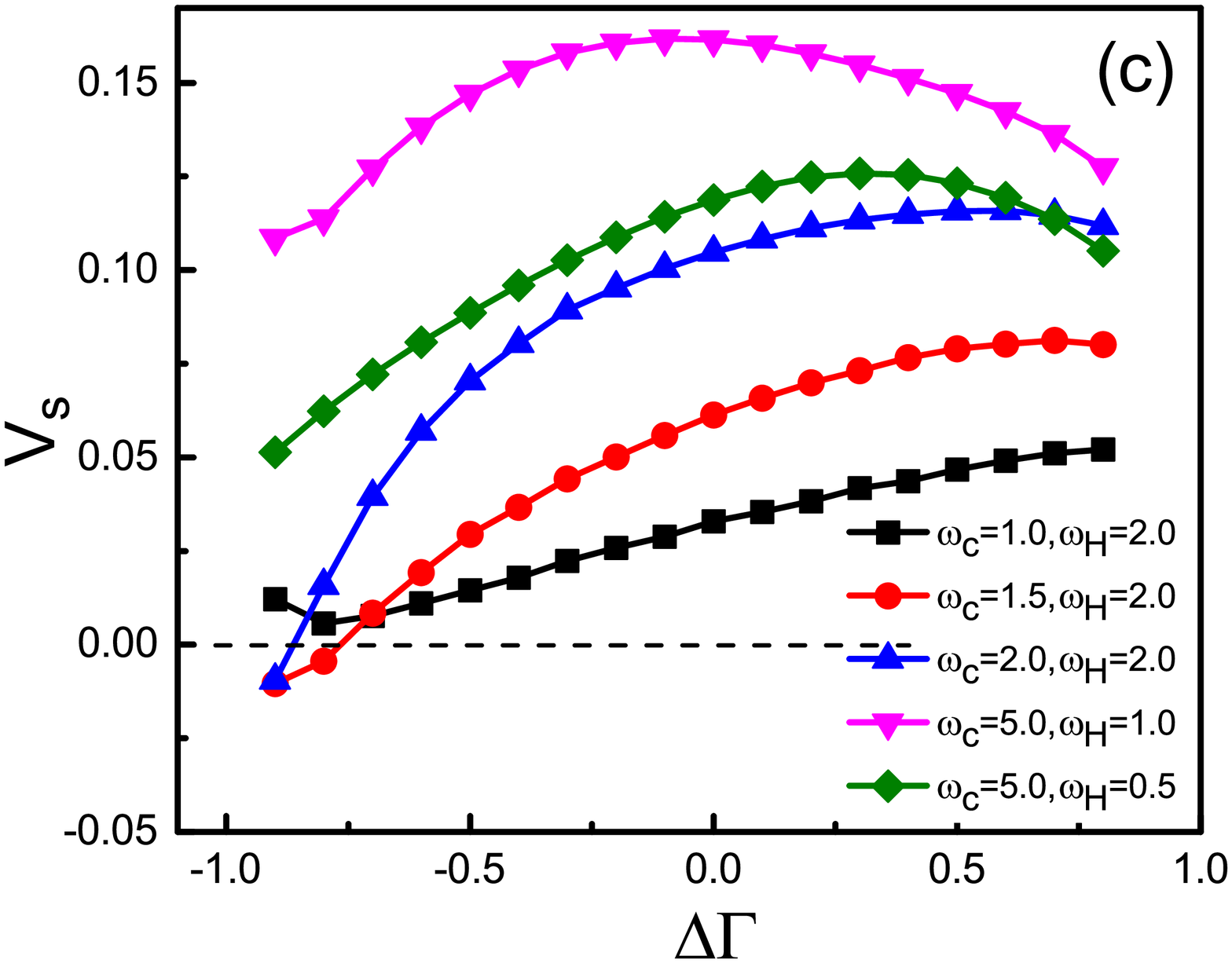} &
    \includegraphics[width=.3\textwidth]{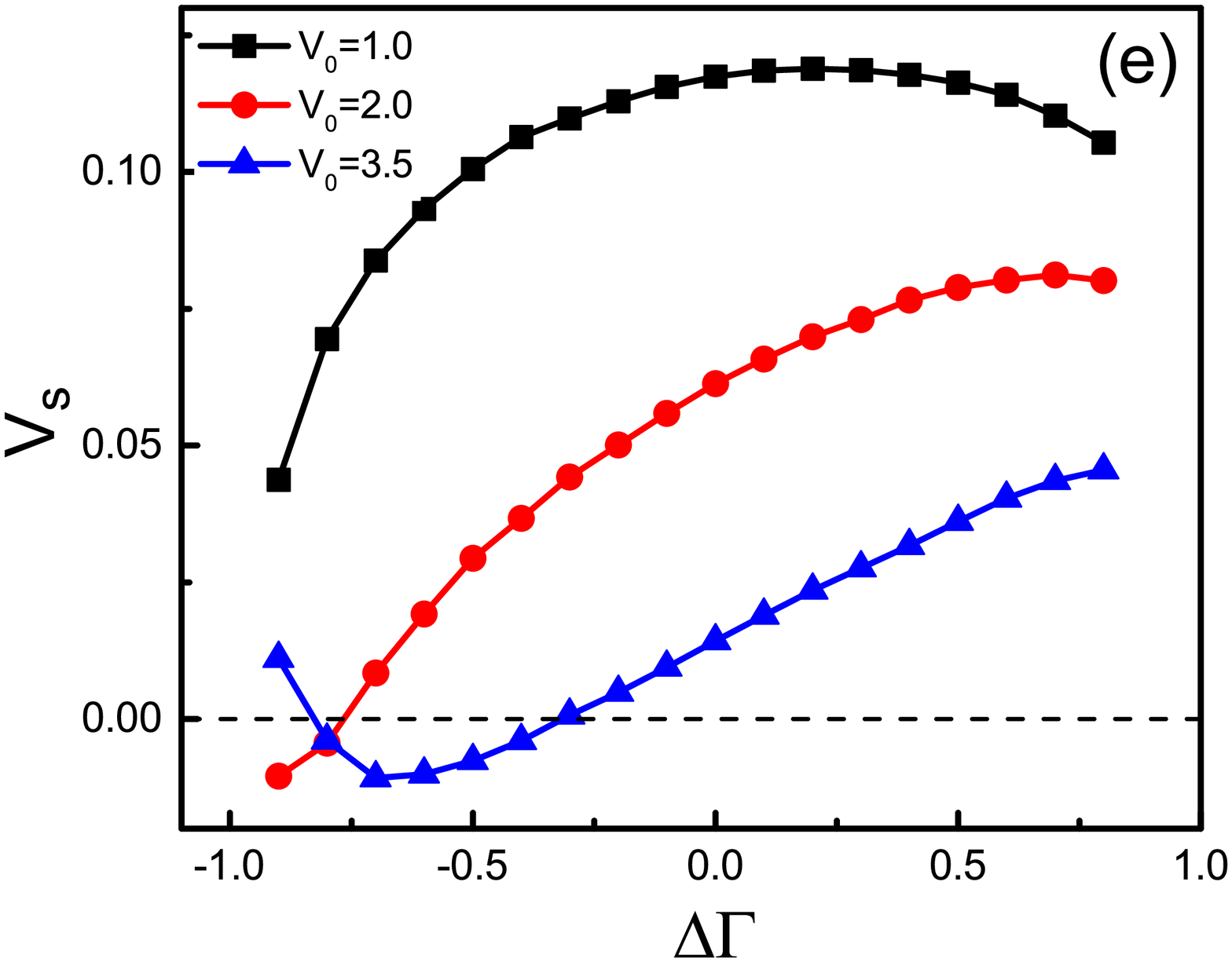} & \\
    \includegraphics[width=.3\textwidth]{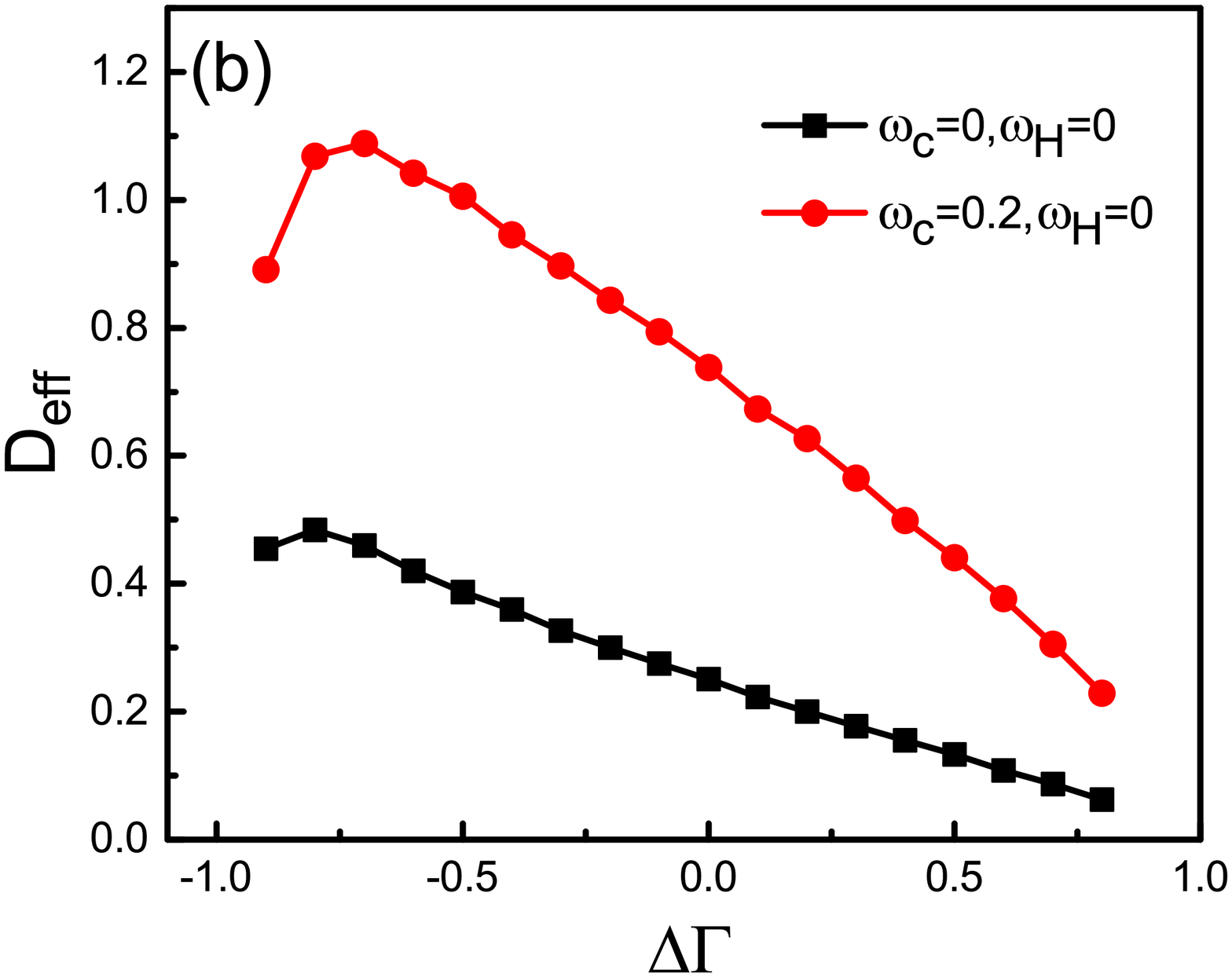} &
    \includegraphics[width=.3\textwidth]{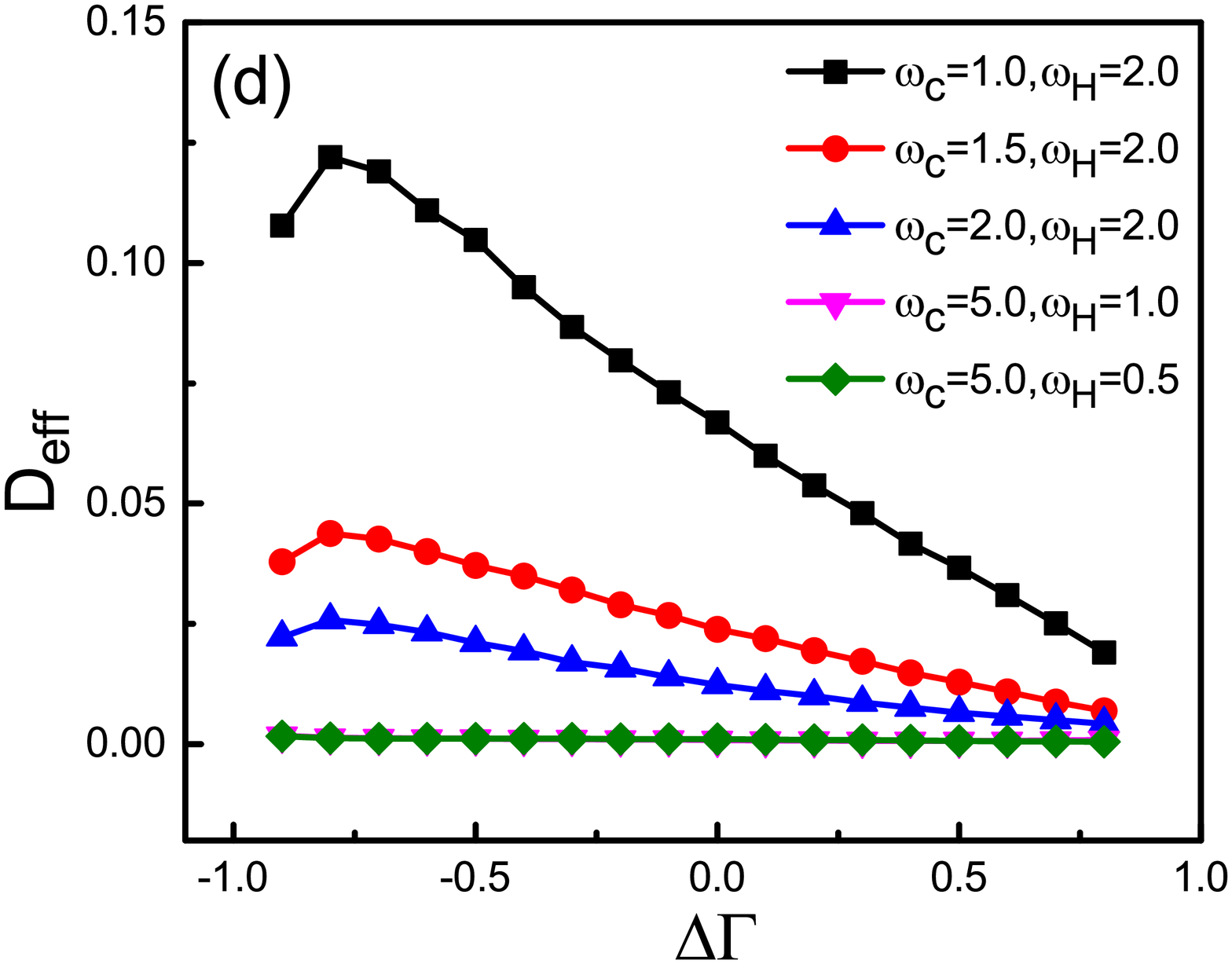} &
    \includegraphics[width=.3\textwidth]{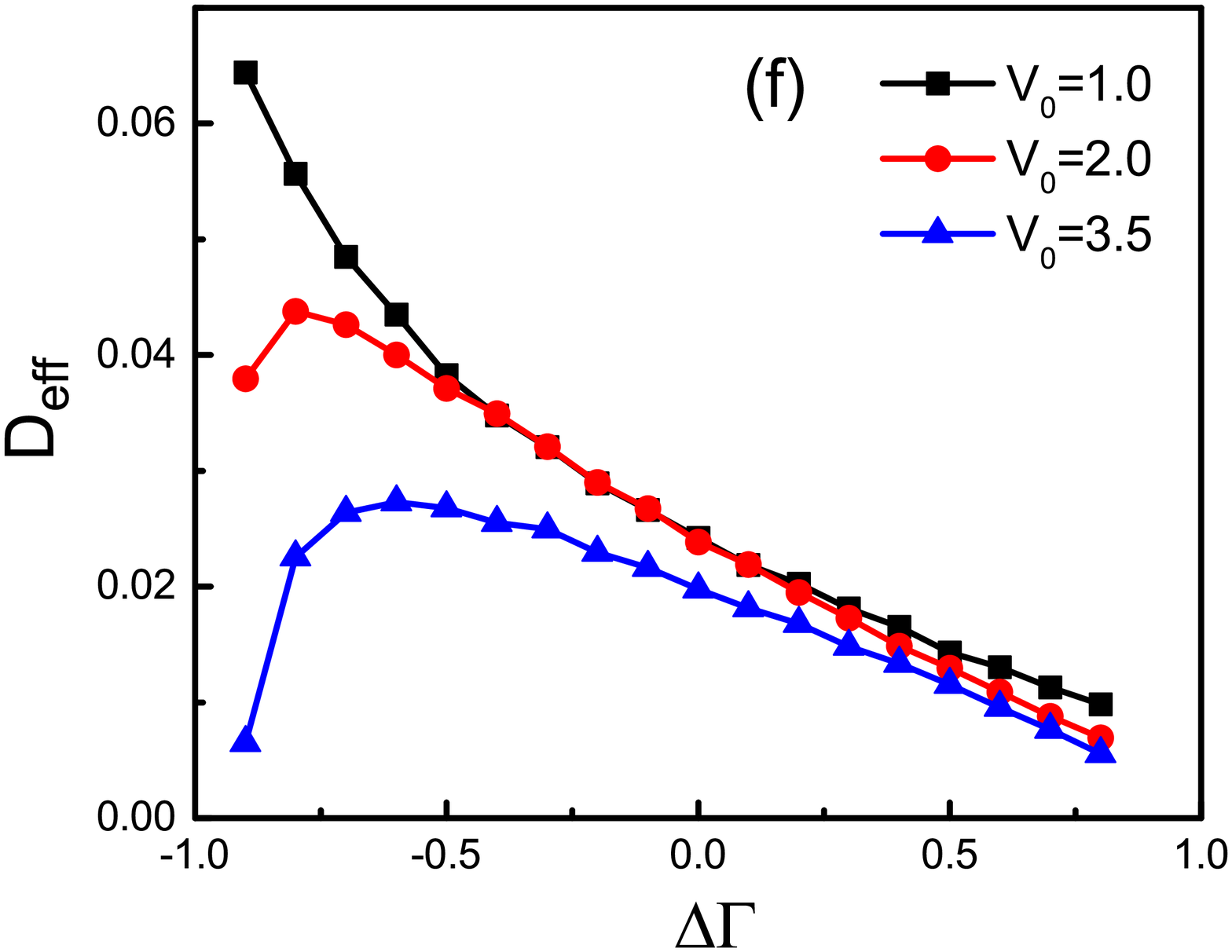} &

  \end{tabular}
  \caption{Scaled average velocity ${\it V}_{{\it s}}$ (a), (c) and the effective diffusion coefficient $D_{eff}$ (b), (d) as functions of the anisotropic parameter $\Delta \Gamma$ of active particles for different values of $\omega _{c} $ and $\omega _{H} $ at $v_{0} =2.0$. Scaled average velocity ${\it V}_{{\it s}}$ (e) and the effective diffusion coefficient $D_{eff}$ (f) as functions of the anisotropic parameter $\Delta \Gamma$ of active particles for different values of $v_{0} $ at $\omega _{c} =1.5$ and $\omega _{H} =2.0$.}
\end{figure*}

\begin{figure*}[htb]
\centering
  \begin{tabular}{@{}cccc@{}}
    \includegraphics[width=.3\textwidth]{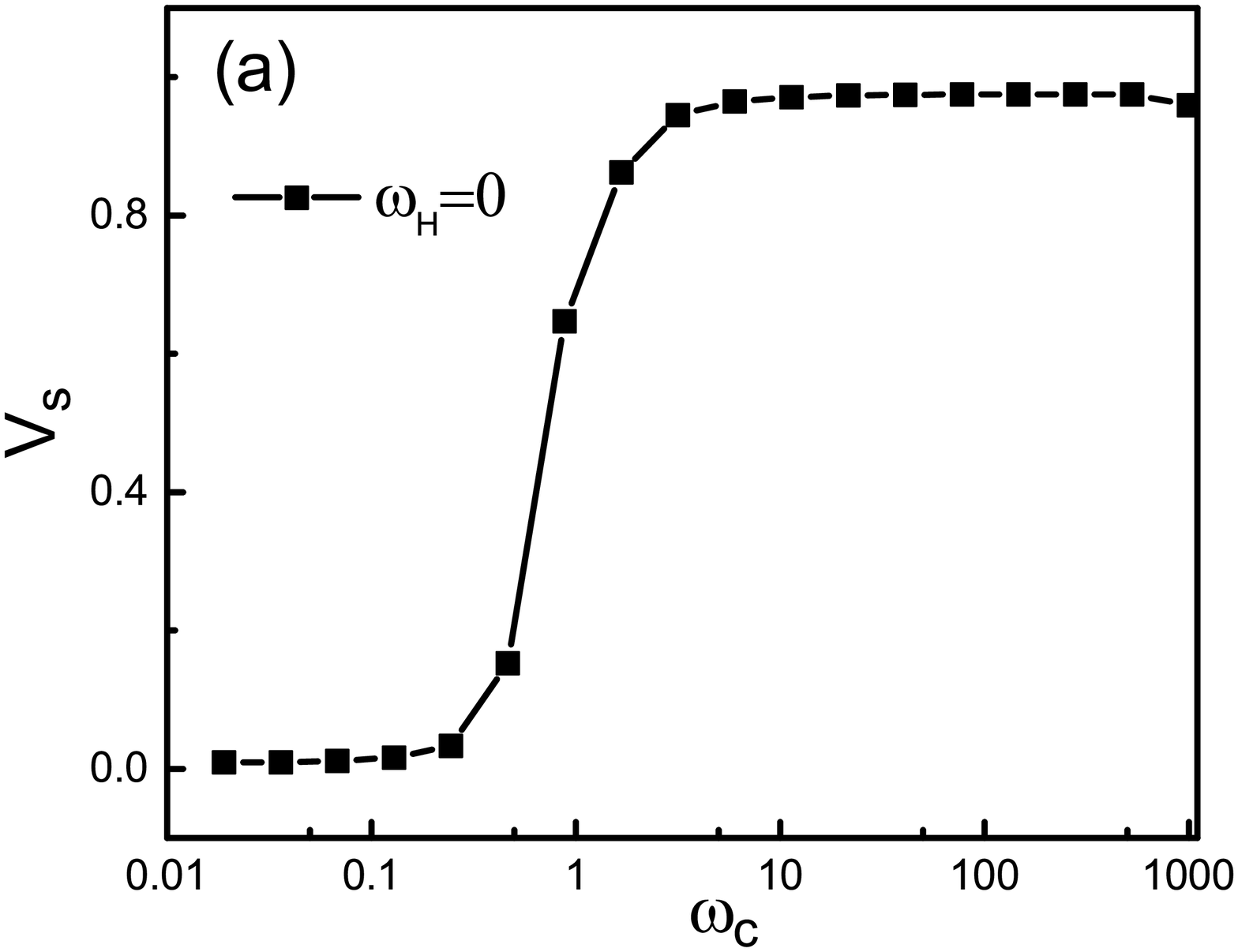} &
    \includegraphics[width=.3\textwidth]{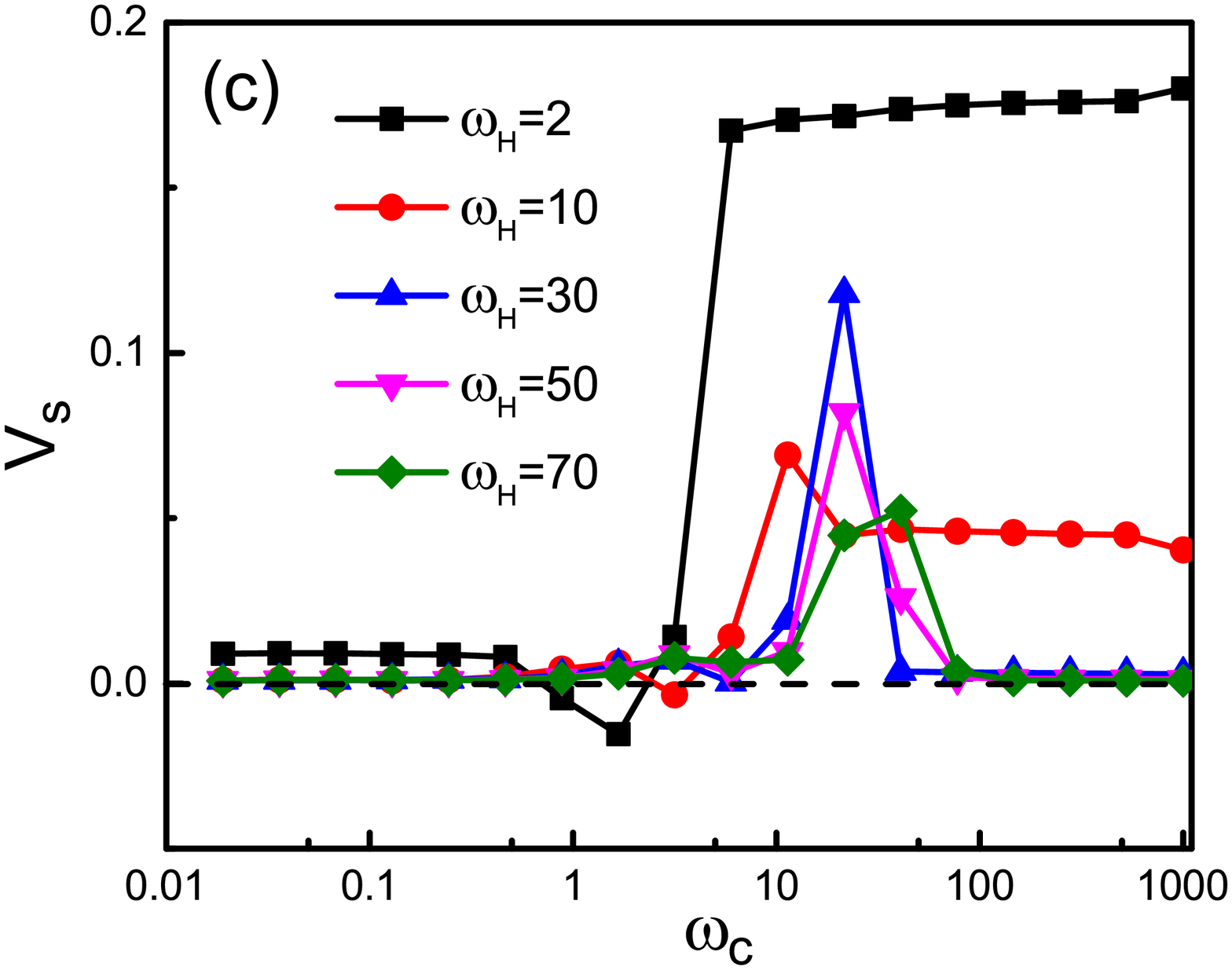} &
    \includegraphics[width=.3\textwidth]{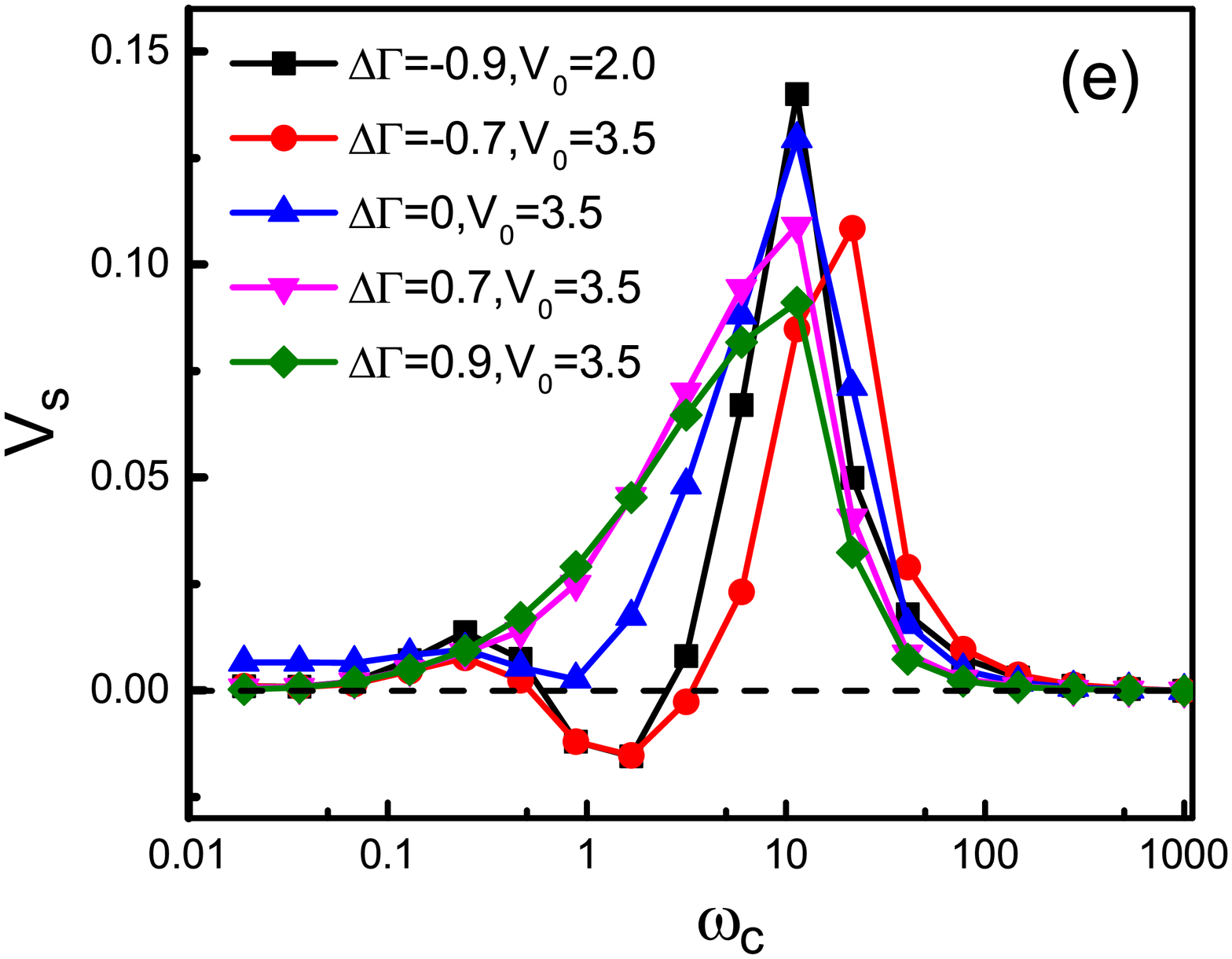} &\\
    \includegraphics[width=.3\textwidth]{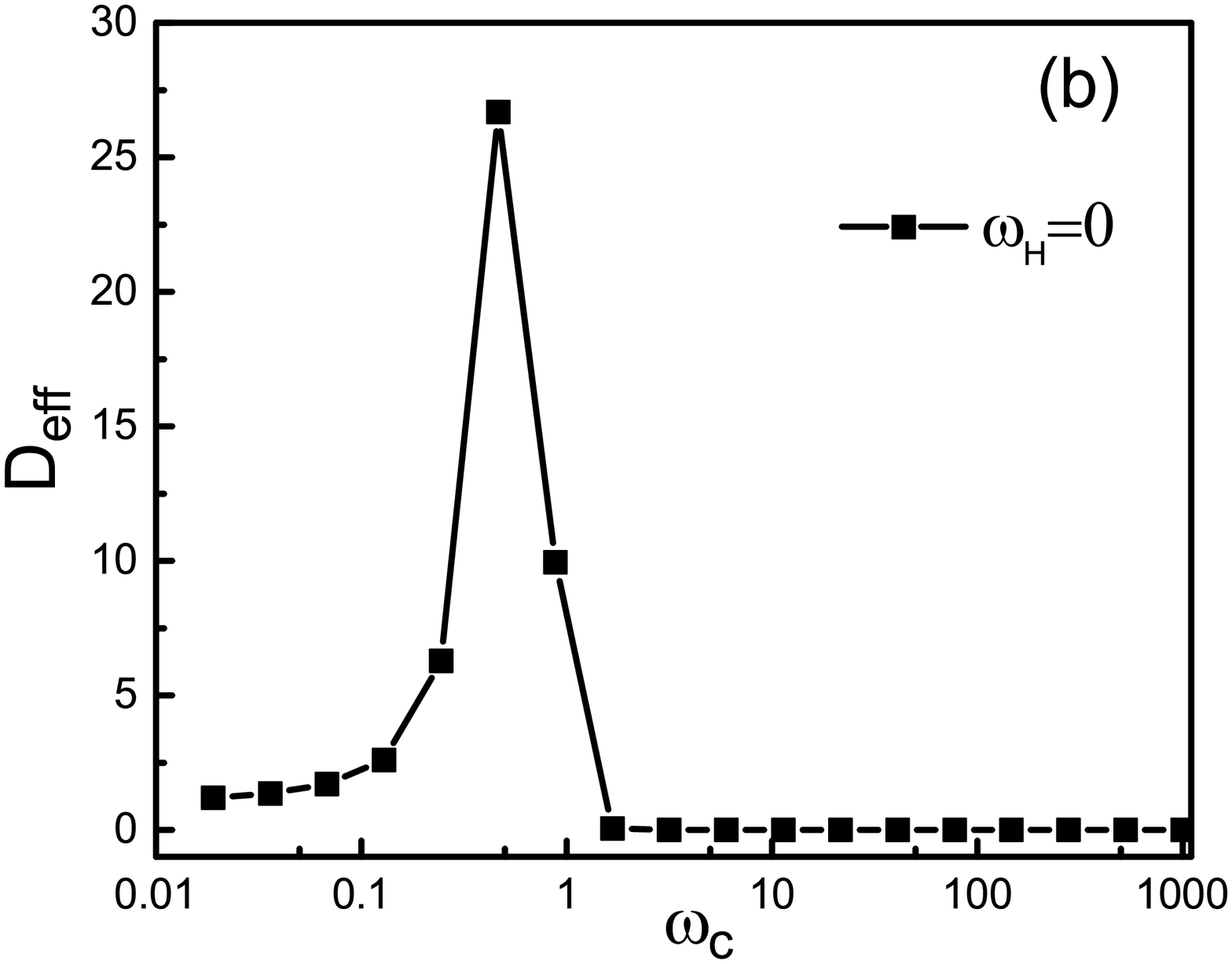} &
    \includegraphics[width=.3\textwidth]{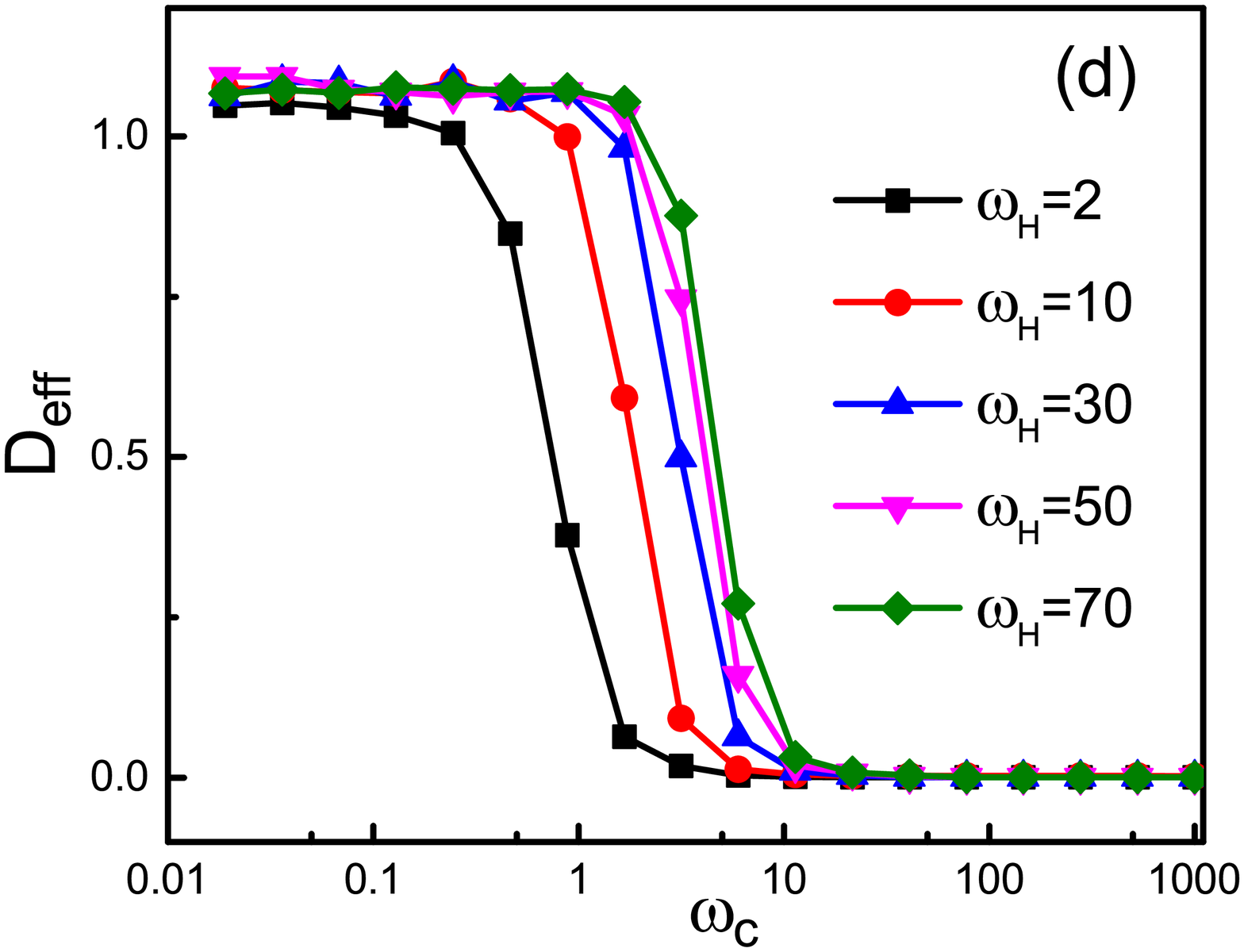} &
    \includegraphics[width=.3\textwidth]{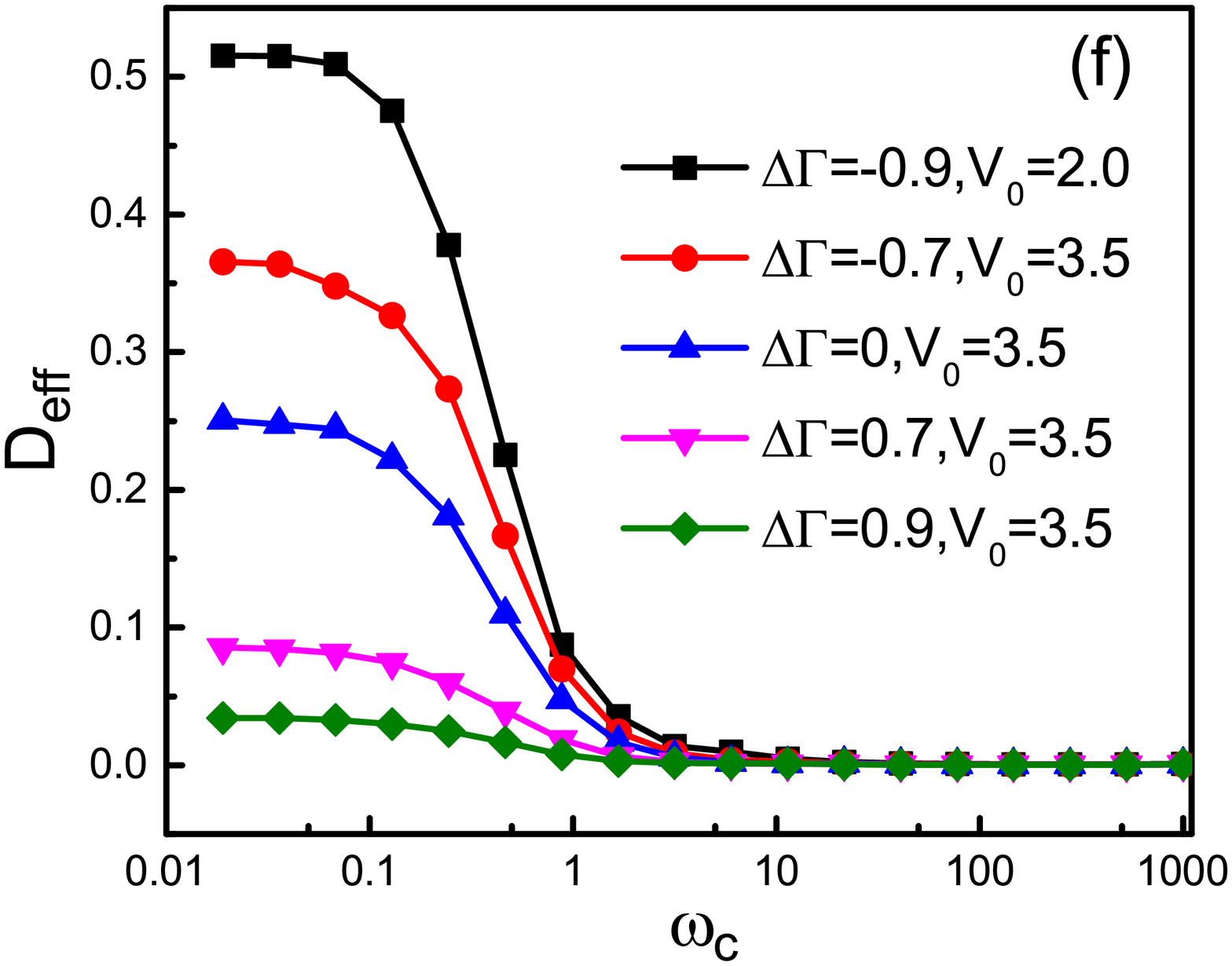} &

  \end{tabular}
  \caption{Scaled average velocity ${\it V}_{{\it s}}$ (a), (c) and the effective diffusion coefficient $D_{eff}$ (b), (d) as functions of the critical frequency $\omega _{c}$ of active particles for different magnetic frequency $\omega _{H} $ at $\Delta \Gamma =-0.7$. Scaled average velocity ${\it V}_{{\it s}}$ (e) and the effective diffusion coefficient $D_{eff}$ (f) as functions of the critical frequency $\omega _{c}$ of active particles for different $\Delta \Gamma $ and $v_{0}$ at $\gamma =0.75$.}
\end{figure*}

Results for the scaled average velocity ${\it V}_{{\it s}} $ and the effective diffusion coefficient $D_{eff} $ as functions of the critical frequency $\omega _{c} $ for different magnetic frequencies $\omega _{H} $ are presented in Fig. 3. From Ref. \cite{ref31}, we know that the rotational deterministic dynamics of the ellipsoidal particle are different and depending on the ratio $\gamma \equiv \omega _{c} /\omega _{H} $. For the case  $\gamma >1$, the ellipsoidal particle rotates synchronously with the magnetic field. For the case $\gamma <1$, the ellipsoidal swimmer performs a back-and-forth rotational motion which is considered as an asynchronous state due to the magnetic field, while it is trying to follow the rotation of the magnetic field. From Eq. (\ref{eq10}), we can see that the self-propelled angle $\theta (t)$ is determined by the frequency $\omega _{c} $, the magnetic frequency $\omega _{H} $ and the random noise. When the particles are subject to a static magnetic field ($\omega _{H} =0$), shown in Fig. 3(a), no external field is applied to the particles for $\omega _{c} \to 0$; this means the ratchet disappears and the symmetry of the system cannot be broken, thus ${\it V}_{{\it s}} \to 0$. On increasing $\omega _{c} $, the rotational motion synchronous with the magnetic field enhances the scaled average velocity $V_{s} $, thus $V_{s} $ increases and reaches the maximal value. When the particles are subject to a rotating magnetic field ($\omega _{H} \ne 0$), shown in Fig. 3(c), both the magnetic frequency $\omega _{H} $ and frequency $\omega _{c}$ play the important roles and work together in the transport. When $\omega _{c} $ is small, i.e. $\gamma <1$, the back-and-forth rotational motion reduces the mobility $V_{s} $, thus $V_{s} $ reaches the minimum value which can be seen a valley in the curve. When $\omega _{c} $ increases to $\gamma >1$, the rotational motion synchronous with the magnetic field enhances the mobility $V_{s} $, thus $V_{s} $ reaches the maximal value which can be seen a peak in the curve. When $\omega _{c}\rightarrow \infty$, the self-propelled angle changes very fast and cannot feel the self-propelled driving, so $V_{s}\rightarrow 0$. It is noted that the position of the peak is the approximate value of $\omega _{c} $ at $\gamma \ge 1$ and shifts to large $\omega _{c} $ when $\omega _{H} $ increases.

From Fig. 3(b), it is found the giant acceleration of diffusion is observed at the critical $\omega _{c}$. When $\omega _{c}$ is small, particles become trapped between potential barriers, and the external field dominates the diffusion. In this case, at the critical $\omega _{c}$, particles can be driven out from the minima of the potential and can diffuse quickly through the potential, which leads to a large value of $D_{eff} $. When $\omega _{c} \to \infty $, particles rotate very fast, the influence of the external field can be neglected and particles are trapped in the valley of the potential, so $D_{eff} \to 0$. From Fig. 3(d), the effective diffusion coefficient $D_{eff} $ decreases as the increase of $\omega _{c} $. Namely, $D_{eff} $ for  $\gamma <1$ is larger than that for $\gamma >1$. In other words, the back-and-forth rotational motion facilitates the effective diffusion coefficient while the rotational motion synchronous with the magnetic field suppresses the effective diffusion coefficient. For different magnetic frequencies $\omega _{H} $, the effective diffusion coefficients demonstrate similar behaviors and only exhibit a little difference for the position of the decreasing slope. It is noted that the scaled average velocity and effective diffusion coefficient are much larger when applying a static magnetic field ($\omega _{H} =0$). Figure 3(e) and 3(f) address $V_{s} $ and $D_{eff} $ as functions of $\omega _{c} $ for different $\Delta \Gamma $ at $\gamma =0.75$. We can find the variation of the maximal value of ${\it V}_{{\it s}} $ and $D_{eff} $ as the increasing $\Delta \Gamma $ is consistent with the results of Fig. 2(e) and 2(f). Remarkably, we note that current reversal occurs at $\Delta \Gamma =-0.9,-0.7$ by changing the magnetic frequency $\omega _{c} $ [shown in Figs. 3(c) and 3(e)].

\begin{figure*}[htpb]
\vspace{1cm}
  \label{fig8}\includegraphics[width=0.45\columnwidth]{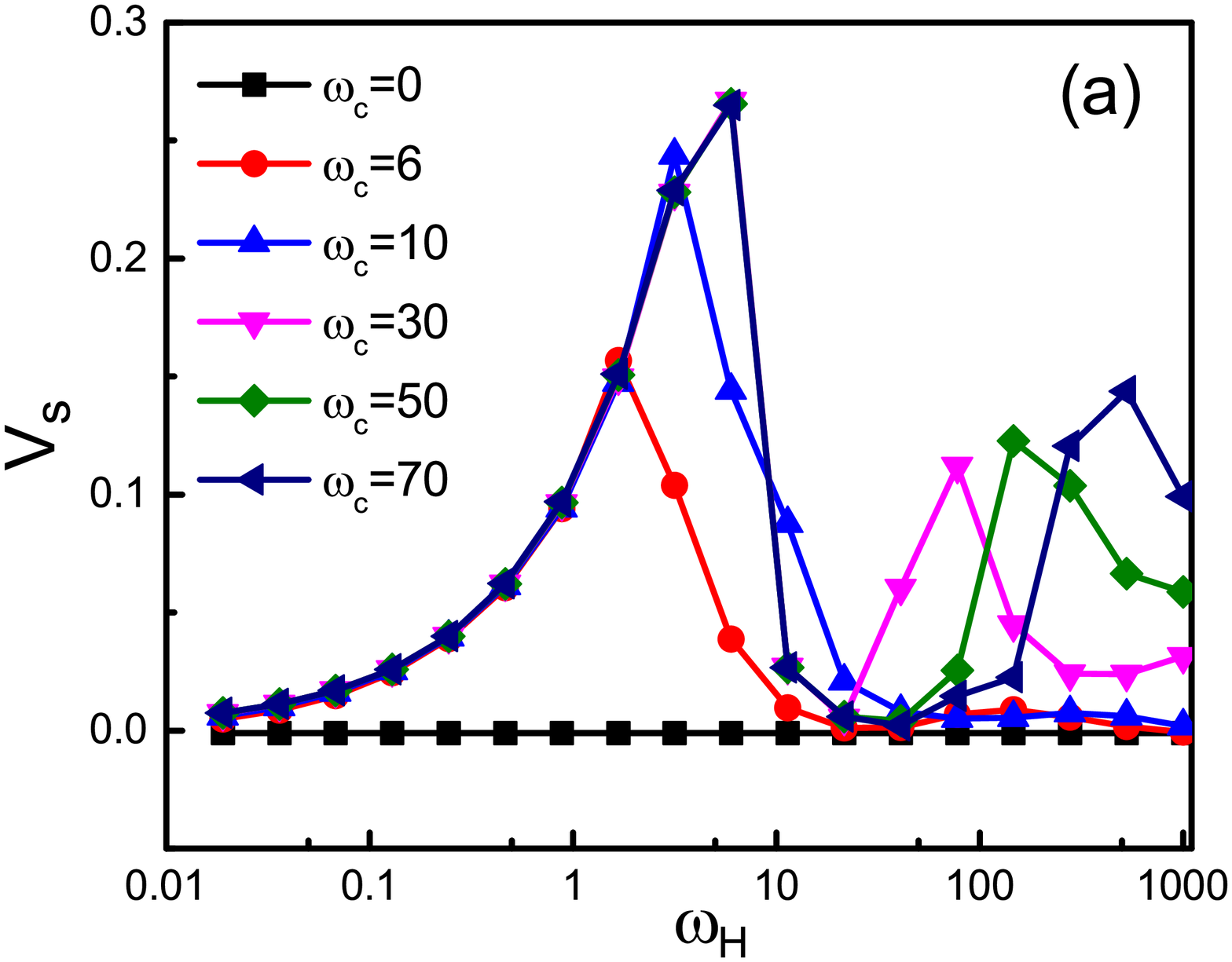}
  \includegraphics[width=0.45\columnwidth]{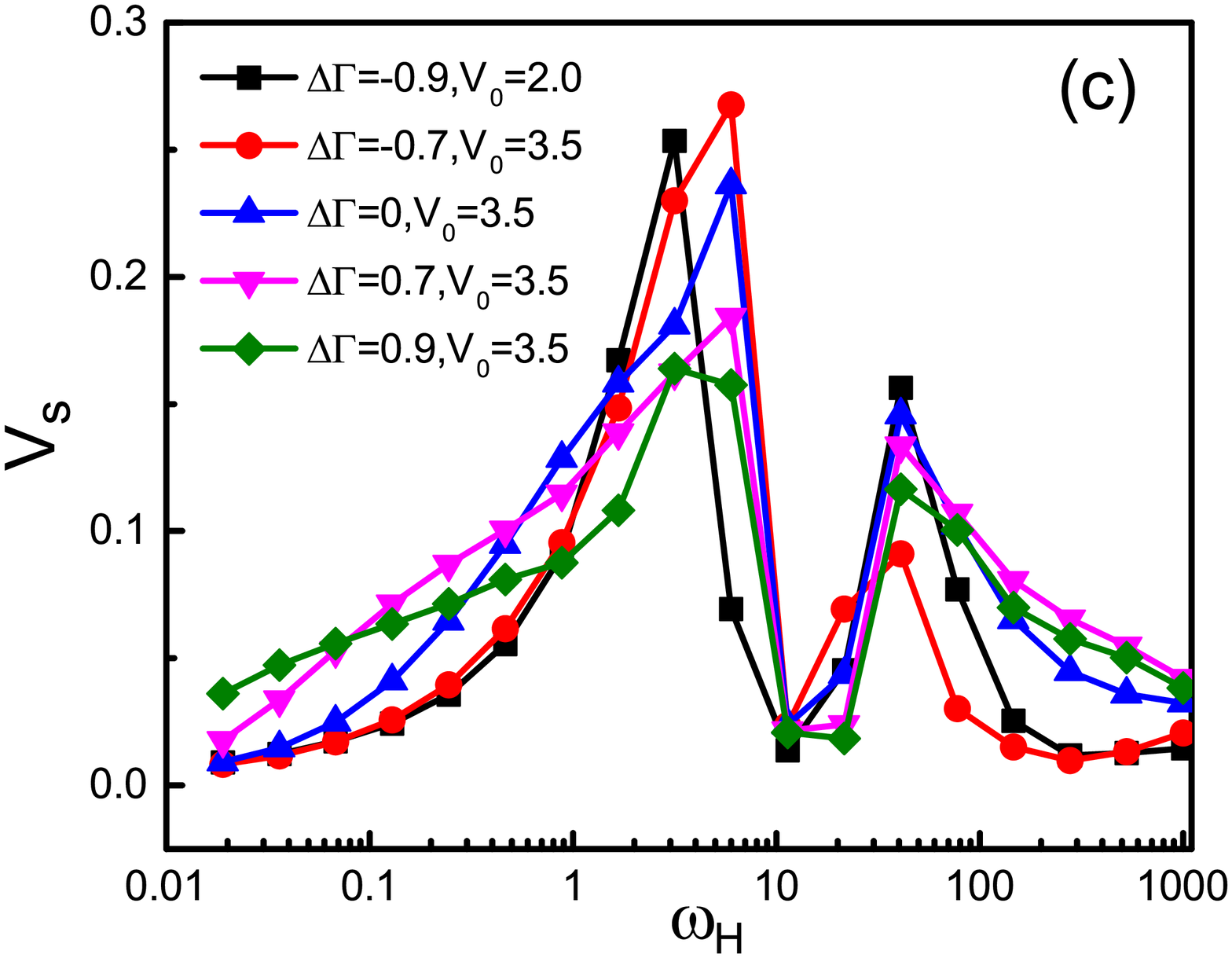}
  \includegraphics[width=0.45\columnwidth]{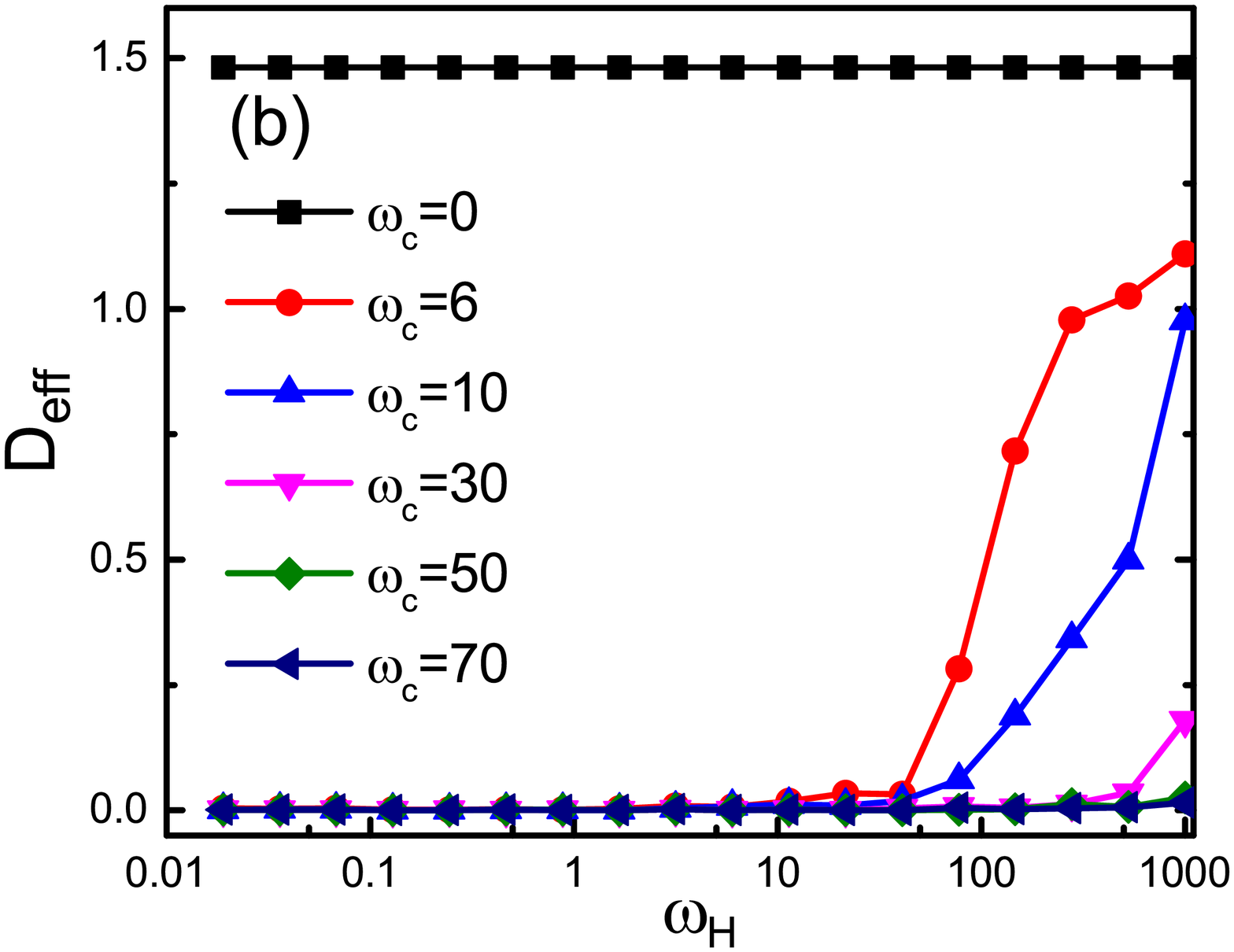}
  \includegraphics[width=0.45\columnwidth]{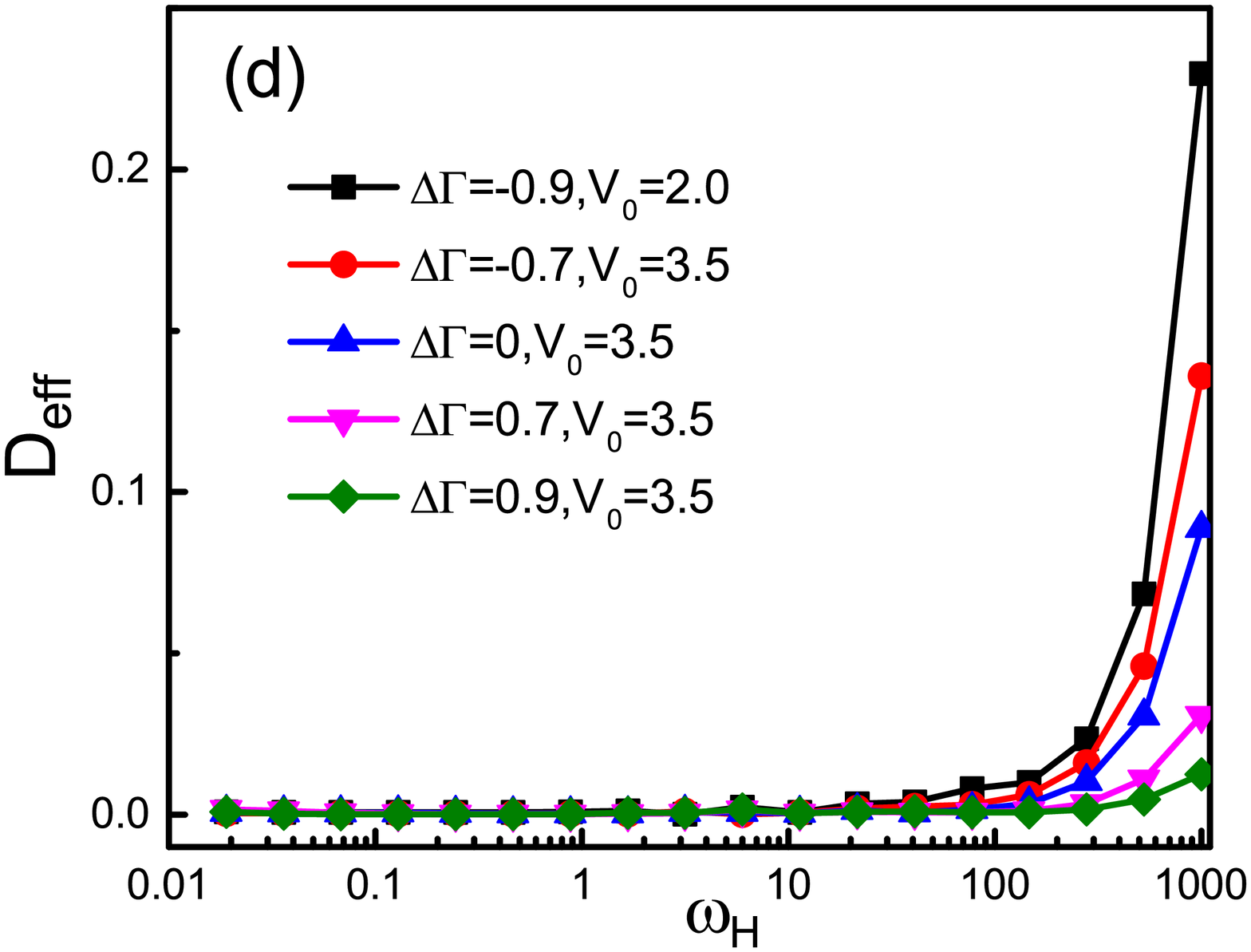}
  \caption{Scaled average velocity ${\it V}_{{\it s}}$ (a) and the effective diffusion coefficient $D_{eff}$ (b) as functions of the magnetic frequency $\omega _{H}$ of active particles for different $\omega _{c}$ at $\Delta \Gamma =-0.7$. Scaled average velocity ${\it V}_{{\it s}}$ (c) and the effective diffusion coefficient $D_{eff}$ (d) as functions of the magnetic frequency $\omega _{H}$ of active particles for different $\Delta \Gamma $ and $v_{0}$ at $\omega _{c} =20.0$.}
\end{figure*}

Figure 4(a) describes the dependence of the scaled average velocity ${\it V}_{{\it s}} $ on the magnetic frequency $\omega _{H} $ for different $\omega _{c} $. Similar to the above results, the back-and-forth rotational motion facilitates the effective diffusion coefficient and reduces the rectification while the rotational motion synchronous with the magnetic field reduces the effective diffusion coefficient and enhances the rectification. From Eqs. (\ref{eq8}) and (\ref{eq10}), we can see that the scaled average velocity ${\it V}_{{\it s}} $ is determined by the self-propelled velocity $v_{0} $, the self-propelled angle $\theta (t)$, the anisotropic parameter $\Delta \Gamma$ and the random noise, and the self-propelled angle $\theta (t)$ is a periodic function of $\omega _{H} $. Thus, there exist two peak values of $\omega _{H} $ at which the scaled average velocity takes its maximal value. As $\omega _{c} $ increases, the peak value of ${\it V}_{{\it s}} $ is bigger and the distance between two peak values is larger. From Fig. 4(b), it is found the effective diffusion coefficient $D_{eff} $ increases as the increasing $\omega _{H} $ which indicates $\gamma $ decreases. Namely, $D_{eff} $ for  $\gamma <1$ is larger than that for $\gamma >1$. For different frequencies $\omega _{c} $, the effective diffusion coefficients demonstrate similar behaviors and exhibit much difference for the position of the increasing slope. We note that the scaled average velocity is zero and the effective diffusion coefficient is constant when the particles are without magnetic field ($\omega _{c} =0$). From Figs. 4(c) and 4(d), we can find the changing as the increasing $\Delta \Gamma $ of the maximal value of ${\it V}_{{\it s}} $ and $D_{eff} $ is consistent with the results of Figs. 2(e) and 2(f).

\begin{figure*}[htb]
\centering
  \begin{tabular}{@{}cccc@{}}
    \includegraphics[width=.3\textwidth]{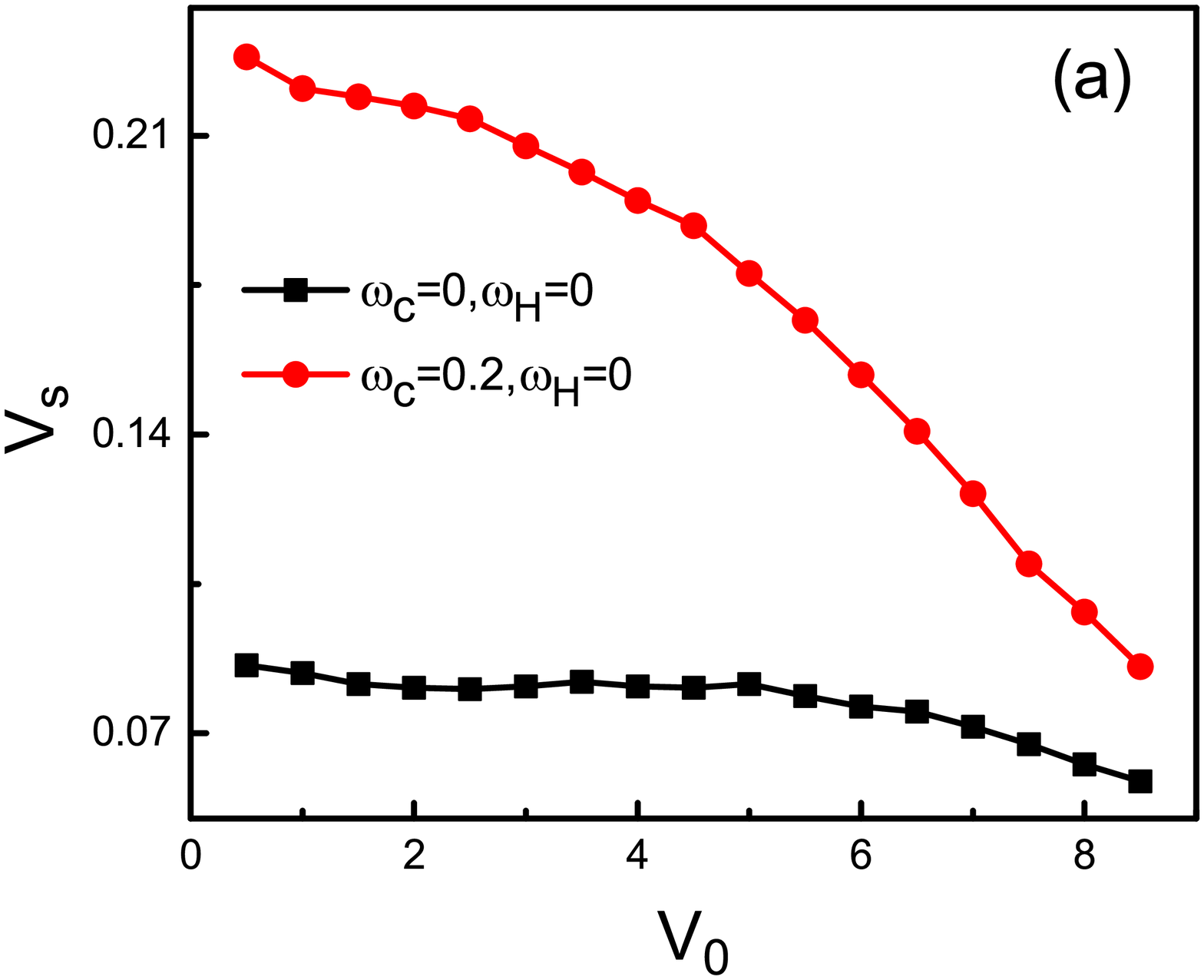} &
    \includegraphics[width=.3\textwidth]{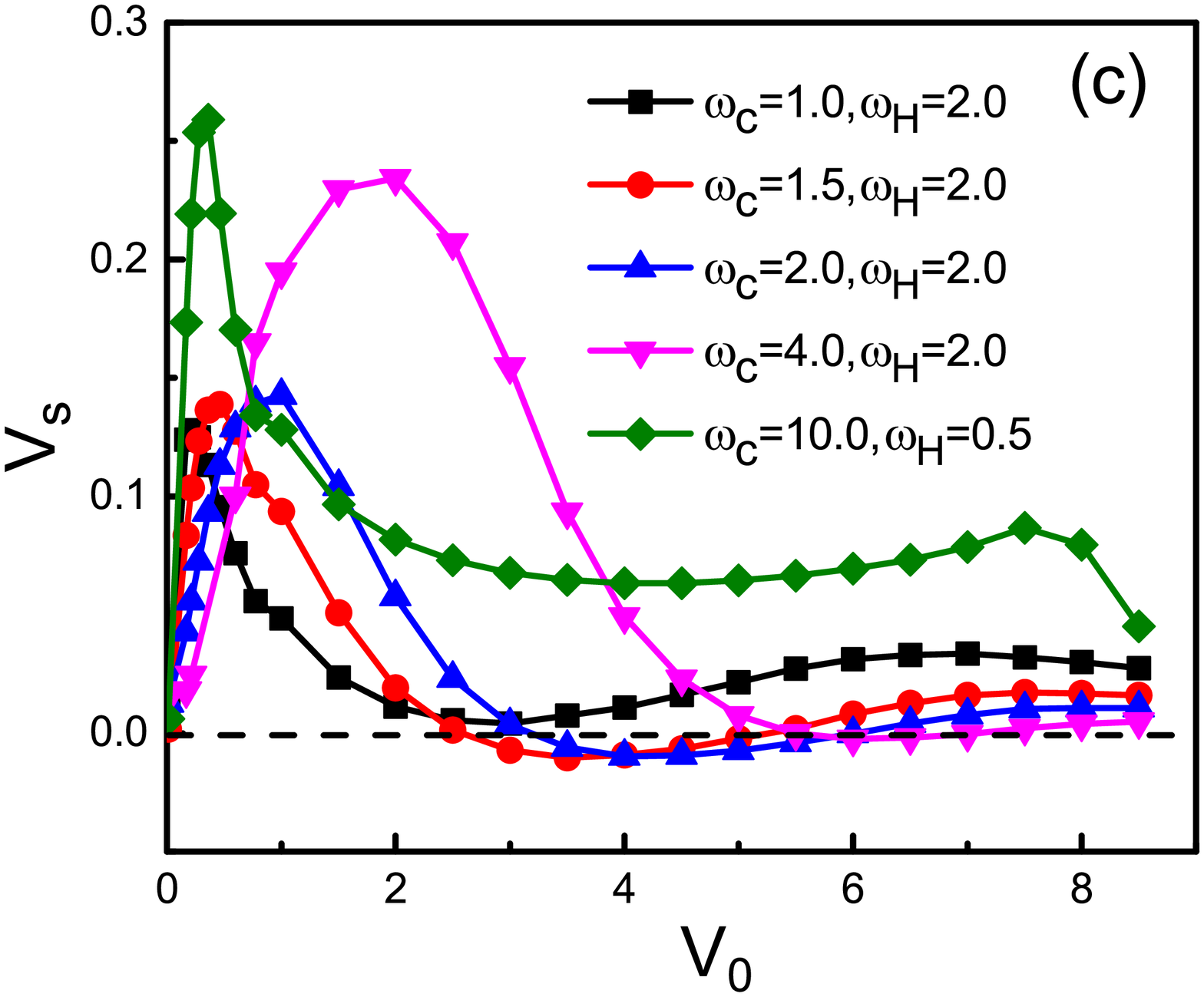} &
    \includegraphics[width=.3\textwidth]{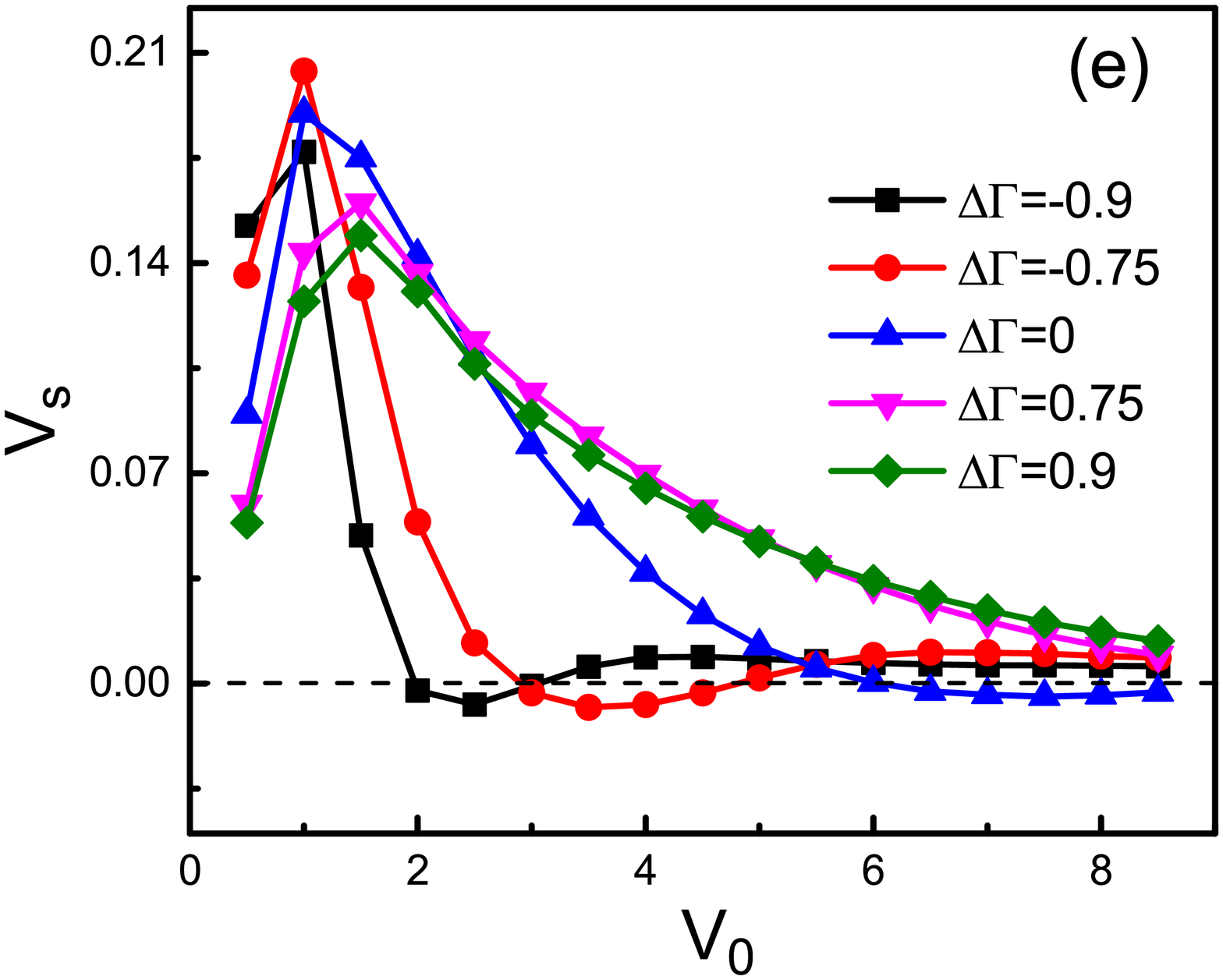} & \\
    \includegraphics[width=.3\textwidth]{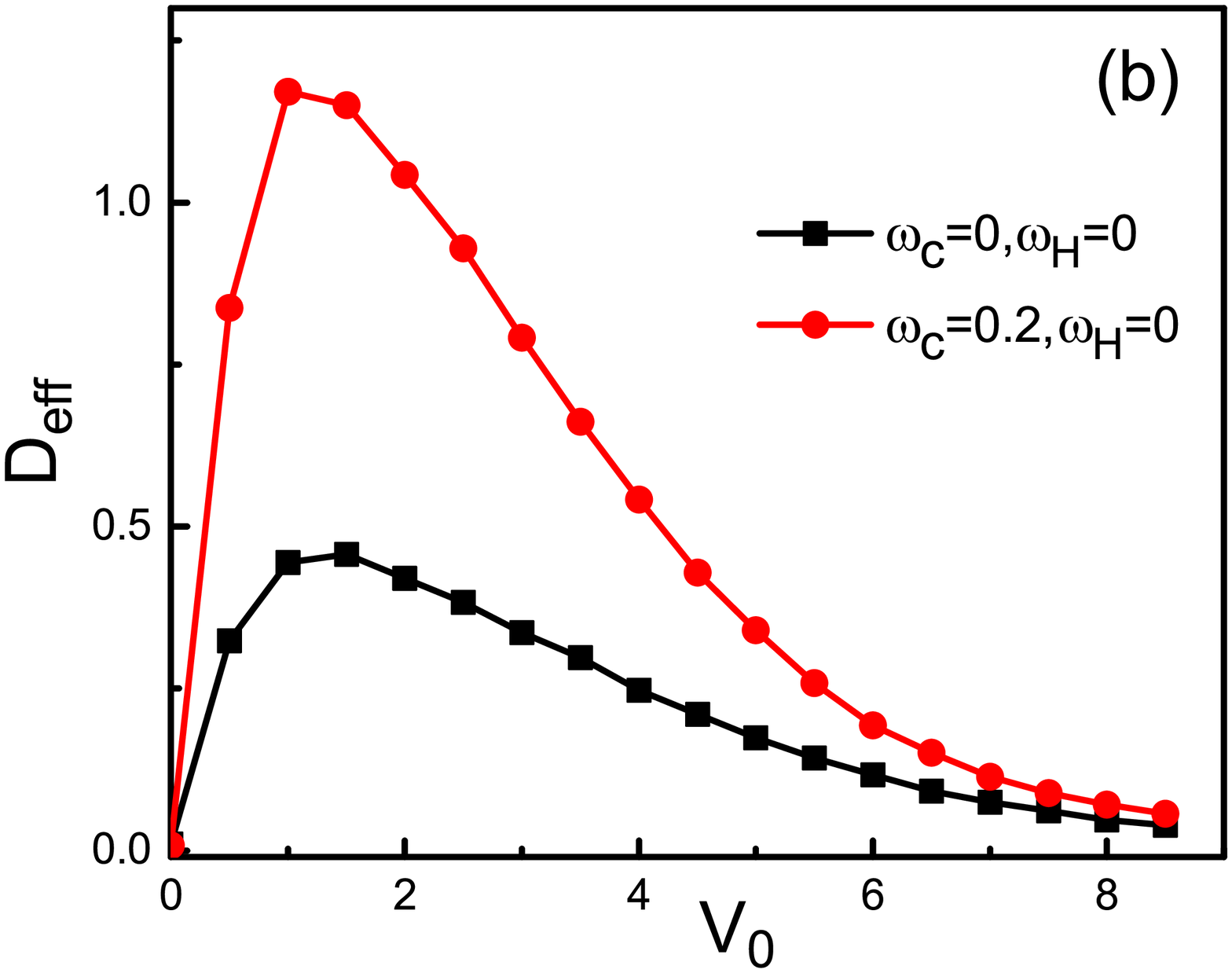} &
    \includegraphics[width=.3\textwidth]{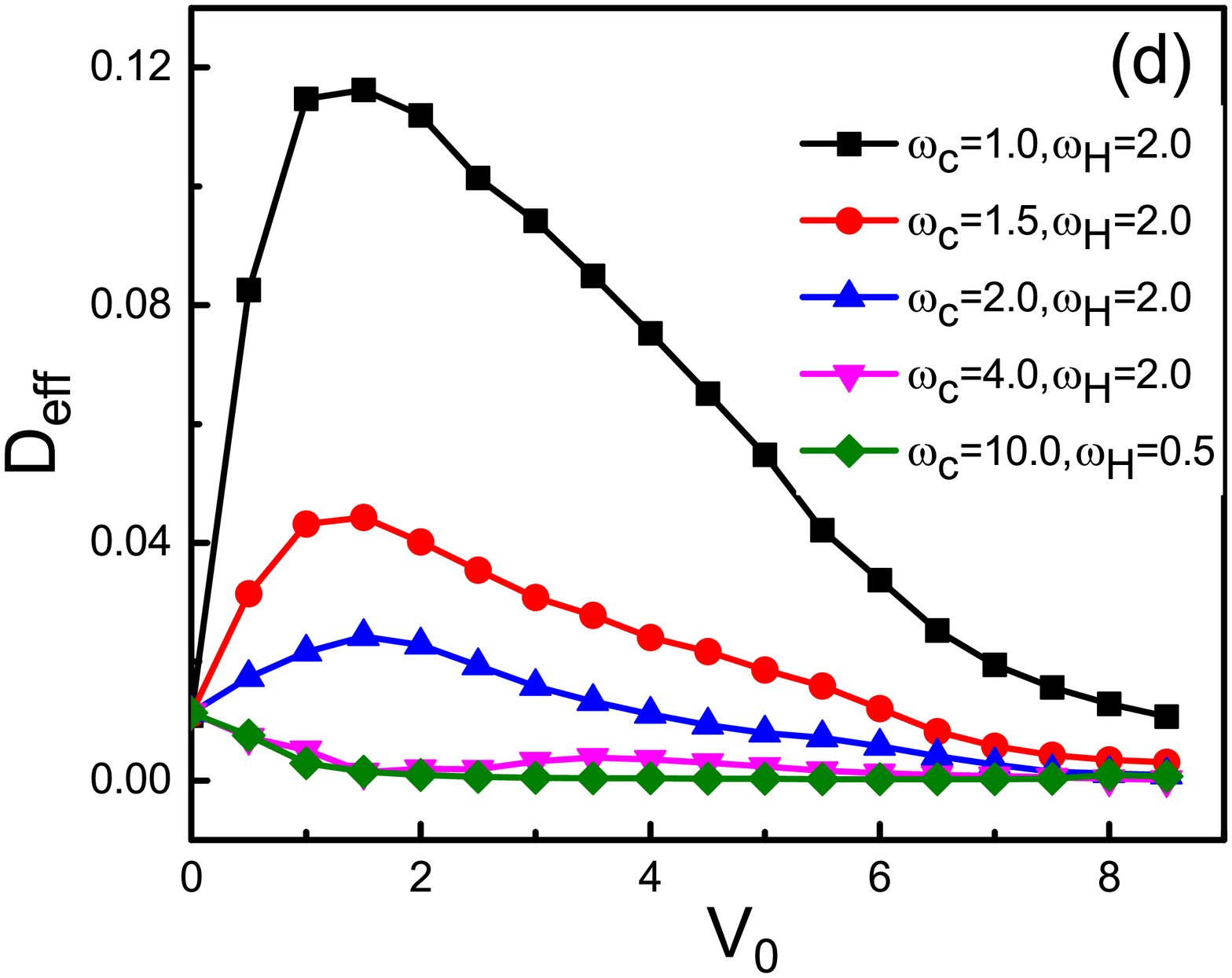} &
    \includegraphics[width=.3\textwidth]{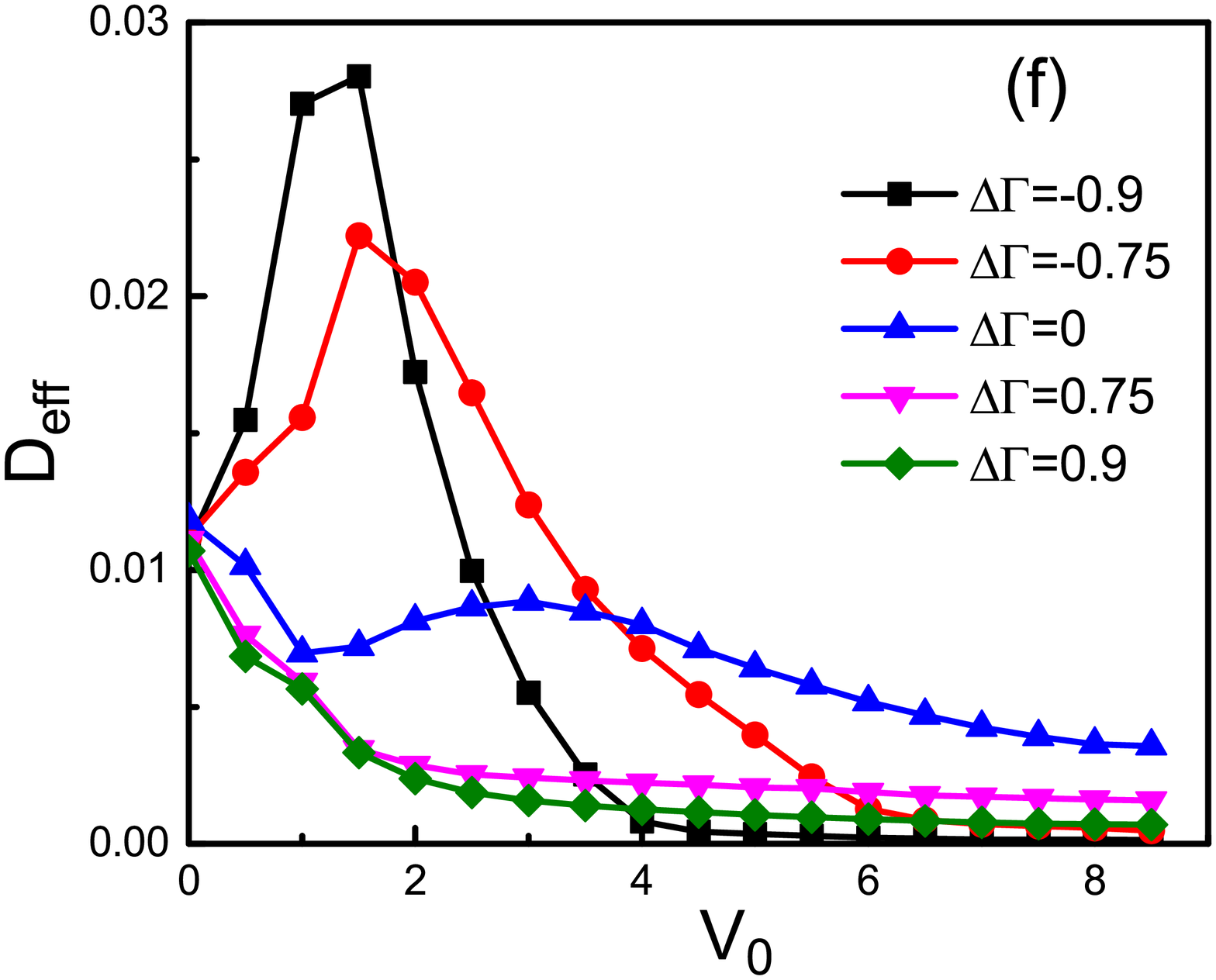} &

  \end{tabular}
  \caption{Scaled average velocity ${\it V}_{{\it s}}$ (a),(c) and the effective diffusion coefficient $D_{eff}$ (b),(d) as functions of the self-propelled velocity $v_{0} $ of active particles for different values of $\omega _{c} $ and $\omega _{H} $ at $\Delta \Gamma=-0.6$. Scaled average velocity ${\it V}_{{\it s}}$ (e) and the effective diffusion coefficient $D_{eff}$ (f) as functions of the anisotropic parameter $\Delta \Gamma$ of active particles for different values of $\Delta \Gamma$ at $\omega _{c} =1.5$ and $\omega _{H} =2.0$.}
\end{figure*}

Figure 5 displays the dependence of the scaled average velocity ${\it V}_{{\it s}} $ and the effective diffusion coefficient $D_{eff} $ on the self-propelled velocity $v_{0} $. When the particles are without a magnetic field ($\omega _{c} =0$) or subject to a static magnetic field ($\omega _{H} =0,$ $\omega _{c} \ne 0$), shown in Fig. 5(a), ${\it V}_{{\it s}} $ decreases with increasing of $v_{0} $. Figure 5(c) shows ${\it V}_{{\it s}} $ as a function of $v_{0} $ for different $\omega _{c} $ and $\omega _{H} $ at $\Delta \Gamma =-0.6$. Figure 5(e) shows ${\it V}_{{\it s}} $ as a function of $v_{0} $ for different $\Delta \Gamma $ at $\omega _{c} =1.5$, $\omega _{c} =2.0$. The term $v_{0} \cos \theta (t)$ in Eq. (\ref{eq8}) can be seen as the external driving force. When $v_{0} \to 0$, the external driving force can be negligible, so $V_{s} \to 0$. For very large values of $v_{0} $, the influence of external field can be negligible, so $V_{s} $ decreases. Therefore, there exists an optimal value of $v_{0} $ at which $V_{s} $ takes its maximal. Remarkably, ${\it V}_{{\it s}} $ is negative for suitable $v_{0} $ when $0.75\le \omega _{c} /\omega _{H} \le 1$ and $\Delta \Gamma =-0.6,-0.75,-0.9$. Therefore, the rectified direction of particles with different shapes and subject to magnetic fields of different frequencies and amplitudes can be reversed by changing $v_{0} $. From Figs. 5(b), 5(d) and 5(f), we can find the effective diffusion coefficient $D_{eff} $ decreases monotonically with increasing of $v_{0} $ when $\Delta \Gamma >0$, while  $D_{eff} $ is a peak function of $v_{0} $ when $\Delta \Gamma \leq 0$. The nonmonotonic behaviors observed in Fig.5 share similarities with negative differential and absolute mobility effects observed in different systems, such as the driven lattice gas models \cite{ref43,ref44,ref45,ref46}, driven particles adverted by laminar flows \cite{ref47,ref48}, and active matter \cite{ref49}.

\begin{figure*}[htb]
\centering
  \begin{tabular}{@{}cccc@{}}
    \includegraphics[width=.3\textwidth]{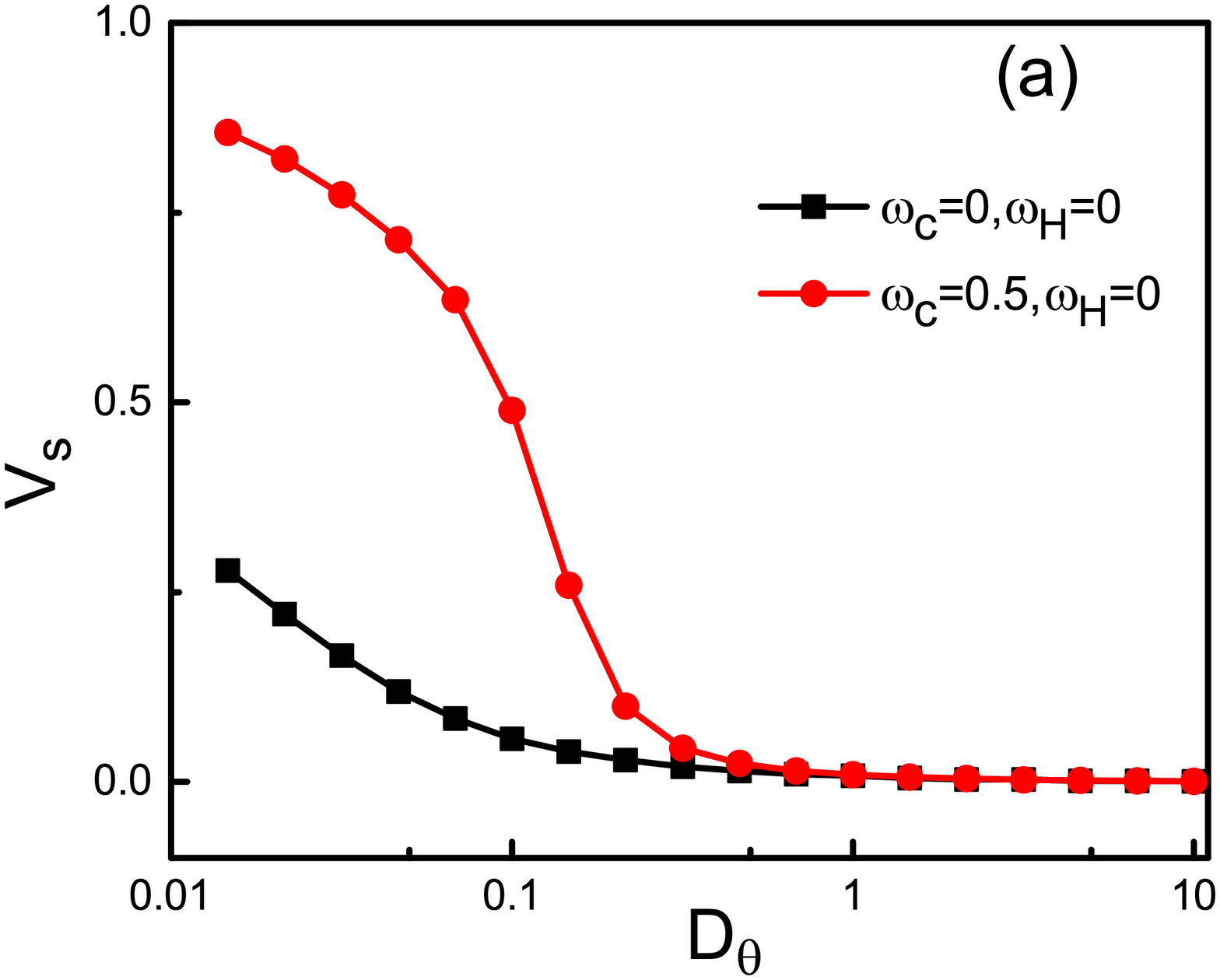} &
    \includegraphics[width=.3\textwidth]{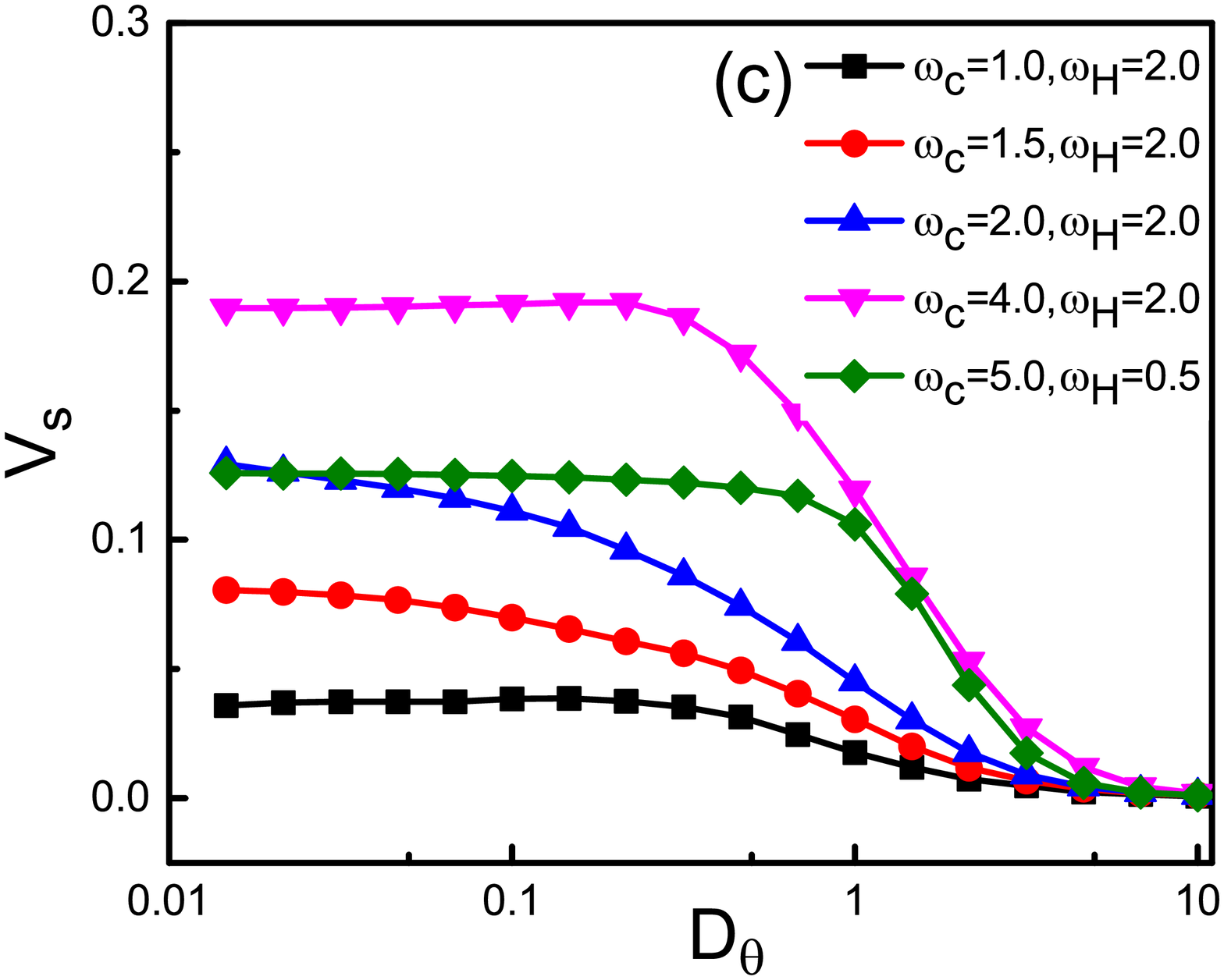} &
    \includegraphics[width=.3\textwidth]{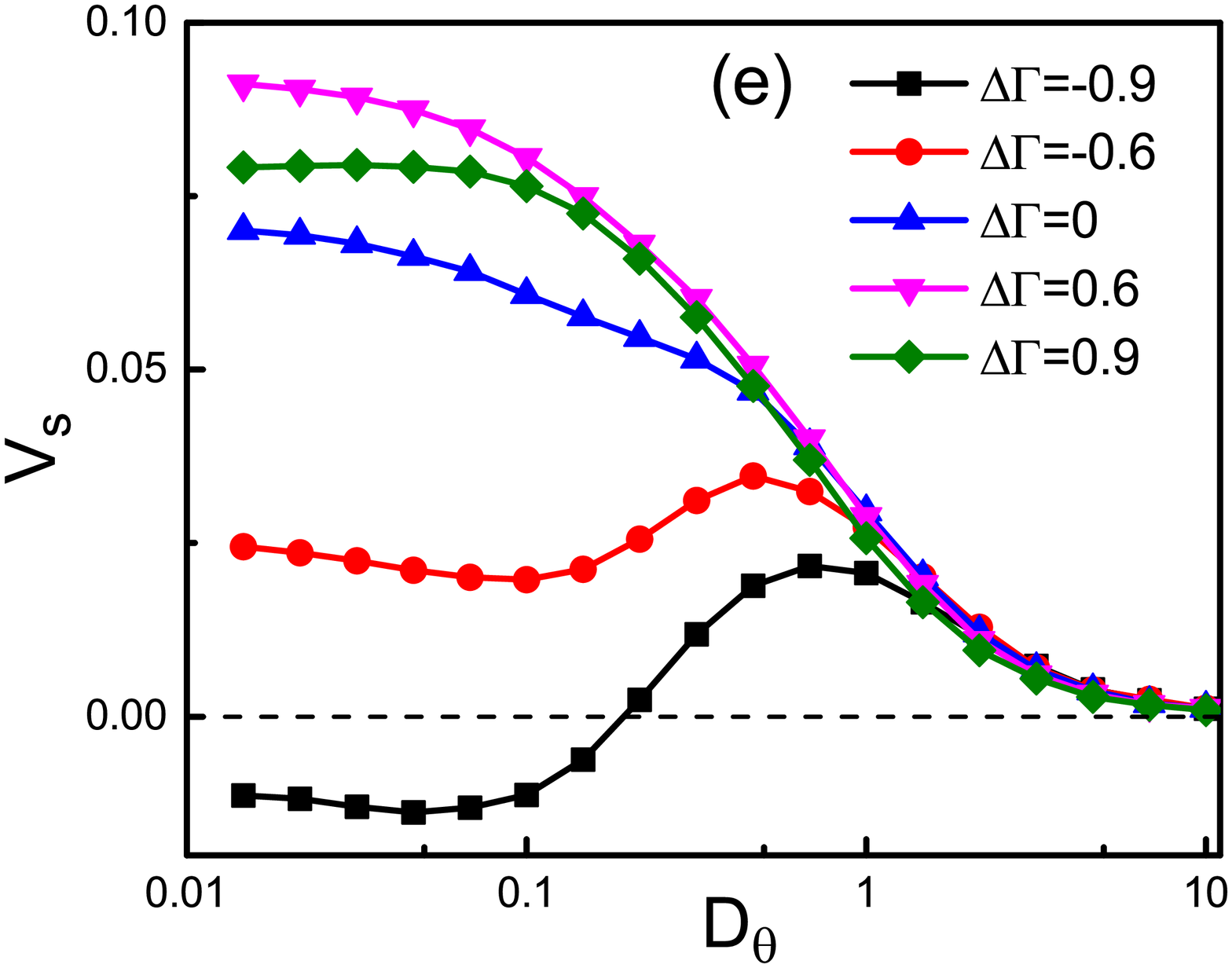} & \\
    \includegraphics[width=.3\textwidth]{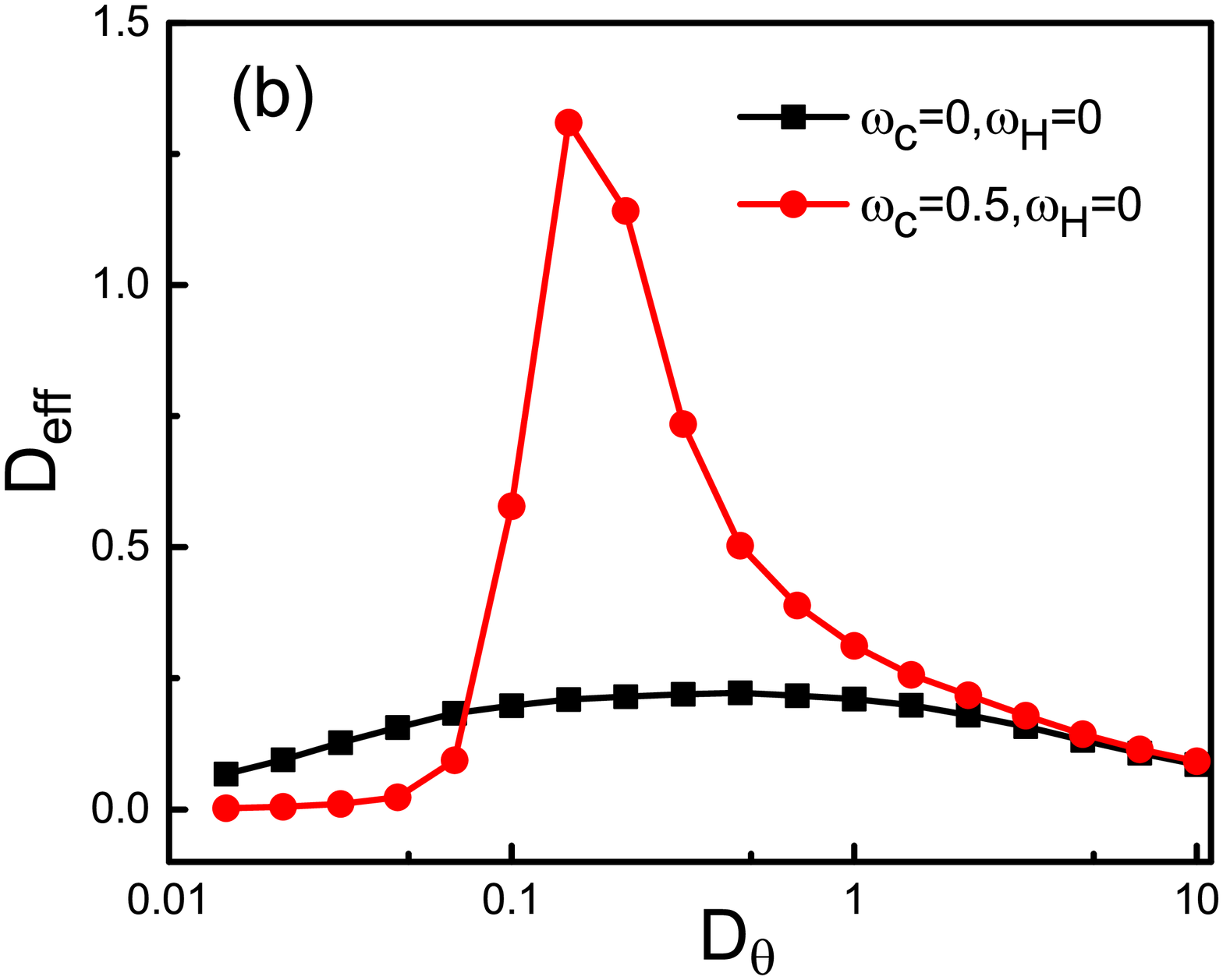} &
    \includegraphics[width=.3\textwidth]{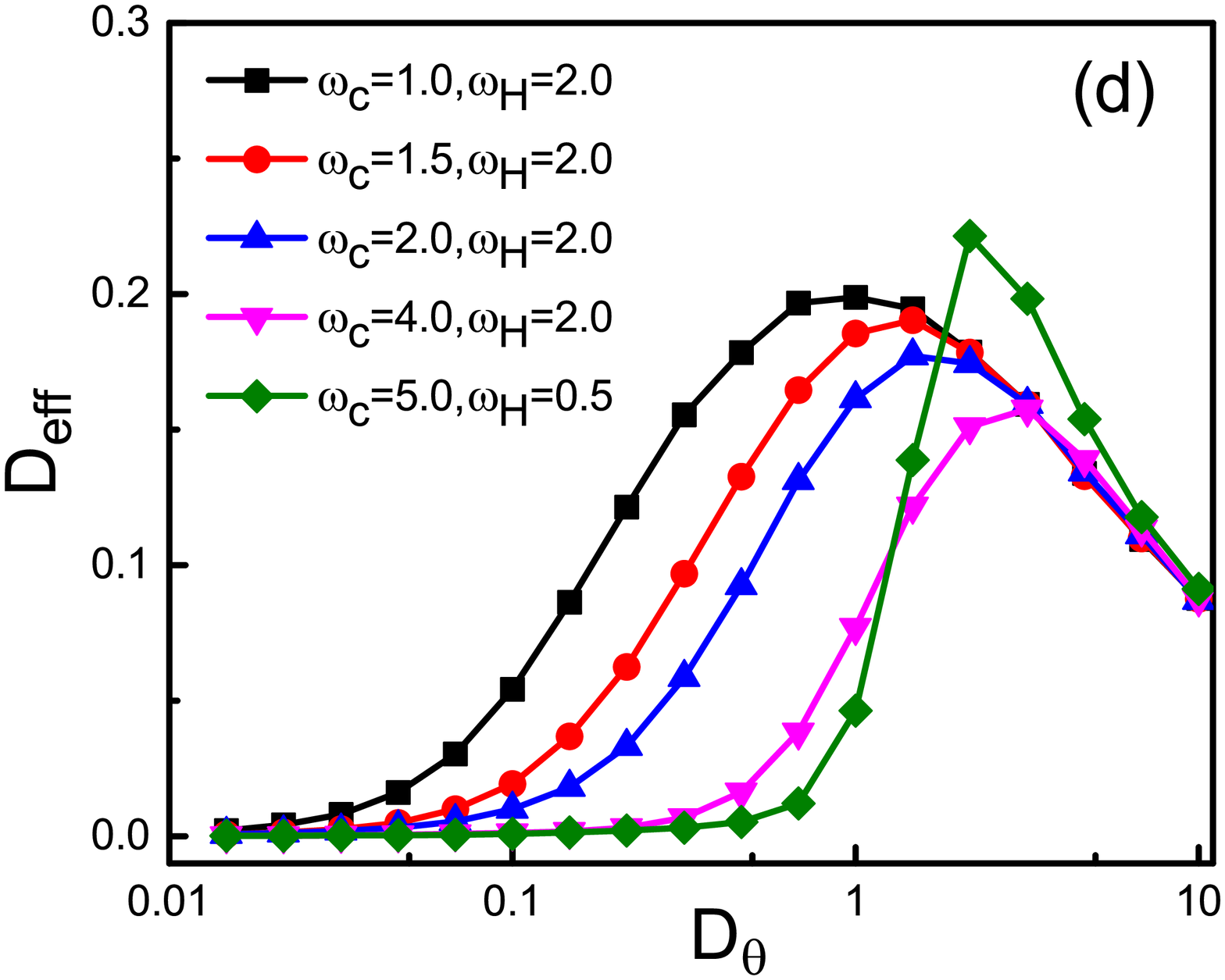} &
    \includegraphics[width=.3\textwidth]{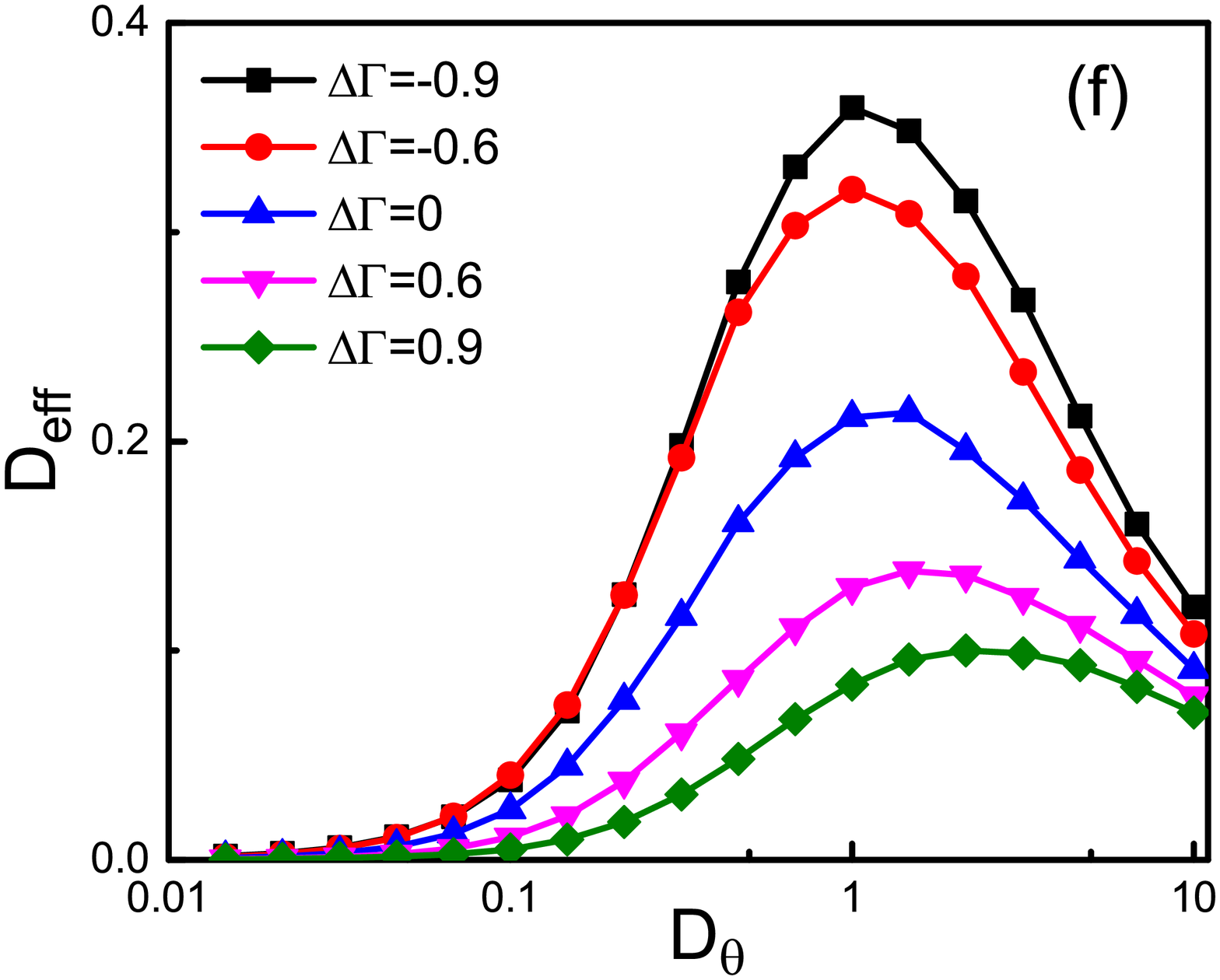} &

  \end{tabular}
  \caption{Scaled average velocity ${\it V}_{{\it s}}$ (a),(c) and the effective diffusion coefficient $D_{eff}$ (b),(d) as functions of the rotational diffusion $D_{\theta }$ of active particles for different values of $\omega _{c} $ and $\omega _{H} $ at $\Delta \Gamma=0.2$ and $v_{0}=2.0$. Scaled average velocity ${\it V}_{{\it s}}$ (e) and the effective diffusion coefficient $D_{eff}$ (f) as functions of the anisotropic parameter $\Delta \Gamma$ of active particles for different values of $\Delta \Gamma$ at $\omega _{c} =1.5$, $\omega _{H} =2.0$, and $v_{0}=2.0$.}
\end{figure*}

The scaled average velocity ${\it V}_{{\it s}} $ and the effective diffusion coefficient $D_{eff} $ as functions of the rotational diffusion $D_{\theta } $ are shown in Fig. 6. From Fig. 6(a), we can find when the particles are without a magnetic field ($\omega _{c} =0$), ${\it V}_{{\it s}} $ decreases slowly with the increase of $D_{\theta } $. When applying a static magnetic field ($\omega _{H} =0,$ $\omega _{c} \ne 0$), the curve is convex. This phenomenon can be easily explained by introducing the two factors: (A) the increase of $\omega _{c} $ enhances the rectification and (B) the increase of $D_{\theta } $ reduces the rectification. For the case without an external field, factor B dominates the transport. When $D_{\theta } \to 0$, the self-propelled angle $\theta $ almost does not change, and ${\it V}_{{\it s}} $ approaches its maximal value. When $D_{\theta } \to \infty $, the particles cannot feel the self-propelled driving and are trapped in the valley of the potential, so $V_{s} \to 0$. For the case of applying a static magnetic field, factor A determines the transport, ${\it V}_{{\it s}} $ reduces slowly, and finally factor B also becomes important, ${\it V}_{{\it s}} $ reduces quickly, so the curve is convex. Besides, the bigger $\omega _{c} $ is, the larger the maximal value ${\it V}_{{\it s}} $ approaches. Figure 6(c) shows ${\it V}_{{\it s}} $ decreases monotonically as $D_{\theta } $ increases and finally tends to be zero when applying a rotating magnetic field. It is also found that the maximal value ${\it V}_{{\it s}} $ is the biggest when $\omega _{c} =4.0$, $\omega _{H} =2.0$. As the above results, the ellipse rotates synchronously with the magnetic field for $\gamma >1$, but performs a back-and-forth rotational motion for $\gamma <1$. In other words, the rotational motion synchronous with the magnetic field facilitates the rectification while the back-and forth rotational motion reduces the rectification. The nearer to $\gamma =1$, the larger the maximal value ${\it V}_{{\it s}} $ is. Figure 6(e) presents ${\it V}_{{\it s}} $ as a function of the rotational diffusion $D_{\theta } $ for different values of $\Delta \Gamma $ at $\omega _{c} =1.5$, $\omega _{H} =2.0$. ${\it V}_{{\it s}} $ decreases monotonically as $D_{\theta } $ increases for $\Delta \Gamma \ge 0$ while there exists an optimal value of $D_{\theta } $ at which ${\it V}_{{\it s}} $ takes its maximal value for $\Delta \Gamma <0$. Because $\Delta \Gamma $ can reduce the growth of ${\it V}_{{\it s}} $ when $\Delta \Gamma <0$ which have been found from Fig. 2. Remarkably, the rectified direction can be reversed by changing $D_{\theta } $ when $\Delta \Gamma =-0.9$. Therefore, for suitable $D_{\theta } $, particles with different shapes will move to the opposite directions and can be separated.

The effective diffusion coefficient $D_{eff} $ as a function of $D_{\theta } $ exhibits a pronounced resonance peak in the curve [shown in Fig. 6(b), 6(d), and 6(f)]. This is due to the mutual interplay between the external field and the rotational diffusion rate. When $D_{\theta } $ is small, the external field dominates the diffusion. When $D_{\theta } \to \infty $, particles rotate very fast, the influence of the external field can be neglected and particles are trapped in the valley of the potential, so $D_{eff} \to 0$. Especially, when applying a static field, particles can be driven out from the minima of the potential and can diffuse quickly through the potential, which leads to a large value of $D_{eff} $. Therefore, $D_{eff} $ is much larger than 1, shown in Fig. 6(b) for $\omega _{H} =0,$ $\omega _{c} \ne 0$, which indicates the giant acceleration of diffusion. In addition, from Fig. 6(d), the amplitude of the peak decreases and the position of the peak shifts to large $D_{\theta } $ when $\omega _{c} /\omega _{H} $ increases. From Fig. 6(f), the maximal value of $D_{eff} $ decreases with increasing $\Delta \Gamma $ which is accordance to Fig. 2.

\subsection{Mobility and diffusion of passive particles ($v_{0} =0$, $f_{0}\ne 0$)}
The results for the mobility $\mu $ and the effective diffusion coefficient $D_{eff} $ depending on the anisotropic parameter $\Delta \Gamma $ of passive particles are presented in Fig. 7. We can find the mobility and the effective diffusion coefficient have similar behaviors. When the particles are subject to a static magnetic field ($\omega _{H} =0,$ $\omega _{c} \ne 0$), $\mu $ and $D_{eff} $ increase with increasing $\Delta \Gamma $. While there exists an optimal value of $\Delta \Gamma $ where $\mu $ and $D_{eff} $ are maximal when the particles are without an external field or subject to a rotating magnetic field. $\mu $ and $D_{eff} $ of anisotropic particles are smaller than that of isotropic particles. Additionally, the mobility $\mu $ and the effective diffusion coefficient $D_{eff} $ change a little when applying rotating magnetic fields of different frequencies.

\begin{figure*}[htpb]
\vspace{1cm}
  \label{fig2}\includegraphics[width=0.45\columnwidth]{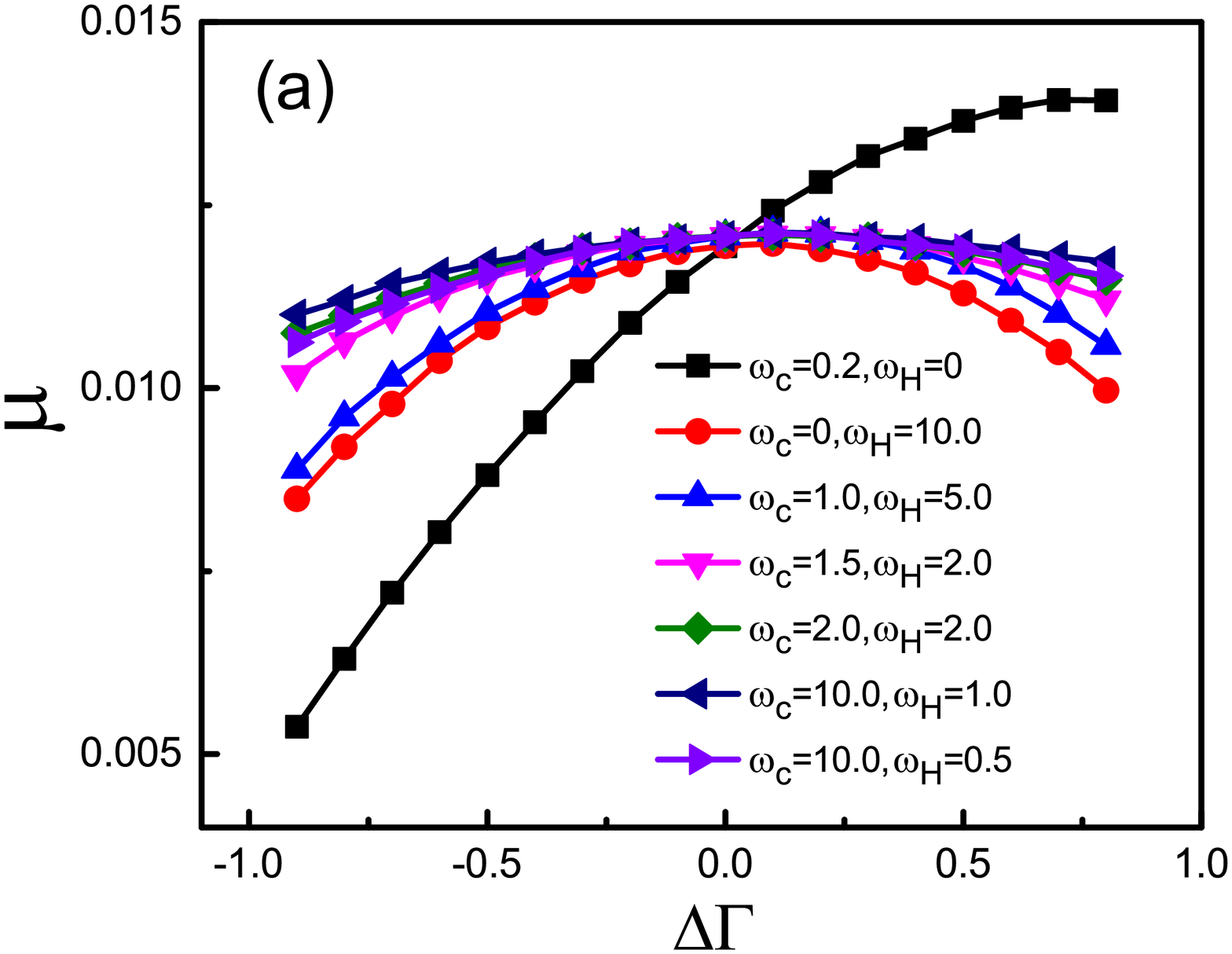}
  \includegraphics[width=0.45\columnwidth]{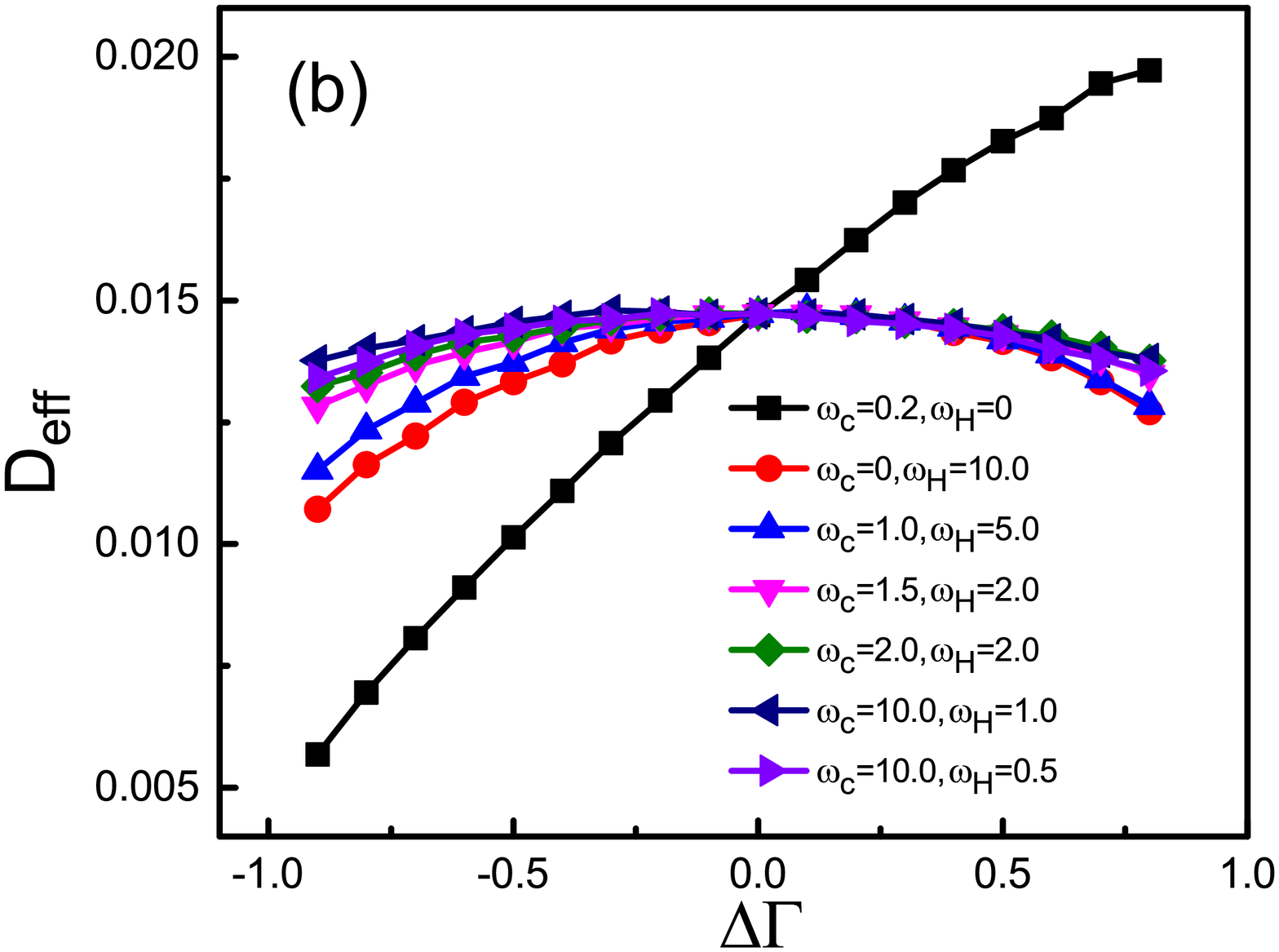}
  \caption{The mobility $\mu $ (a) and the effective diffusion coefficient $D_{eff}$ (b) as functions of the anisotropic parameter $\Delta \Gamma$ of passive particles for different values of $\omega _{c} $ and $\omega _{H}$.}
\end{figure*}

\begin{figure*}[htpb]
\vspace{1cm}
  \label{fig3}\includegraphics[width=0.45\columnwidth]{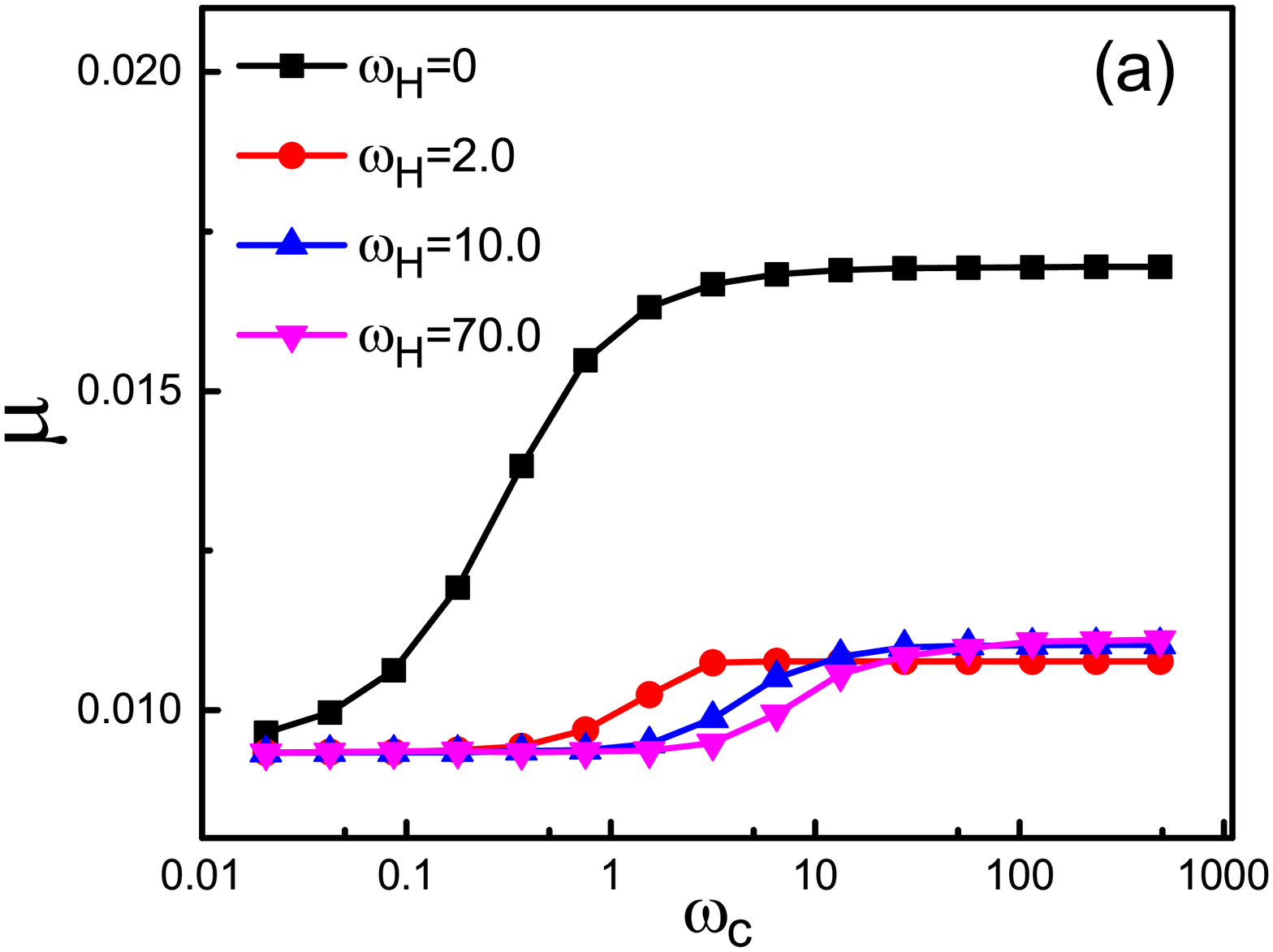}
  \includegraphics[width=0.45\columnwidth]{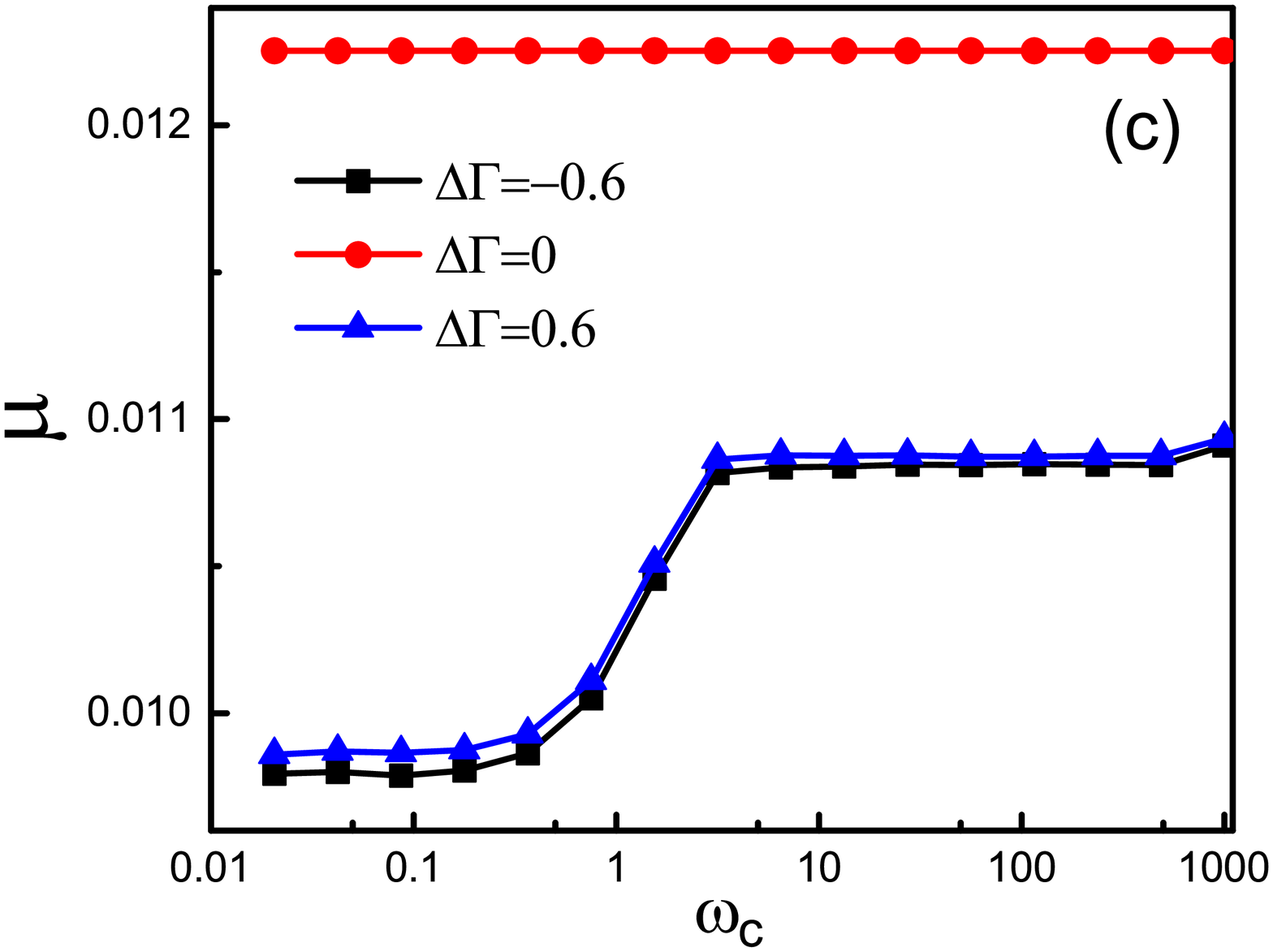}
  \includegraphics[width=0.45\columnwidth]{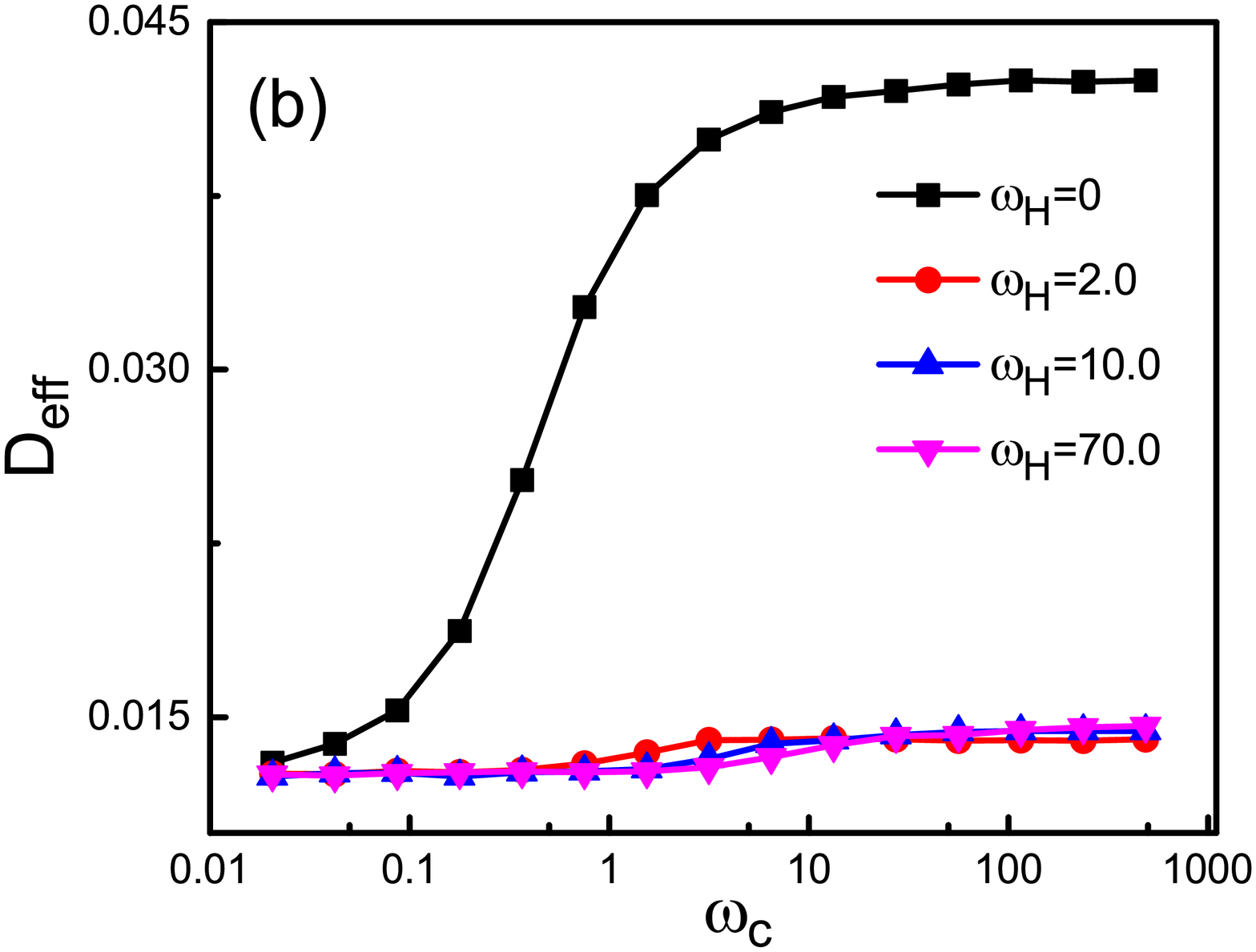}
  \includegraphics[width=0.45\columnwidth]{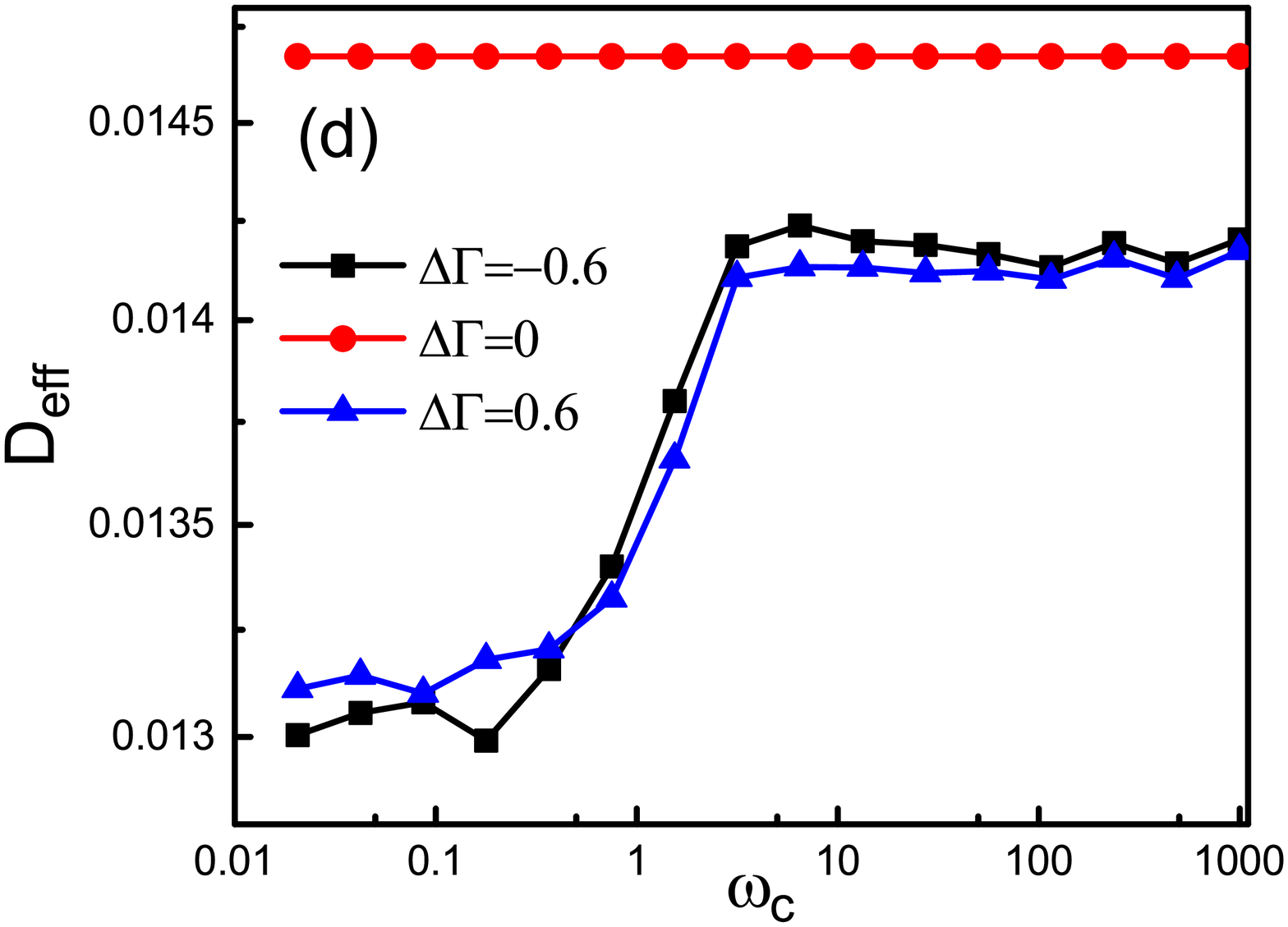}
  \caption{The mobility $\mu $ (a) and the effective diffusion coefficient $D_{eff}$ (b) as functions of the critical frequency $\omega _{c}$ of passive particles for different magnetic frequency $\omega _{H} $ at $\Delta \Gamma =0.7$. The mobility $\mu $ (c) and the effective diffusion coefficient $D_{eff}$ (d) as functions of the critical frequency $\omega _{c}$ of passive particles for different $\Delta \Gamma $ at $\omega _{H}=2.0$.}
\end{figure*}

In Fig. 8, we plot the mobility $\mu $ and the effective diffusion coefficient $D_{eff} $ as functions of the critical frequency $\omega _{c} $ of passive particles. It is found the mobility and the effective diffusion coefficient have also similar behaviors and increase as the increasing $\omega _{c} $ and then remain constant values. Namely, the frequency $\omega _{c} $ facilitates the mobility and the effective diffusion coefficient first, and plays little influence finally. As we know, for the case  $\gamma >1$, the ellipsoidal particle rotates synchronously with the magnetic field. For the case $\gamma <1$, the ellipsoidal swimmer performs a back-and-forth rotational motion which is considered as an asynchronous state. Especially, $\mu $ and $D_{eff} $ of the particles which are subject to a static field ($\omega _{H} =0$) are much larger than that of the particles which are subject to a rotating field ($\omega _{H} \ne 0$) from Figs. 8(a) and 8(b). In the other hand, $\mu $ and $D_{eff} $ only exhibit a little difference when changing the value of $\omega _{H} $. The position of reaching the maximum is the approximate value of $\omega _{c} $ at $\gamma \ge 1$ and shifts to large $\omega _{c} $ when $\omega _{H} $ increases. From Figs. 8(c) and 8(d), we can find $\mu $ and $D_{eff} $ remain constant values for isotropic particles ($\Delta \Gamma =0$), and are always larger than that of anisotropic particles which can be also demonstrated in Fig. 7.

\begin{figure*}[htpb]
\vspace{1cm}
  \label{fig4}\includegraphics[width=0.45\columnwidth]{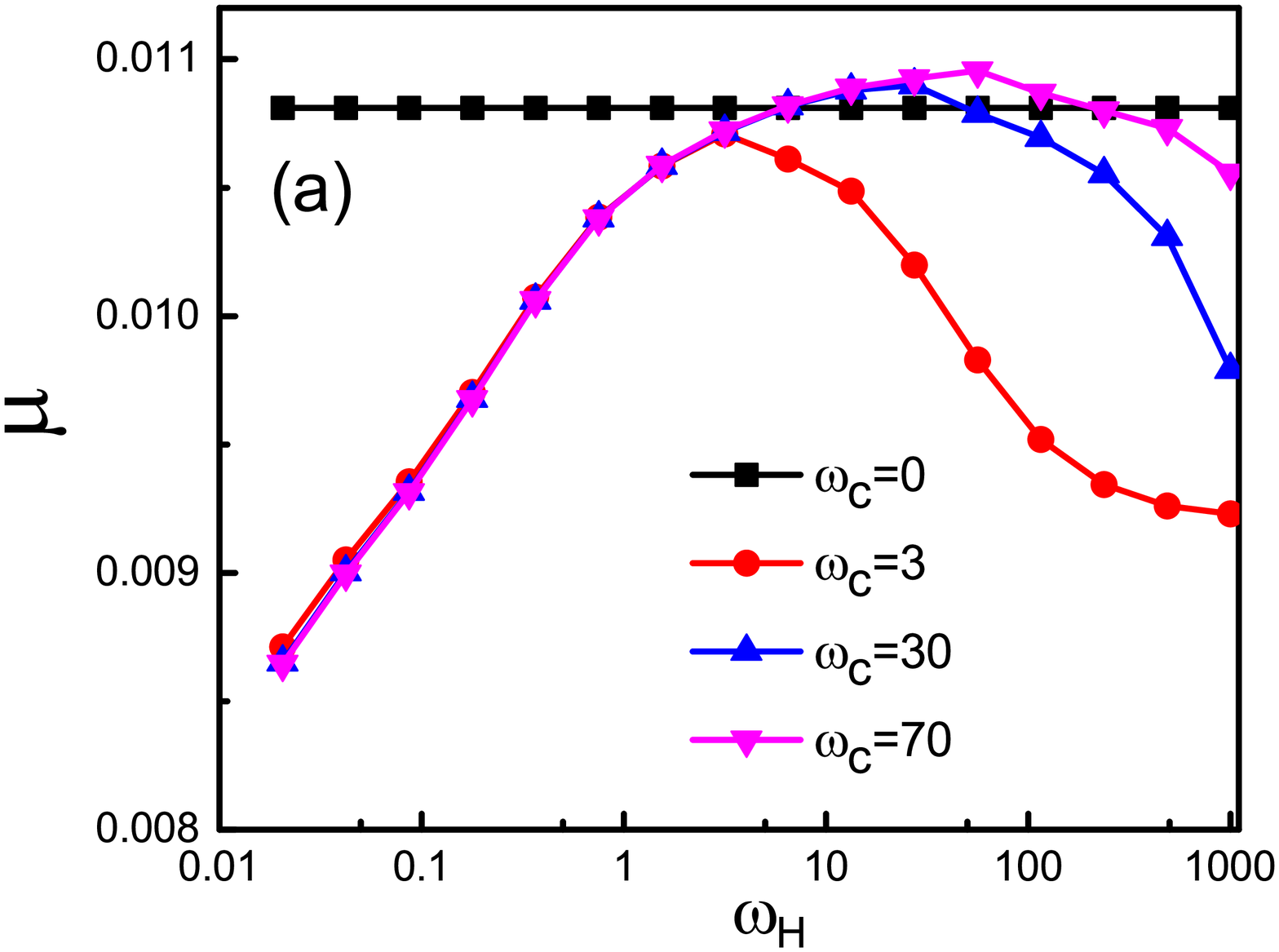}
  \includegraphics[width=0.45\columnwidth]{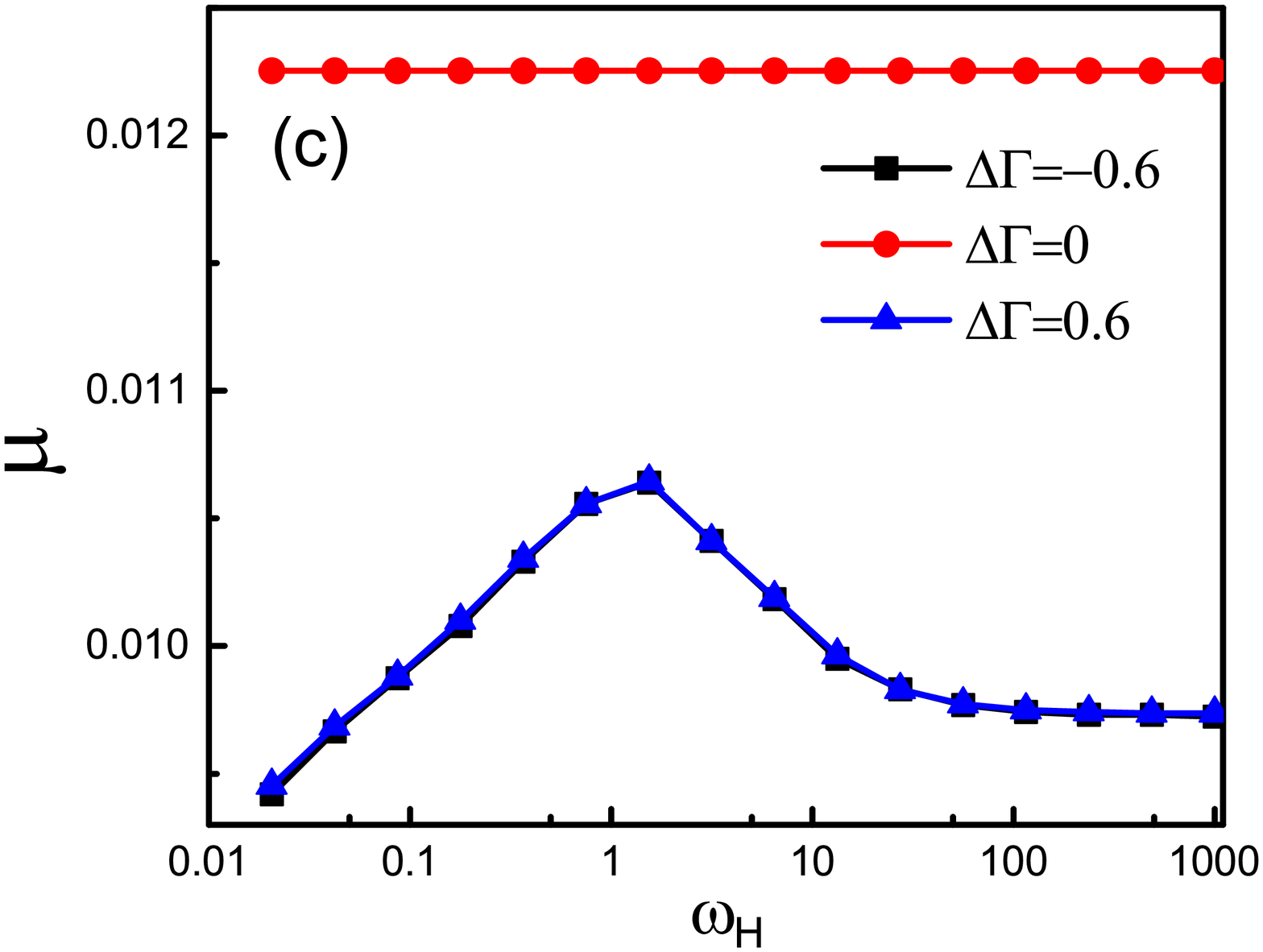}
  \includegraphics[width=0.45\columnwidth]{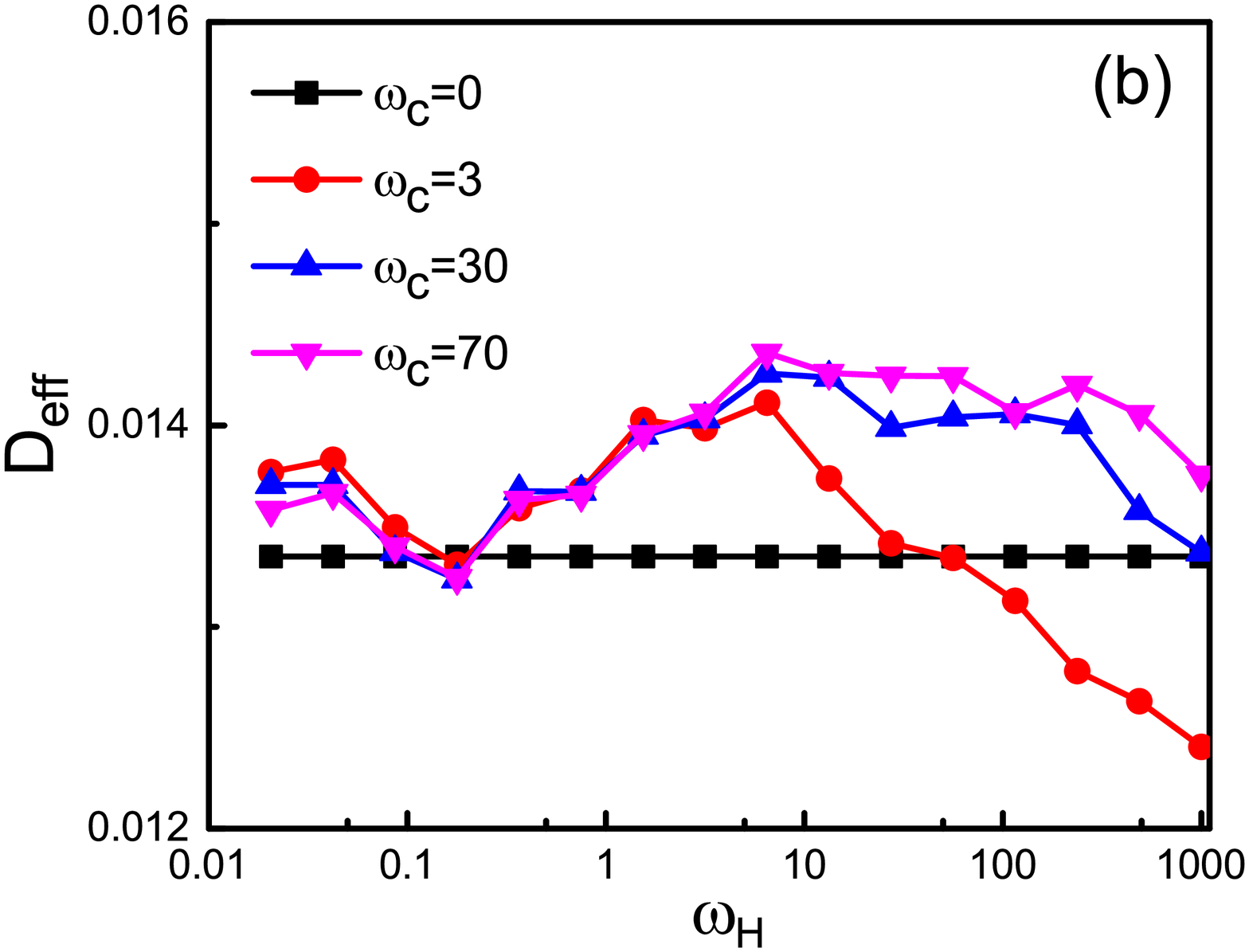}
\includegraphics[width=0.45\columnwidth]{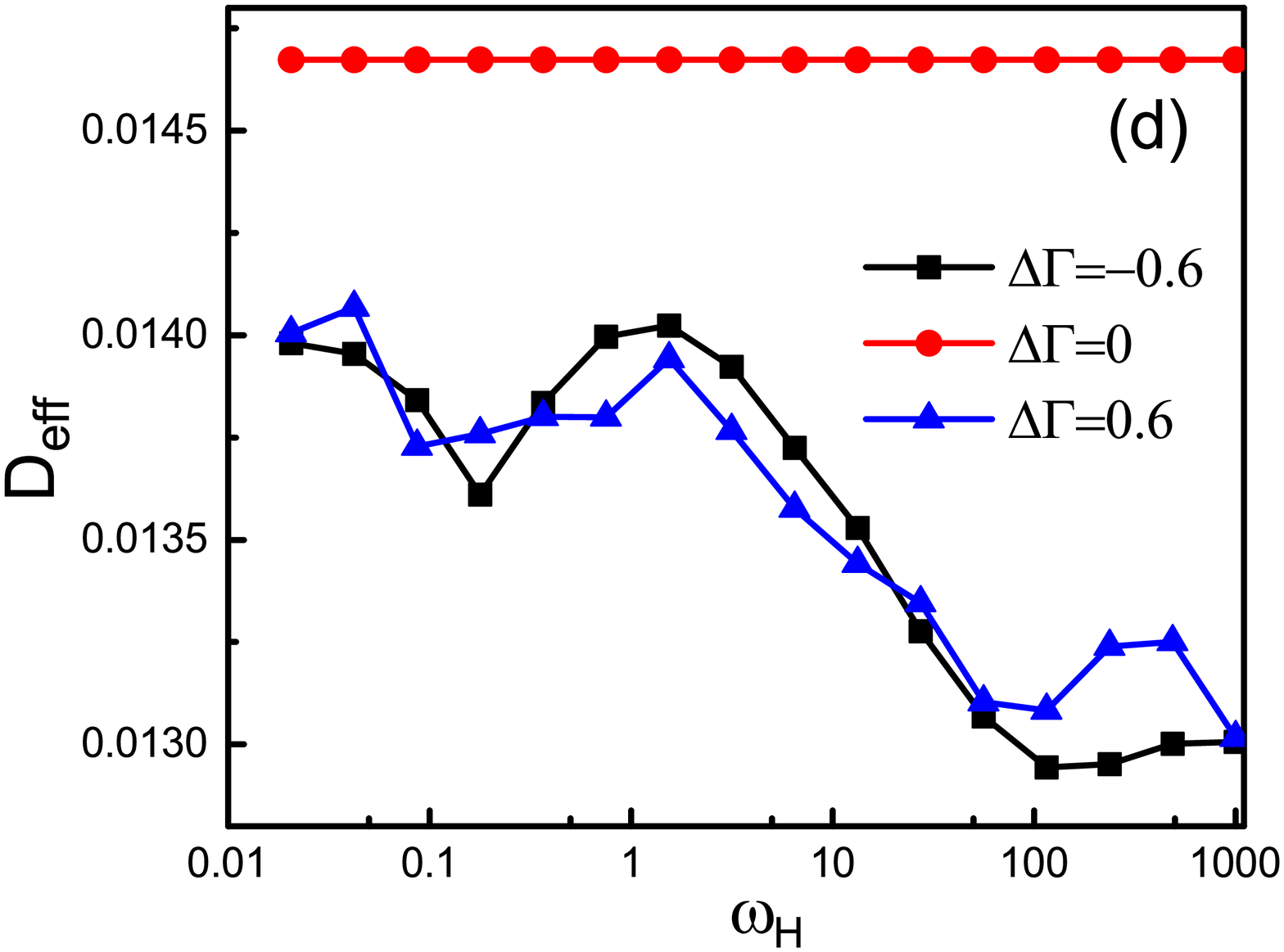}
  \caption{The mobility $\mu $ (a) and the effective diffusion coefficient $D_{eff}$ (b) as functions of the magnetic frequency $\omega _{H}$ of passive particles for different $\omega _{c}$ at $\Delta \Gamma =0.7$. The mobility $\mu $ (c) and the effective diffusion coefficient $D_{eff}$ (d) as functions of the magnetic frequency $\omega _{H}$ of passive particles for different $\Delta \Gamma $ at $\omega _{c} =2.0$.}
\end{figure*}

From Fig. 9, we can find the mobility $\mu $ and the effective diffusion coefficient $D_{eff} $ depend on the magnetic frequency $\omega _{H} $ of passive particles. When the particles are without an external field ($\omega _{c} =0$), $\mu $ and $D_{eff} $ remain unchanged. When the particles are subject to a rotating field, the rotational motion synchronous with the magnetic field for $\gamma >1$ enhances the mobility first, then the back-and-forth rotational motion for $\gamma <1$ reduces the mobility. Thus, there exists one optimal value of $\omega _{H} $ at which the mobility takes its maximal value. The position of the peak is nearly to $\omega _{H} \leq \omega _{c} $. However, the effective diffusion coefficient decreases first, then increases, and finally decreases to a constant value, shown in Fig. 9(b). There exist two peak values of $\omega _{H} $. From Figs. 9(c) and 9(d), we can also find $\mu $ and $D_{eff} $ remain constant values for isotropic particles ($\Delta \Gamma =0$), and are always larger than that of anisotropic particles which are similar to the above results.

\begin{figure*}[htpb]
\vspace{1cm}
  \label{fig5}\includegraphics[width=0.45\columnwidth]{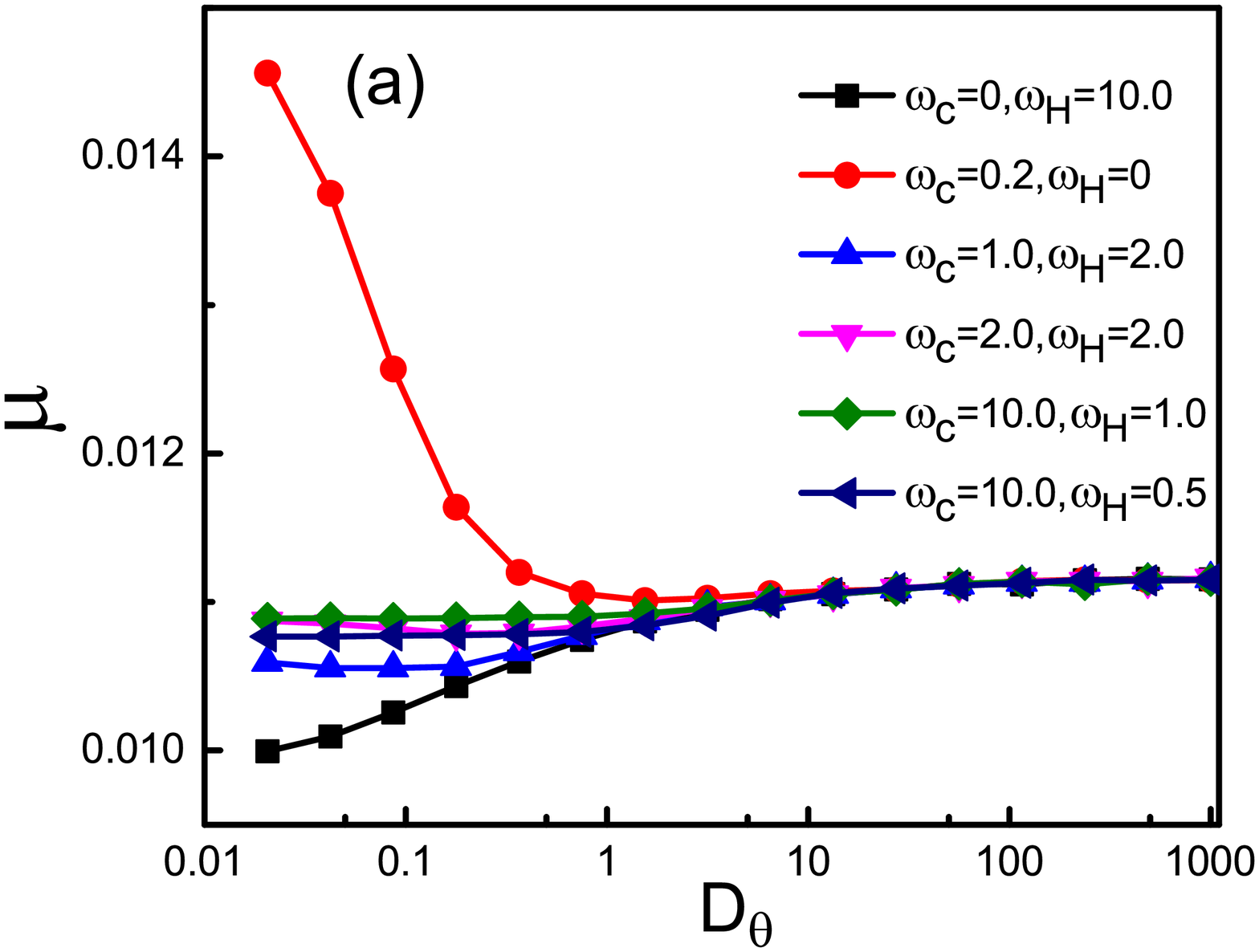}
  \includegraphics[width=0.45\columnwidth]{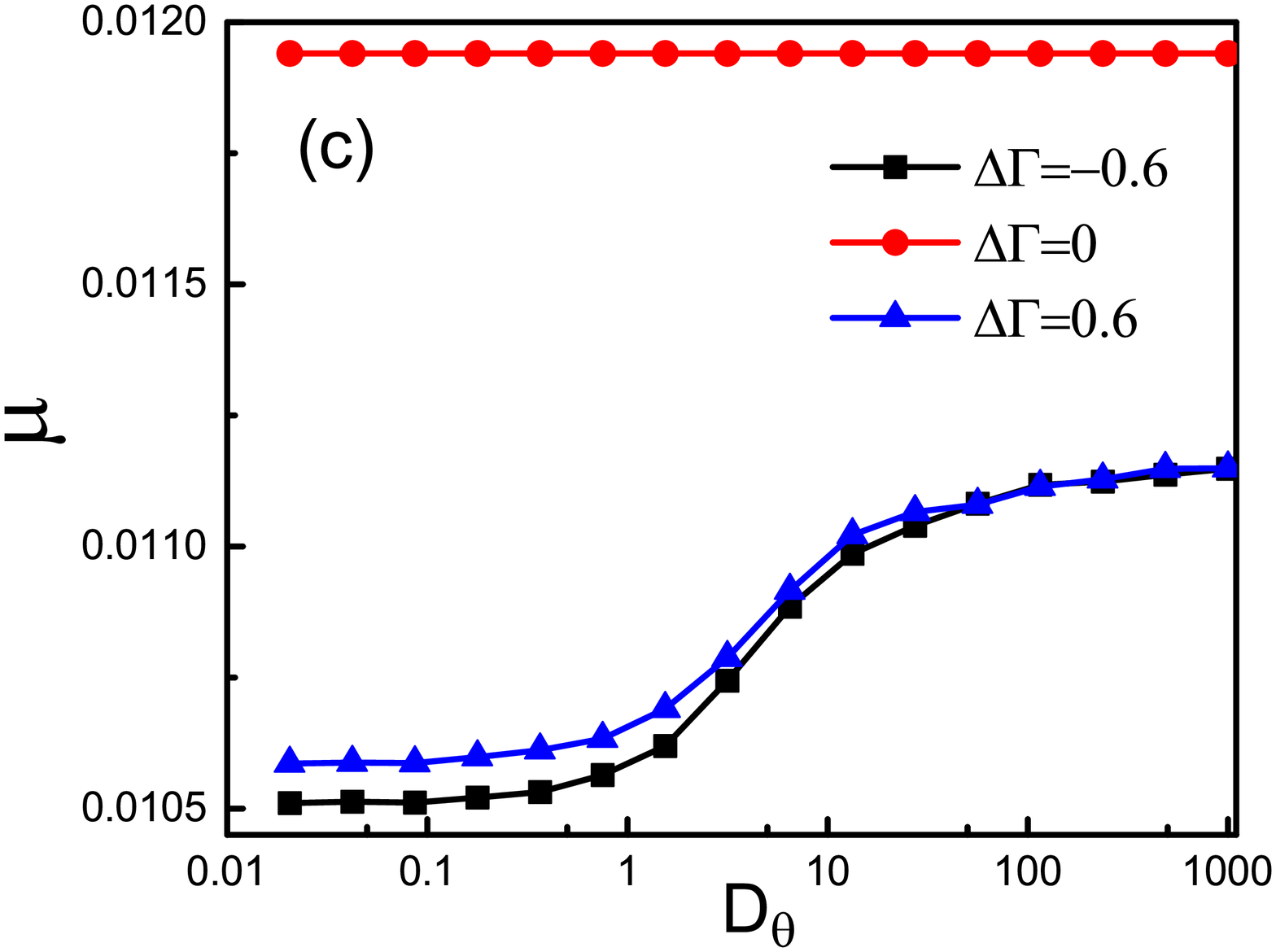}
\includegraphics[width=0.45\columnwidth]{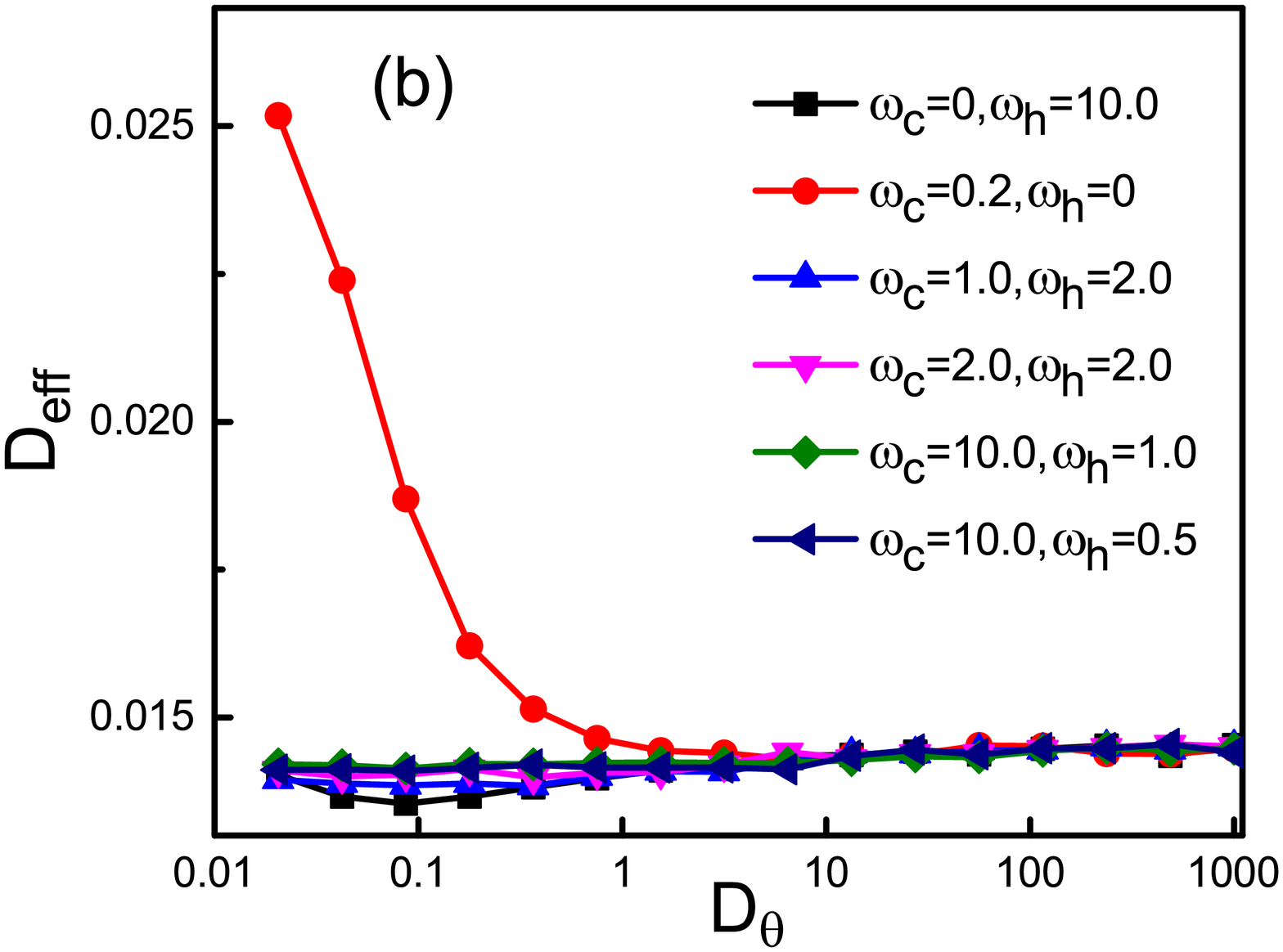}
\includegraphics[width=0.45\columnwidth]{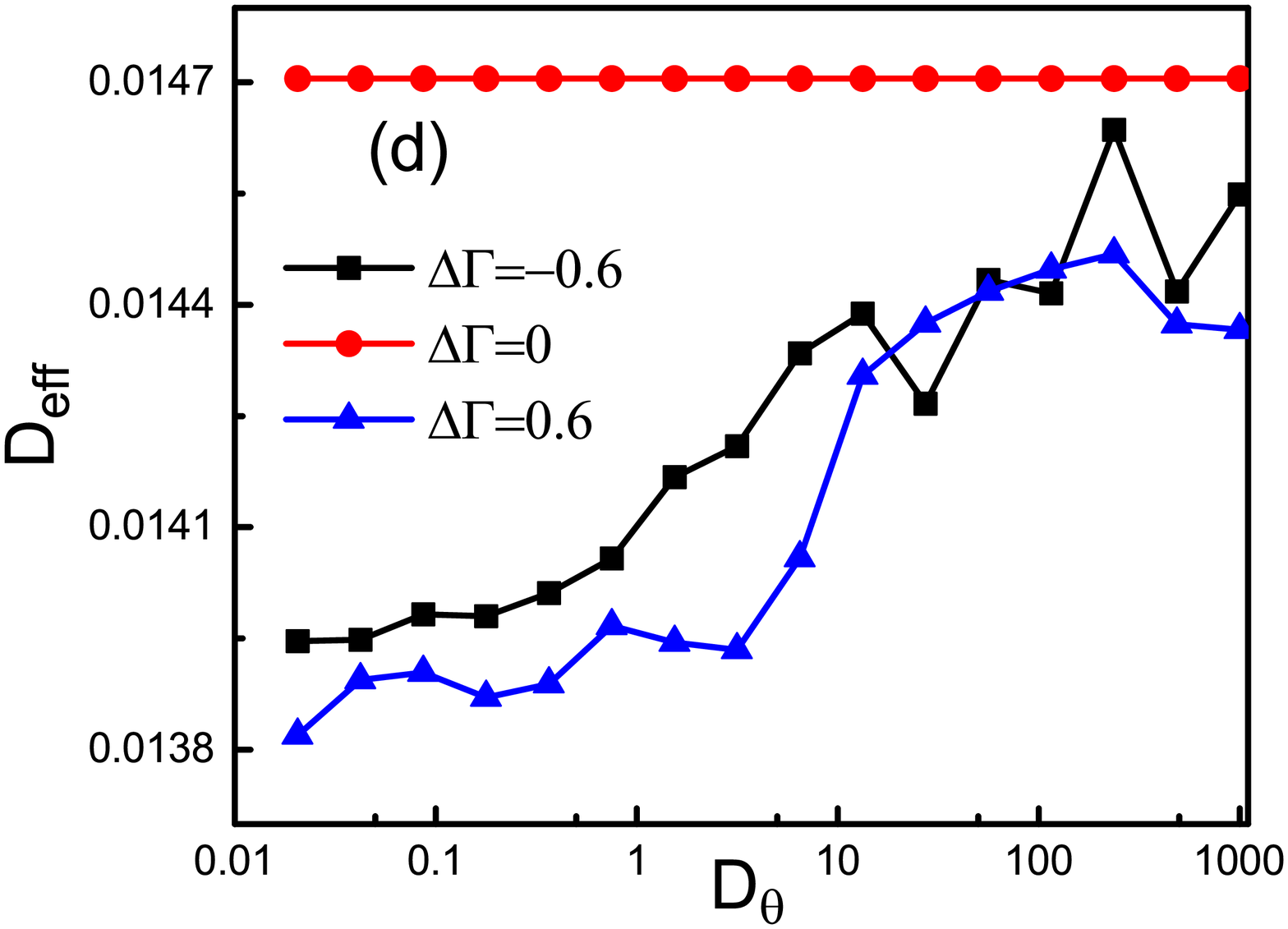}
  \caption{The mobility $\mu $ (a) and the effective diffusion coefficient $D_{eff}$ (b) as functions of the rotational diffusion $D_{\theta }$ of passive particles for different values of $\omega _{c} $ and $\omega _{H} $ at $\Delta \Gamma=0.5$. The mobility $\mu $ (c) and the effective diffusion coefficient $D_{eff}$ (d) as functions of the rotational diffusion $D_{\theta }$ of passive particles for different values of $\Delta \Gamma$ at $\omega _{c} =10.0$ and $\omega _{H} =0.5$.}
\end{figure*}
Figure 10 addresses the mobility $\mu $ and the effective diffusion coefficient $D_{eff} $ as functions of the rotational diffusion $D_{\theta } $ of passive particles. From Figs. 10(a) and 10(b), the behaviors of the mobility $\mu $ and the effective diffusion coefficient $D_{eff} $ for the case of a static magnetic field ($\omega _{H} =0,$ $\omega _{c} \ne 0$) are different from that for other cases. When the particles are subject to a static magnetic field, $\mu $ and $D_{eff} $ decrease with the increase of $D_{\theta } $. Because the rotational motion synchronous with the magnetic field enhances the mobility and the effective diffusion coefficient. When $D_{\theta } \to 0$, the external field plays a key role in the mobility and the effective diffusion coefficient, then make the value of $\mu $ and $D_{eff} $ reach the maximum. When $D_{\theta } $ becomes large, particles rotate faster, the effect of the external field can be reduced, so $\mu $ and $D_{eff} $ finally go to a constant. When the particles are without a magnetic field or subject to a rotating field, $\mu $ and $D_{eff} $ increase with the increase of $D_{\theta } $ and remain nearly unchanged when $\gamma $ is large. This is due to the competition between the external field and the rotational diffusion. From Figs. 10(c) and 10(d), we can find the mobility and the effective diffusion coefficient remain constant values for isotropic particles ($\Delta \Gamma =0$), and are always larger than that of anisotropic particles. When $D_{\theta } \to 0$, the particle maintains its direction for a very long time and it is dominated by the translational diffusion. When $D_{\theta } \to \infty $, the particle rotates very fast and the influence of the particle anisotropy disappears, thus the mobility and the effective diffusion coefficient of anisotropic particles exhibit the same behaviors as that of isotropic particles finally.

\begin{figure*}[htpb]
\vspace{1cm}
  \label{fig5}\includegraphics[width=0.45\columnwidth]{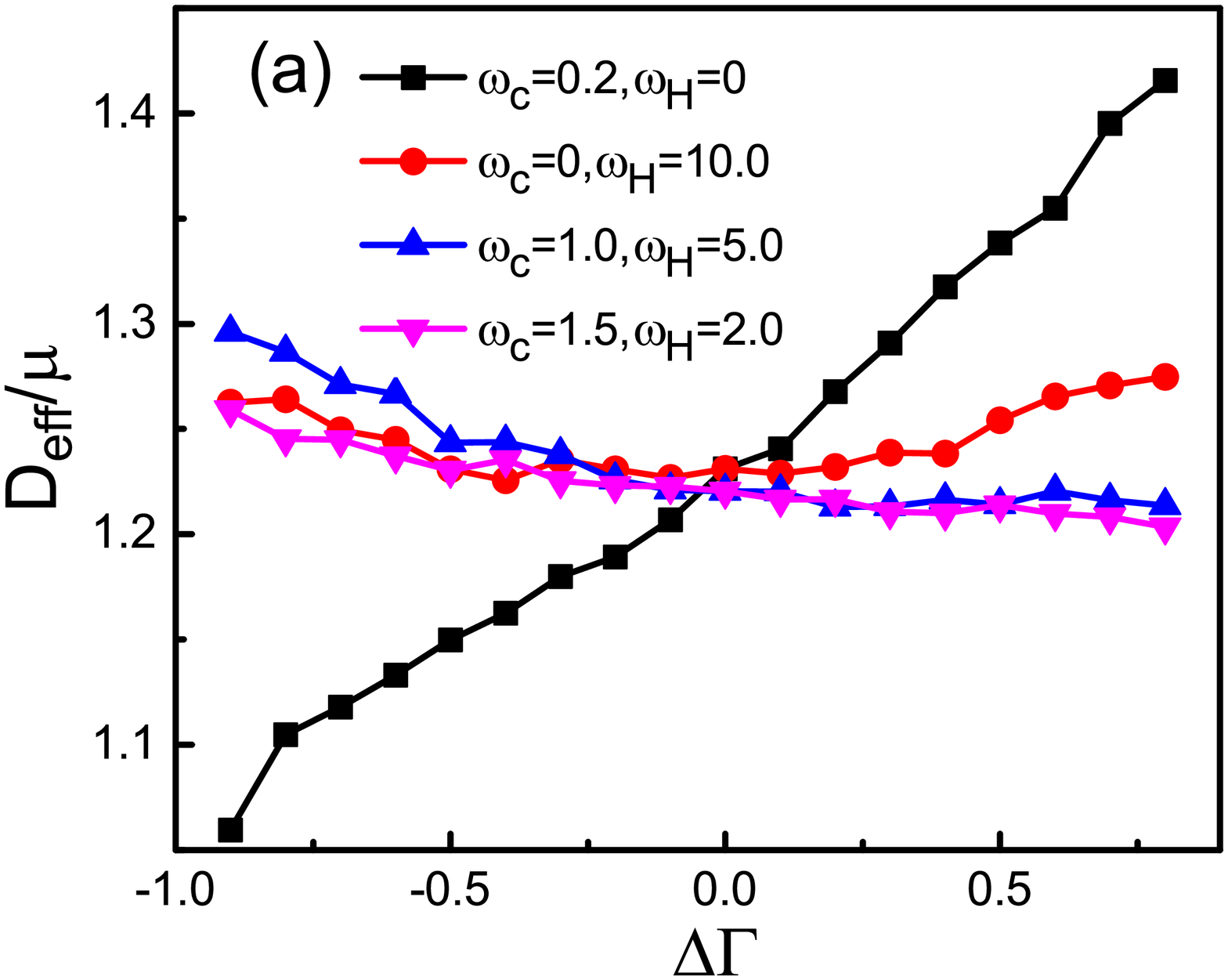}
  \includegraphics[width=0.45\columnwidth]{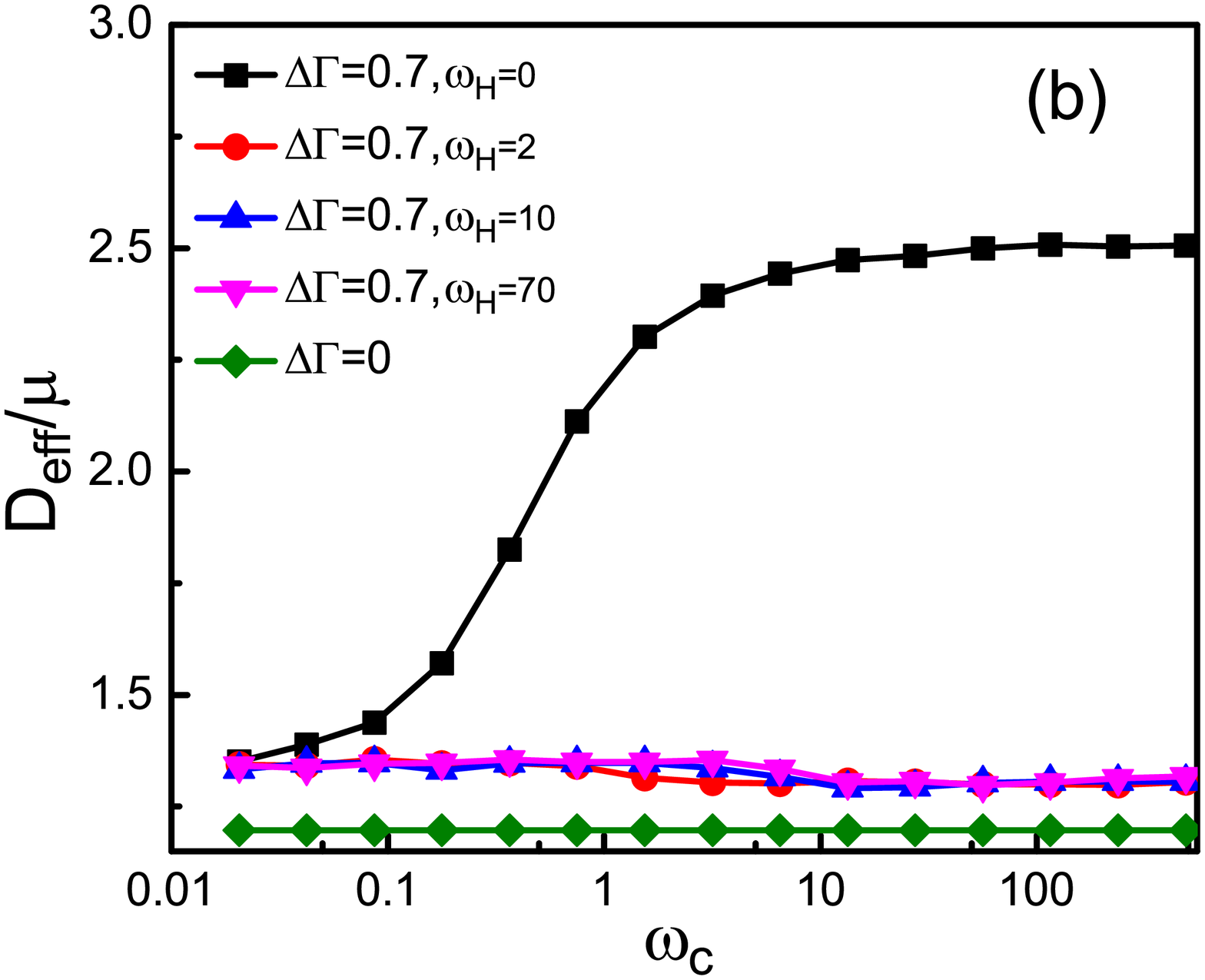}
  \includegraphics[width=0.45\columnwidth]{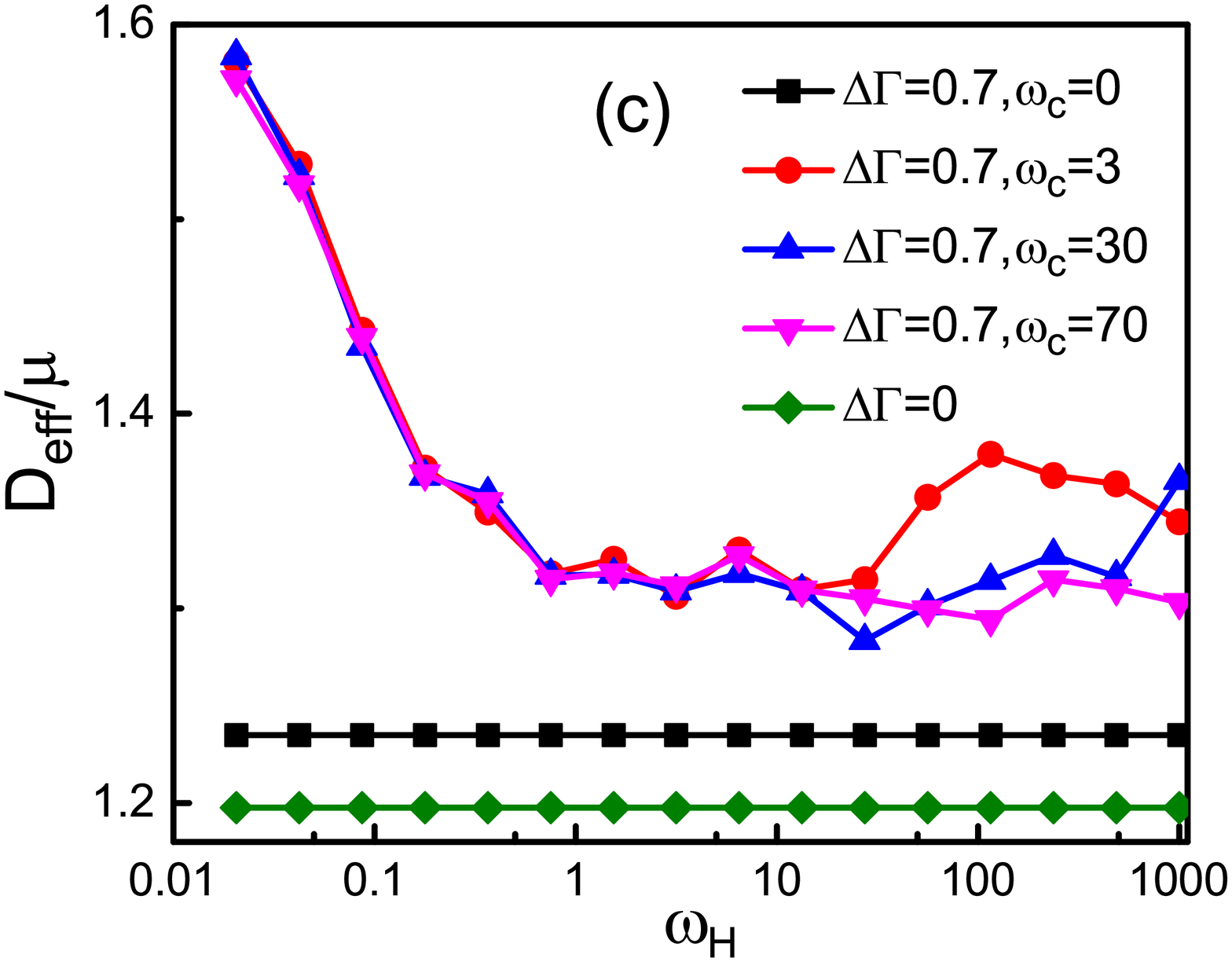}
  \includegraphics[width=0.45\columnwidth]{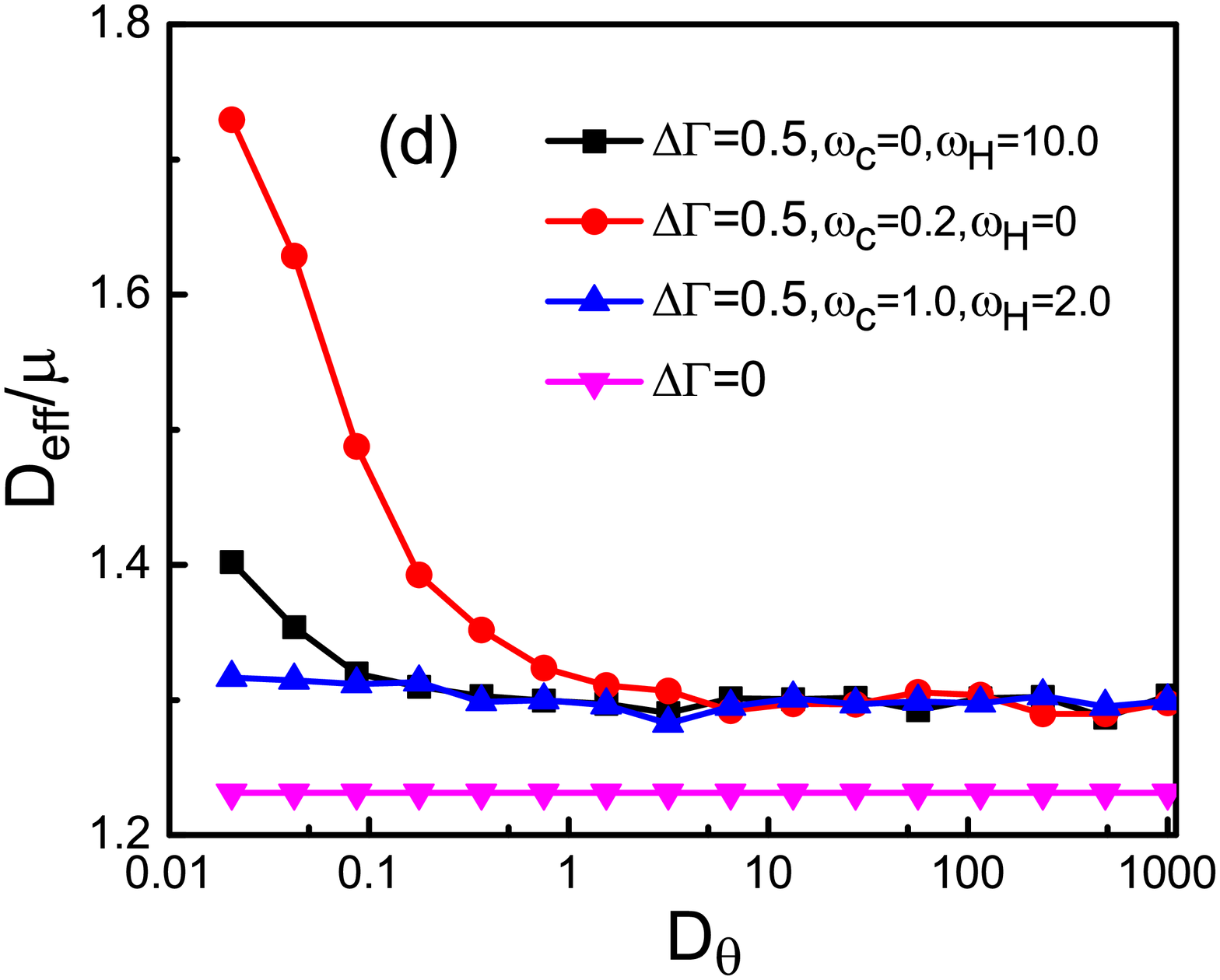}
  \includegraphics[width=0.45\columnwidth]{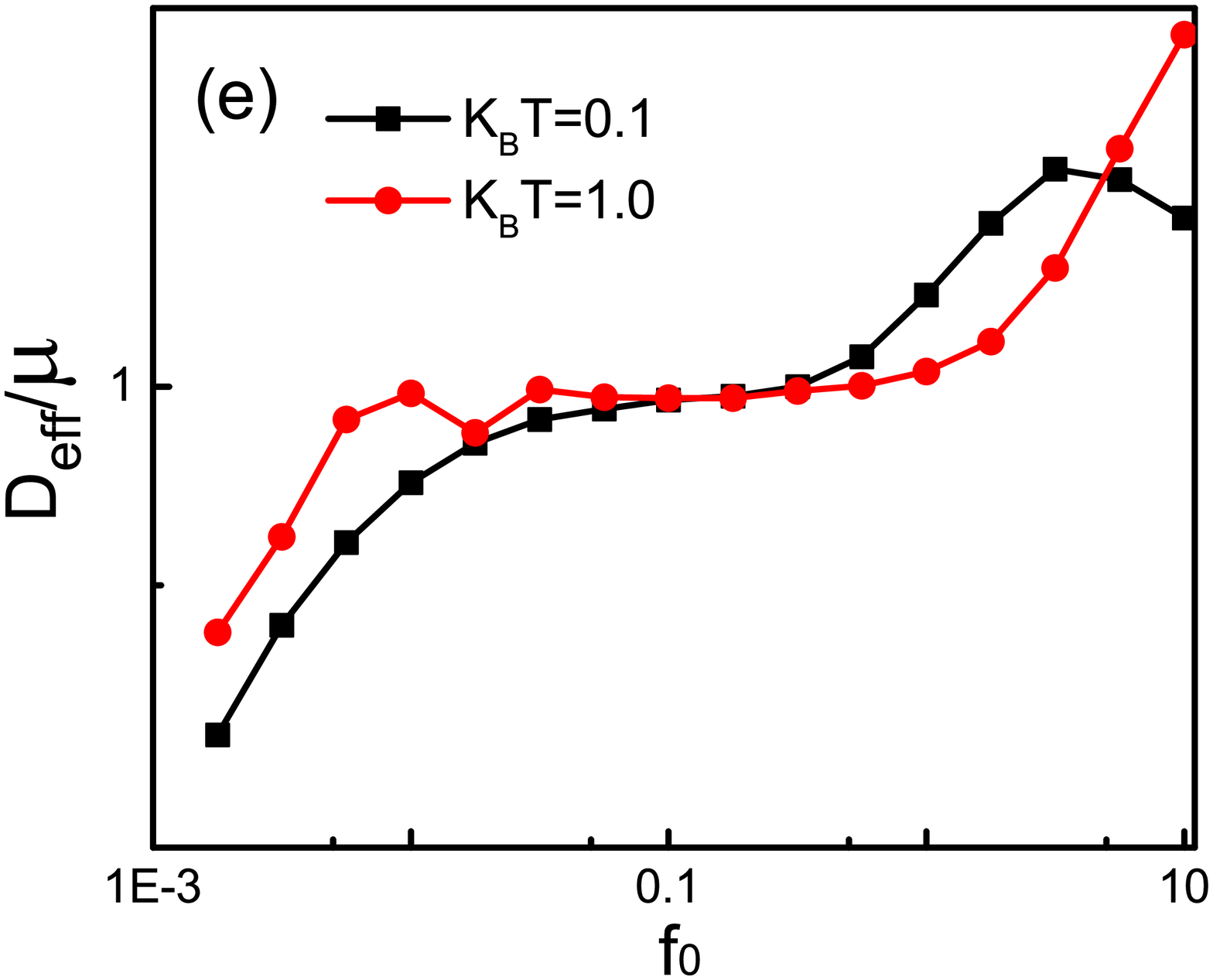}
  \includegraphics[width=0.45\columnwidth]{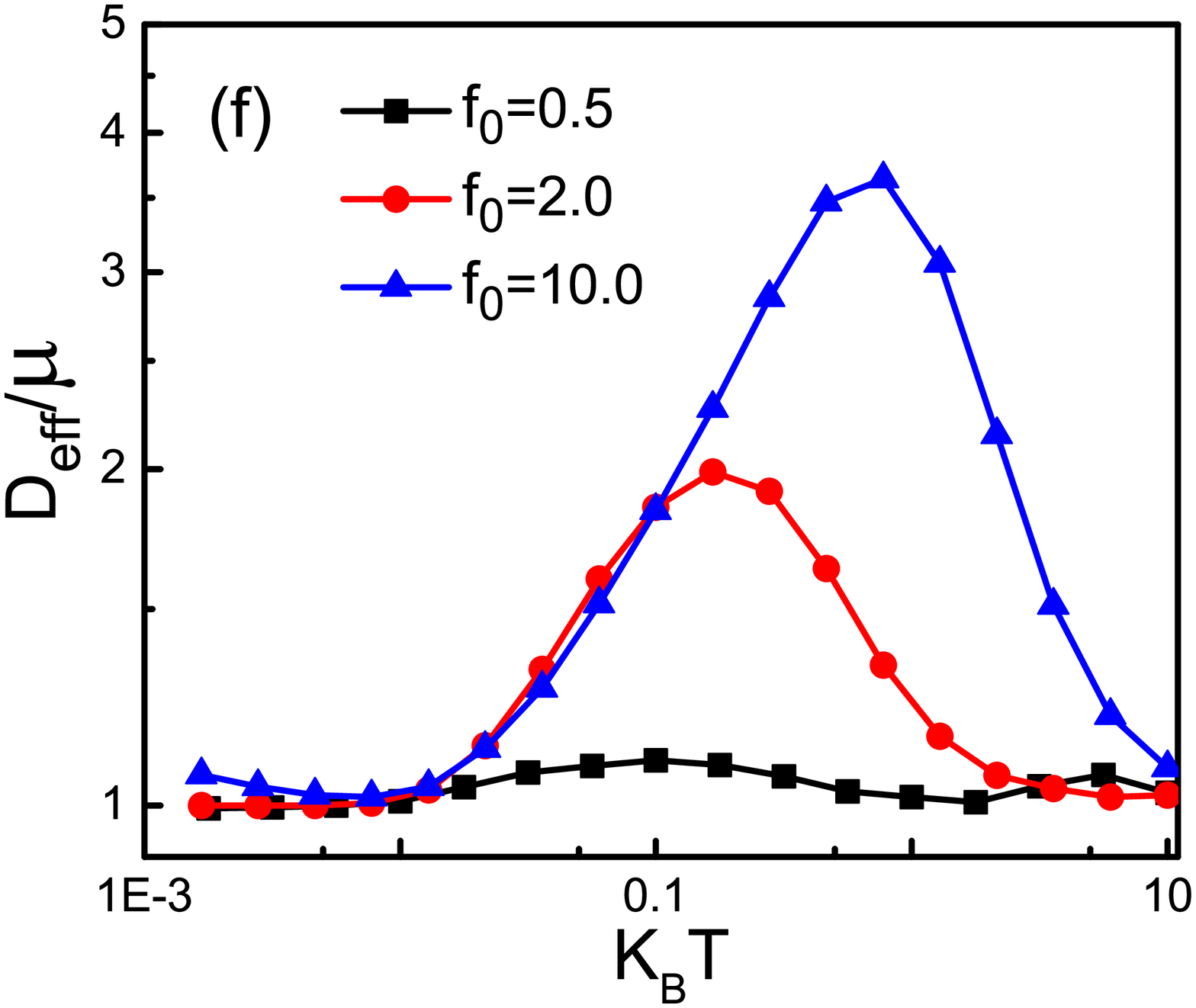}
  \caption{The ratio of the effective diffusion coefficient $D_{eff}$ to the mobility $\mu $ of passive particles (a) as a function of $\Delta \Gamma$ for different values of $\omega _{c} $ and $\omega _{H}$ at $k_B T=1.0$; (b) as a function of $\omega _{c}$ for different values of $\Delta \Gamma$ and $\omega _{H}$ at $k_B T=1.0$; (c) as a function of $\omega _{H}$ for different values of $\Delta \Gamma$ and $\omega _{c}$ at $k_B T=1.0$; (d) as a function of $D_\theta$ for different values of $\Delta \Gamma$, $\omega _{c}$ and $\omega _{H}$ at $k_B T=1.0$; (e) as a function of $f_0$ for different values of $k_B T$ at $\Delta \Gamma=0$; (f) as a function of $k_B T$ for different values of $f_0$ at $\Delta \Gamma=0$.}
\end{figure*}

By comparing with active particles, we can find that the mobility and the effective diffusion coefficient of passive particles have similar behaviors. In other words, the self-propelled velocity $v_{0} $ plays a key role in the difference between the transport and diffusive behavior. Additionally, the mobility and the effective diffusion coefficient of passive particles change a little when applying rotating magnetic fields of different frequencies. And the transport direction of passive particles can not be reversed by applying rotating magnetic fields.

Now we discuss the validity of the Einstein relation between diffusion and mobility in the case of passive particles. In equilibrium conditions, and in the linear regime, the Einstein relation predicts a proportionality between the diffusivity and the mobility, via the temperature \cite{ref51},
\begin{equation} \label{eq18}
D_x \equiv \mu k_B T.
\end{equation}
We plot the ratio of the effective diffusion coefficient $D_{eff}$ to the mobility $\mu$ of passive particles as a function of $\Delta \Gamma$, $\omega _{c} $, $\omega _{H}$, $D_\theta$, $f_0$, and $k_B T$ as shown in Figs. 11(a)-11(f), respectively. From Figs. 11(a)-11(e), it is found that the system is in the nonlinear regime, the ratio of $D_{eff}$ to $\mu$ is changing in most cases while the ratio remains constant in the cases of $\Delta \Gamma=0$ in Figs. 11(b)-11(d) and $\omega _{c}=0$ in Fig. 11(c). From Fig. 11(f), we can find when $f_0$ and $k_B T$ is not small, the ratio of $D_x$ to $\mu$ is not proportional to $k_B T$ and the system is in the nonlinear regime. However, the curves are near to be straight when $f_0$ and $k_B T$ is very small, the ratio of $D_x$ to $\mu$ is nearly proportional to $k_B T$, the effect of $f_0$ and $k_B T$ is negligible, the system can be considered at equilibrium and satisfies the Einstein relation Eq. (\ref{eq18}).

Finally, we discuss the possibility of realizing our model in experimental setups. Consider a system of paramagnetic ellipsoidal particles moving in a two-dimensional channel at room temperature. Paramagnetic ellipsoidal particles are obtained by adapting the method of Ref. \cite{ref42} to commercially available magnetite-doped un-cross-linked polystyrene microspheres (Micromod GmbH, Germany). The channel is asymmetric and chosen to be corrugated structure. A rotating magnetic field in the particle plane is achieved by connecting two custom-made coils perpendicular to each other with the main axis along the $(x, y)$ directions with a wave generator (TTi-TGA1244) feeding a power amplifier (IMG STA-800) \cite{ref19,ref20}. Due to the upper-lower asymmetry of the channel, paramagnetic ellipsoidal particles in a rotating magnetic field can produce the directed transport. To measure the motion of the paramagnetic ellipsoidal particles, we can image the particles by a high-speed camera, from which the average velocity and effective diffusion coefficient can be calculated. In the experimental setup, we can conveniently control the strength and frequency of rotating magnetic fields.

\section{Concluding Remarks}
In this paper, we have numerically studied transport and diffusion of paramagnetic ellipsoidal particles under the action of a rotating magnetic field in a two-dimensional channel. It is found that paramagnetic ellipsoidal particles in a rotating magnetic field can be rectified in the upper-lower asymmetric channel. The transport and diffusion are sensitively influenced by the external magnetic field. For active particles, the rectification and the effective diffusion coefficient are much more different and complicated. The back-and-forth rotational motion facilitates the effective diffusion coefficient and reduces the rectification while the rotational motion synchronous with the magnetic field suppresses the effective diffusion coefficient and enhances the rectification. Applying different magnetic fields (static or rotating) or without magnetic field, the rectification and the effective diffusion coefficient of active particles with different shapes, self-propulsion velocities, and rotational diffusion rates exhibit different behaviors. It is due to the competition among the self-propelled velocity $v_{0} $, the critical frequency $\omega _{c} $, the magnetic frequency $\omega _{H} $, the anisotropic parameter $\Delta \Gamma $, and the rotational diffusion rate $D_{\theta } $. There exist optimized values of the parameters (the anisotropic degree, the amplitude and frequency of magnetic field, the self-propelled velocity, and the rotational diffusion rate) at which the average velocity and diffusion take their maximal values. In addition, by applying rotating magnetic fields of suitable amplitude and frequency, three particle separation ways are presented: (1) shape separation: particles with ${\rm -}0.4\le \Delta \Gamma \le {\rm -}0.9$ move to the left, whereas other particles move to the right; (2) self-propelled velocity separation: particles with $2.0\le v_{0} \le 5.0$ move to the left, whereas others move to the right; (3) rotational diffusion rate separation: particles with $D_{\theta } \le 0.2$ move to the left, whereas others move to the right. Therefore, we can separate particles with different shapes, self-propelled speeds or rotational diffusion rates. For passive particles, the mobility and the effective diffusion coefficient have similar behaviors and change a little when applying rotating magnetic fields of different frequencies. The perfect sphere particle can facilitate the mobility and the effective diffusion coefficient, while the needlelike particle suppresses the mobility and the effective diffusion coefficient. The rotational motion synchronous with the magnetic field enhances the mobility and the effective diffusion coefficient, while the back-and-forth rotational motion reduces the mobility and the effective diffusion coefficient. There exists an optimal value of $\omega _{H} $ at which the mobility takes its maximal value. The position of the peak is nearly to $\omega _{H} \leq \omega _{c} $.

The results we have presented can characterize transport and diffusion of paramagnetic ellipsoidal particles in the previous experiment \cite{ref19,ref20}. It may open up the possibility of changing the frequency or strength of the applied magnetic field to separate particles, orient the particles along any direction at will during motion and control the particle diffusion in several applications, such as drug release and migration of contaminants in porous media. Furthermore, the anisotropic particles, like our paramagnetic ellipsoids, can be used as force sensors, microstirrers, active components in constrained geometries, microrheological probes or externally actuated micropropellers \cite{ref13}. Our analysis can be easily extended to more complex situations in the future, such as in the underdamped regime and consideration of particle interactions.

\section*{Acknowledgments}
This work was supported in part by the National Natural Science Foundation of
China (Grants No. 11575064, No. 61762046 and No. 11704164), the Natural Science Foundation of Guangdong Province, China (Grants
No. 2014A030313426 and No. 2017A030313029), the Natural Science Foundation of Jiangxi Province,
 China (Grants No. GJJ161580 and No. GJJ160624) , the Innovation Project of Graduate School of South China Normal University, and the Foreign Joint Training Program for PhD of South China Normal University.


\begin{thebibliography}{}
	\bibitem{ref1} X. Ao, P. K. Ghosh, Y. Li, G. Schmid, P. H\"{a}nggi, and F. Marchesoni, Europhysics Letters \textbf{109}, 10003 (2015).
	\bibitem{ref2} P. S. Burada, P. H\"{a}nggi , F. Marchesoni , G. Schmid , and P. Talkner , Chemphyschem \textbf{10}, 45 (2009).
	\bibitem{ref3} P. K. Ghosh, P. H\"{a}nggi, F. Marchesoni, and F. Nori, Phys. Rev. E \textbf{89}, 062115 (2014).
	\bibitem{ref4} A. E. Antipov, A. V. Barzykin, A. M. Berezhkovskii, Y. A. Makhnovskii, V. Y. Zitserman, and S. M. Aldoshin, Phys. Rev. E \textbf{88}, 054101 (2013).
	\bibitem{ref5} X. Wang, and G. Drazer, Physics of Fluids \textbf{21}, 102002 (2009).
	\bibitem{ref6} Y. Li, P. K. Ghosh, F. Marchesoni , and B. Li, Phys. Rev. E \textbf{90}, 062301 (2014).
	\bibitem{ref7} B. Q. Ai, and L. G. Liu, Phys. Rev. E \textbf{74}, 051114 (2006).
	\bibitem{ref8} K. Lindenberg, J. M. Sancho, A. M. Lacasta, and I. M. Sokolov, Physical Review Letters \textbf{98}, 020602 (2007).
	\bibitem{ref9} P. Reimann, and R. Eichhorn, Phys. Rev. Lett. \textbf{101}, 180601 (2008).
	\bibitem{ref10} M. Khoury, A. M. Lacasta, J. M. Sancho, and K. Lindenberg, Phys. Rev. Lett. \textbf{106}, 090602 (2011).
	\bibitem{ref11} P. Tierno, P. Reimann, T. H. Johansen, and F. Sagu\'{e}s, Phys. Rev. Lett. \textbf{105}, 230602 (2010).
	\bibitem{ref12} R. Dreyfus, J. Baudry, M. L. Roper, M. Fermigier, H. A. Stone, and J. Bibette, Nature \textbf{437}, 862 (2005).
	\bibitem{ref13} P. Tierno, Physical chemistry chemical physics \textbf{16}, 23515 (2014).
	\bibitem{refAC} A. C\={e}bers and M. Ozols, Phys. Rev. E \textbf{73}, 021505 (2006).
    \bibitem{ref14} Nicolas Waisbord, C. Lef\`{e}vre, Lyd\'{e}ric Bocquet, Christophe Ybert, and C\'{e}cile Cottin-Bizonne, arXiv:1603.00490  (2016).
	\bibitem{ref15} D. M. Fanlong Meng, and Ramin Golestanian, arXiv:1710.08339  (2017).
	\bibitem{ref16} T. Chen, X. B. Wang, and T. Yu, Phys. Rev. E \textbf{90}, 022101 (2014).
    \bibitem{ref38} R. M. Erb, J. J. Martin, R. Soheilian, C. Pan, and J. R. Barber, Advanced Functional Materials \textbf{26}, 3859 (2016).
	\bibitem{ref22} A. Snezhko, M. Belkin, I. S. Aranson, and W. K. Kwok, Phys. Rev. Lett. \textbf{102}, 118103 (2009).
	\bibitem{ref23} A. Snezhko, Journal of Physics Condensed Matter An Institute of Physics Journal \textbf{23}, 153101 (2011).
	\bibitem{ref24} A. Ghanbari, M. Bahrami, and M. R. H. Nobari, Physical Review E \textbf{83}, 046301 (2011).
	\bibitem{ref25} P. J. Vach, D. Walker, P. Fischer, P. Fratzl, and D. Faivre, Journal of Physics D Applied Physics \textbf{50} 11LT03 (2017).
	\bibitem{ref26} S. Babel, H. L\"{o}wen, and A. M. Menzel, Europhysics Letters \textbf{113}, 58003 (2016).
	\bibitem{ref27} F. Meshkati, and H. C. Fu, Phys. Rev. E \textbf{90}, 063006 (2014).
	\bibitem{ref28} I. S. M. Khalil, A. F. Tabak, A. Klingner, and M. Sitti, Applied Physics Letters \textbf{109}, 033701 (2016).
    \bibitem{ref18} J. E. Martin, Phys. Rev. E \textbf{79}, 011503 (2009).
	\bibitem{ref29} R. Marino, R. Eichhorn, and E. Aurell, Phys. Rev. E \textbf{93}, 012132 (2016).
	\bibitem{ref30} O. G\"{u}ell, P. Tierno, and F. Sagu{\'e}s, The European Physical Journal Special Topics \textbf{187}, 15 (2010).
	\bibitem{ref31} Wai-TongLouis Fan, O. S. Pak, and M. Sandoval, Physical Review E \textbf{95}, 032605 (2017).
    \bibitem{ref39} D. Matsunaga, F. Meng, A. Zottl, R. Golestanian, and J. M. Yeomans, Phys. Rev. Lett. \textbf{119}, 198002 (2017).
	\bibitem{ref17} F. Liu, L. Jiang, H. M. Tan, A. Yadav, P. Biswas, J. R. van der Maarel, C. A. Nijhuis, and J. A. van Kan, Biomicrofluidics \textbf{10}, 064105 (2016).
    \bibitem{ref21} W. Gao, D. Kagan, O. S. Pak, C. Clawson, S. Campuzano, E. Chuluunerdene, E. Shipton, E. E. Fullerton, L. Zhang, and E. Lauga, Small \textbf{8}, 460 (2012).
    \bibitem{ref40} J. K. Hamilton, P. G. Petrov, C. P. Winlove, A. D. Gilbert, M. T. Bryan, and F. Y. Ogrin, Sci. Rep. \textbf{7}, 44142 (2017).
    \bibitem{ref32} T. Petit, L. Zhang, K. E. Peyer, B. E. Kratochvil, and B. J. Nelson, Nano Letters \textbf{12}, 156 (2012).
	\bibitem{ref33} P. Fischer, and A. Ghosh, Nanoscale \textbf{3}, 557 (2011).
	\bibitem{ref34} O. S. Pak, W. Gao, J. Wang, and E. Lauga, Soft Matter \textbf{7}, 8169 (2011).
    \bibitem{ref19} P. Tierno, J. Claret, F. Sagues, and A. C\={e}bers, Phys. Rev. E \textbf{79}, 021501 (2009).
	\bibitem{ref20} P. Tierno, R. Albalat, and F. Sagues, Small \textbf{6}, 1749 (2010).
	\bibitem{ref35} B. ten Hagen, S. van Teeffelen, and H. L\"{o}wen, J. Phys. Condens. Matter \textbf{23}, 194119 (2011).
	\bibitem{ref36} R. Grima, and S. N. Yaliraki, J. Chem. Phys. \textbf{127}, 084511 (2007).
	\bibitem{ref37} Y. Han, A. M. Alsayed, M. Nobili, J. Zhang, T. C. Lubensky, and A. G. Yodh, Science \textbf{314}, 626 (2006).
    \bibitem{ref41} S. Denisov, P. H\"{a}nggi, and J. L. Mateos, American Journal of Physics \textbf{77}, 602 (2009).
    \bibitem{ref43} O. B{\'e}nichou, P. Illien, G. Oshanin, A. Sarracino, and R. Voituriez, Phys. Rev. Lett. \textbf{113}, 268002 (2014)
    \bibitem{ref44} O. B{\'e}nichou, P. Illien, G. Oshanin, A. Sarracino, and R. Voituriez, Phys. Rev. E \textbf{93}, 032128 (2016)
    \bibitem{ref45} S. Leitmann and T. Franosch, Phys. Rev. Lett. \textbf{118}, 018001 (2017)
    \bibitem{ref46} P. Illien, O. B{\'e}nichou, G. Oshanin, A. Sarracino, and R. Voituriez, Phys. Rev. Lett. \textbf{120}, 200606 (2018)
    \bibitem{ref47} A. Sarracino, F. Cecconi, A. Puglisi, and A. Vulpiani, Phys. Rev. Lett. \textbf{117}, 174501 (2016)
    \bibitem{ref48} F. Cecconi, A. Puglisi, A. Sarracino, and A. Vulpiani, Eur. Phys. J. E \textbf{40}, 81 (2017)
    \bibitem{ref49} C. Reichhardt and C. J. O. Reichhardt, J. Phys. : Condens. Matter \textbf{30}, 015404 (2017)
    \bibitem{ref51} A. Puglisi, A. Sarracino, and A. Vulpiani, Physics Reports \textbf{709-710}, 1-60 (2017)
    \bibitem{ref42} J. A. Champion, Y. K. Katare, and S. Mitragotri, Proc. Natl. Acad. Sci. U.S.A. \textbf{104}, 11901 (2007).





\end{thebibliography}
\end{document}